\documentclass[prx,aps,twocolumn,showpacs,superscriptaddress,nofootinbib,longbibliography]{revtex4-1}
\usepackage{graphicx}
\usepackage{dcolumn}
\usepackage{bm}
\usepackage[bbgreekl]{mathbbol}
\usepackage{amsmath}
\usepackage{color}
\usepackage{textgreek}
\usepackage{amsfonts}
\def\ii{{\rm i}}  
\def\GG{{\bf G}}
  
\def\gb{{\bf g}} 
\def\db{\boldsymbol{\wp}}  
\def\dbu{\hat{\boldsymbol{\wp}}}  
\def\Eb{{\bf E}}  
\def\rb{{\bf r}}

\def\pb{{\bf p}}

\def\hge{\hat{\sigma}_{ge}}  
\def\heg{\hat{\sigma}_{eg}}

\def\kg{k_{\rm 1D}}
\def\ga{\Gamma_{\rm 1D}}

\def\bra#1{\mathinner{\langle{#1}|}}
\def\ket#1{\mathinner{|{#1}\rangle}}
\def\braket#1{\mathinner{\langle{#1}\rangle}}
\def\kb{\textbf{k}}

\def\qb{\textbf{q}}
\newcommand{\Braket}[2]{\langle{#1}|{#2}\rangle}
 
\def\Qb{\textbf{Q}} 
\def\Li{\textrm{Li}}
\def\ln{\textrm{ln}}
\def\pl{{||}}

\begin{document}
\title{Exponential improvement in photon storage fidelities using subradiance and ``selective radiance'' in atomic arrays}
\author{A. Asenjo-Garcia}
\email{ana.asenjo@caltech.edu}
\affiliation{Norman Bridge Laboratory of Physics MC12-33, California Institute of Technology, Pasadena, CA 91125, USA}
\affiliation{Institute for Quantum Information and Matter, California Institute of Technology, Pasadena, CA 91125, USA}
\author{M. Moreno-Cardoner}
\author{A. Albrecht}
\affiliation{ICFO-Institut de Ciencies Fotoniques, The Barcelona Institute of Science and Technology, 08860 Castelldefels, Barcelona, Spain}
\author{H. J. Kimble}
\affiliation{Norman Bridge Laboratory of Physics MC12-33, California Institute of Technology, Pasadena, CA 91125, USA}
\author{D. E. Chang}
\affiliation{ICFO-Institut de Ciencies Fotoniques, The Barcelona Institute of Science and Technology, 08860 Castelldefels, Barcelona, Spain}

\date{\today}
\begin{abstract}
A central goal within quantum optics is to realize efficient, controlled interactions between photons and atomic media. A fundamental limit in nearly all applications based on such systems arises from spontaneous emission, in which photons are absorbed by atoms and then re-scattered into undesired channels. In typical theoretical treatments of atomic ensembles, it is assumed that this re-scattering occurs independently, and at a rate given by a single isolated atom, which in turn gives rise to standard limits of fidelity in applications such as quantum memories for light or photonic quantum gates. However, this assumption can in fact be dramatically violated. In particular, it has long been known that spontaneous emission of a collective atomic excitation can be significantly suppressed through strong interference in emission between atoms. While this concept of ``subradiance" is not new, thus far the techniques to exploit the effect have not been well-understood. In this work, we provide a comprehensive treatment of this problem. First, we show that in ordered atomic arrays in free space, subradiant states acquire an elegant interpretation in terms of optical modes that are guided by the array, which only emit due to scattering from the ends of the finite system. We also go beyond the typically studied regime of a single atomic excitation, and elucidate the properties of subradiant states in the many-excitation limit. Finally, we introduce the new concept of ``selective radiance." Whereas subradiant states experience a reduced coupling to all optical modes, selectively radiant states are tailored to simultaneously radiate efficiently into a desired channel while scattering into undesired channels is suppressed, thus enabling an enhanced atom-light interface. We show that these states naturally appear in chains of atoms coupled to nanophotonic structures, and we analyze the performance of photon storage exploiting such states. We find numerically that selectively radiant states allow for a photon storage error that scales exponentially better with number of atoms than previously known bounds.
\end{abstract}
\maketitle

\section{Introduction}
The ability to achieve controlled, deterministic interactions between photons and atomic media constitutes an important resource in applications ranging from quantum information processing to metrology. As single photons and atoms typically do not interact efficiently, a common approach has been to employ atomic ensembles, where the interaction probability with a given optical mode is enhanced via a large number of atoms~\cite{HSP10}. Atomic ensembles have enabled a number of spectacular proof-of-principle demonstrations of quantum protocols, such as coherent photon storage and quantum memories for light \cite{HHD99,FIM05,HSP10}, entanglement generation between light and atomic spins \cite{JKP01}, nonlinear interactions between photons at the level of individual quanta \cite{PWA13,PFL12,GTS14,TBS14}, and quantum-enhanced metrology \cite{WJK10,LSV10,SKN12}. It has also been proposed that such systems could lead to exotic many-body physics, such as strongly correlated photon ``gases'' \cite{NA16}.

A fundamental limitation in nearly all such possibilities arises from spontaneous emission, wherein photons in a desired optical mode~(\textit{e.g.}, a Gaussian input beam) that facilitate the process are absorbed by the atoms and then re-scattered into other inaccessible modes or channels. Within the context of quantum light-matter interfaces based upon atomic ensembles, it is typically assumed that spontaneous emission occurs independently, and at the same rate given by a single, isolated atom. In that case, the infidelity or error arising from spontaneous emission for a desired process typically decreases with the ``optical depth'' $D$ of the medium as $1/D$ or slower. The optical depth is given by $D\sim (\lambda_0^2/A_\text{eff})N$, where $N$ is the atom number, and $\lambda_0^2/A_\text{eff}$ represents the interaction probability between a single atom and a single photon in the preferred optical mode~($\lambda_0$ being the wavelength associated with the atomic transition and $A_\text{eff}$ the beam area). Intuitively, the $1/D$~(or $1/N$) scaling directly reflects the fact that a given atom is assumed to succeed or fail independently, and that the success is enhanced by the number of atoms involved.

Technically, however, the assumption of independent emission cannot strictly be correct. In particular, as scattering is a wave phenomenon, the emission into other directions may exhibit collective interference. In fact, the possibility that an atomic ensemble can experience a significantly enhanced radiation rate via interference (``superradiance'') was already pointed out in the seminal work of Dicke \cite{D1954}, and has been thoroughly studied for decades \cite{GH1982}. The complementary phenomenon of subradiance, in which photon emission becomes highly suppressed, has also been theoretically studied \cite{PG00,SCS10,JR12,POR15,BGA15,BGA16,FJR16,SR16,ZR10,ZR11,KSP16}, and even observed in recent experiments \cite{DB96,GAK16,SBF17}. Clearly the possibility to enhance atom-light interfaces by suppressing unwanted emission is a tantalizing one, and has started to gain theoretical interest \cite{POR15}. However, finding protocols where subradiance clearly improves the scaling of errors remains an elusive task, in part because techniques to efficiently address subradiant states remain poorly developed.

In this paper, we provide a comprehensive description of subradiance in the case where atoms form ordered arrays. We also present an explicit construction of a protocol exploiting suppressed emission into undesired channels, which enables an exponential improvement in infidelity as a function of atom number over previously known bounds. Our main results are summarized as follows:

\begin{itemize}
\item We first consider infinite 1D or 2D arrays of atoms, which consist of an electronic ground state $\ket{g}$ and excited state $\ket{e}$ that couple to light through a dipole transition. Examining the case of a single collective excitation, we find that a set of perfectly subradiant states with zero decay rate emerge, which can be interpreted as optical ``guided modes.'' Specifically, in exact analogy to guided modes of conventional optical fibers or photonic structures, the spin-wave excitations that constitute these subradiant states have associated wave vectors that are mismatched from free-space radiation fields, which consequently prevents the decay of energy from these states.
\item In the case of a finite array, a set of single-excitation collective atomic modes can exhibit decay rates which are polynomially suppressed with atom number $N$. The finite decay rate can be understood as emerging from scattering of guided excitations into free space through the boundaries of the array.
\item We go beyond the most frequently studied case of subradiance within a single-excitation manifold, and investigate the nature of multi-excitation subradiant modes. Specifically, we show that subradiance is largely destroyed when excitations spatially overlap, as the scattering of two excitations generates many wave vectors that couple to free-space radiation due to the ``hard core'' nature of spins. In 1D arrays, we find that a ``fermionic'' ansatz works well to describe multi-excitation subradiant modes, where these multi-excitation states are constructed from anti-symmetric combinations of single-excitation subradiant modes in order to enforce a spatial repulsion~(\textit{i.e.}, ``Pauli exclusion'') of excitations. These states preserve the same polynomial suppression of decay rate with atom number for any low density of excitations.
\item Having elucidated the salient properties of subradiant states in free-space atomic arrays, we introduce the new concept of ``selectively radiant'' states. In particular, while subradiant states couple weakly to all electromagnetic modes, to realize an efficient atom-light interface it is instead desirable to construct states that are simultaneously superradiant to a preferred photonic mode and subradiant to all the others. We show that one natural way to achieve such a scenario is by coupling an atomic array to the guided modes of a nanophotonic structure, such as an optical nanofiber \cite{VRS10,GCA12,GSO15,polzik,CGC16}. As the wave vectors of the guided modes of the structure itself are mismatched from free-space radiation, it becomes possible for a set of atomic spin waves to efficiently couple to these guided modes, while retaining a suppressed coupling to free-space modes. We analyze the specific protocol of photon storage \cite{L03,FIM05} using an atomic array coupled to a nanofiber \cite{SCA15,GMN15}, and find numerically a storage infidelity that is exponentially small in the atom number or optical depth, $\sim \exp(-D)$. This scaling represents an exponential improvement over the best previously established error bound of $\sim 1/D$~\cite{GAF07,GAL07}, derived assuming that emission into undesired modes is independent.\end{itemize}

This article is structured as follows. In Sec. II we begin by introducing a theoretical framework for atom-light interactions that does not invoke the typical assumption of independent atomic emission, and instead formally and exactly describes collective emission and interactions of atoms via photon fields. In Sec. III we apply this formalism to investigate single- and multi-excitation subradiant states in atomic arrays, with the main results having already been summarized above. In Sec. IV we present the idea of selectively radiant states, and analyze the efficiency of a quantum memory consisting of a chain of atoms close to a nanofiber. Finally, in Sec. V we discuss possible implementations and other photonic platforms for observing subradiant physics. An outlook is provided in Sec. VI.

\section{Spin model}
Here we introduce a theoretical formalism to describe the fully quantum interaction between atoms and radiation fields, which is valid in the presence of any linear, isotropic, dielectric media. This rather general formalism will enable us to equally treat the case of atomic arrays in free space (Sec. III), or interacting via the guided modes of an optical fiber (Sec. IV). In particular, we present a model in which the field is integrated out and the dynamics of the atomic internal (``spin") degrees of freedom follow a master equation that only depends on atomic operators. Moreover, once the time evolution of the atoms is solved for, one can recover the field at any point in space by means of a generalized input-output equation. 

The first step to describe how atoms couple to radiation is to quantize the electromagnetic field. The traditional approach involves explicitly finding a normal mode decomposition of the fields, and associating bosonic annihilation and creation operators to each mode. This is well-suited to cases where a limited number of modes are assumed to be relevant (such as a high-Q cavity). In our case, though, as we want to exactly capture collective effects in spontaneous emission involving all modes, such an approach becomes unwieldy (as in free space) \cite{GH1982,BS06} or impossible, such as for complex dielectric structures. We require a more general technique that allows us to treat these situations. Such a framework was developed by Welsch and coworkers \cite{GW96,DKW02,BW07,BI12}, and is based on the classical electromagnetic Green's function (or Green's tensor). 

The Green's function $\GG(\rb,\rb',\omega)$ is the fundamental solution of the electromagnetic wave equation, and obeys \cite{NH06}:
\begin{align}\label{gf}
\bm{\nabla}\times\bm{\nabla}\times \GG(\rb,\rb',\omega)-\frac{\omega^2}{c^2}\epsilon(\rb,\omega)\, \GG(\rb,\rb',\omega)=\delta(\rb-\rb')\mathbb{1},
\end{align}
where $\epsilon(\rb,\omega)$ is the position- and possibly frequency-dependent relative permittivity of the medium. The Green's function physically describes the field at point $\rb$ due to a normalized, oscillating dipole at $\rb'$. $\GG_{\alpha\beta}$ is a tensor quantity ($\{\alpha,\beta\}=\{x,y,z\}$), as $\alpha$ and $\beta$ refer to the possible orientations of the field and dipole, respectively. Here, we will deal with cases where the Green's function can be solved analytically, but for more complex structures it is also possible to obtain it numerically ~\cite{ORI10,RH07,HGA16}. In the following, we introduce a prescription of how to write down an equation that relates the field and the atomic coherence operators, built upon the intuition provided by classical physics. For a more formal derivation of the field quantization, we refer the reader to Refs.~\cite{GW96,DKW02,BW07,BI12,AHC17}.

In the frequency domain, the analogous classical problem that one would like to solve is to find the total field $\Eb(\rb,\omega)$ at point $\rb$, given a known input field $\Eb_{\rm p}(\rb,\omega)$ and a collection of $N$ polarizable dipoles $\pb_j(\omega)$ located at $\rb_j$, which are excited by the fields and re-scatter light themselves. The values of $\pb_j(\omega)$ are not known a priori, since they depend on the polarizability and the total field at $\rb_j$ [solving for $\pb_j(\omega)$ will be discussed in following steps]. As the field at any given point in space is just the sum of the external or driving field and the field re-scattered by the dipoles, we find $\Eb(\rb,\omega)=\Eb_{\rm p}(\rb,\omega)+\mu_0\omega^2\,\sum_{j=1}^N\GG(\rb,\rb_j,\omega)\cdot\pb_j(\omega)$, where $\mu_0$ is the vacuum permeability. 

The question is how to translate this classical equation into an equation for quantum operators. In fact, the quantum nature of the field is inherited from the quantum properties (\textit{e.g.}, correlations and fluctuations) of the sources, while the field propagation remains the same as both the quantum and classical fields obey Maxwell's equations. Therefore, the above equation is valid for quantum fields, but replacing $\pb_j(\omega)$ by the dipole moment operator $\hat{\pb}_j(\omega)$, and $\Eb(\rb,\omega)$ by the field operator $\hat{\Eb}^+(\rb,\omega)$, where the superscript refers to the positive-frequency component. In the case that the quantum dipoles are atoms, one can make a further approximation, taking advantage of the fact that an atom only has a significant optical response in a narrow bandwidth around its resonance frequency $\omega_0$. Thus, one is able to approximate $\GG(\rb,\rb_j,\omega)$ by $\GG(\rb,\rb_j,\omega_0)$, which allows the Fourier transform of the equation to become local in time. Then, one arrives to the generalized input-output equation in the time domain, which reads \cite{DKW02,CMS15,XF15}
\begin{align}\label{fielddef}
\hat{\Eb}^+(\rb)&=\hat{\Eb}_{\rm p}^+(\rb)+\mu_0 \omega_0^2\sum_{j=1}^N \GG(\rb,\rb_j,\omega_0)\cdot\db\,\hge^j.
\end{align}
To obtain the above expression, we have made use of the fact that $\hat{\pb}_j = \db^* \, \hat{\sigma}^j_{eg} + \db \, \hat{\sigma}^j_{ge}$, where $\heg^j=\ket{e_j}\bra{g_j}$ is the atomic coherence operator between the ground and excited states of atom $j$, and  $\db$ is the dipole matrix element associated with that transition. This equation is valid in the Markovian regime, where the dispersion in the Green's function can be neglected and the replacement of $\GG(\rb,\rb_j,\omega)$ by $\GG(\rb,\rb_j,\omega_0)$ is well-founded. For this to be true, two conditions have to be fulfilled. First, the retardation arising from the physical distance between atoms can be ignored \cite{CJG12,SCC15}. For atoms in free space, this means that they should sit much closer than the length of a spontaneously emitted photon ($\lesssim 1$ meter). Second, the electromagnetic environment itself should not have very narrow-bandwidth features (\textit{e.g}., one must avoid the strong coupling regime of cavity QED \cite{TRK92}).

What remains now is to solve for the dipoles (in this case, $\hge$) themselves. We do so by  writing down Heisenberg-Langevin equations for the atomic internal degrees of freedom, starting from the full atom-field Hamiltonian. Intuitively, the atomic spin $\heg^{i}$ will be driven by the quantum field at position $\rb_i$. However, as the field itself only depends on other atoms via the input-output equation, the atomic dynamics can be fully derived from an equivalent master equation of the form $\dot{\hat{\rho}}_{\rm A}=-(\ii/\hbar)\,[\mathcal{H},\hat{\rho}_{\rm A}]+\mathcal{L}[\hat{\rho}_{\rm A}]$ \cite{MS1990}, where $\hat{\rho}_{\rm A}$ is the atomic density matrix, and the Hamiltonian and Lindblad operators read
\begin{subequations}
\begin{equation}\label{ham}
\mathcal{H}=\hbar\omega_0\sum_{i=1}^N\hat{\sigma}_{ee}^i+\hbar\sum_{i,j=1}^N J^{ij}\hat{\sigma}_{eg}^i\hat{\sigma}_{ge}^j,
\end{equation}
\begin{equation}\label{lind}
\mathcal{L}[\hat{\rho}_{\rm A}]=\sum_{i,j=1}^N\frac{\Gamma_{ij}}{2}\,\left(2\hat{\sigma}_{ge}^j\hat{\rho}_{\rm A}\hat{\sigma}_{eg}^i-\hat{\sigma}_{eg}^i\hat{\sigma}_{ge}^j\hat{\rho}_{\rm A}-\hat{\rho}_{\rm A}\hat{\sigma}_{eg}^i\hat{\sigma}_{ge}^j\right).
\end{equation}
\end{subequations}
In the above expressions, the rates for coherent and dissipative interactions between atoms $i$ and $j$ are respectively given by
\begin{subequations}\label{rates}
\begin{align}\label{shiftrate}
J^{ij}&=-\frac{\mu_0\omega_0^2}{\hbar}\,\db^*\cdot\text{Re}\,\mathbf{G}(\rb_i,\rb_j,\omega_0)\cdot\db, 
\end{align}
\begin{equation}\label{dissiprate}
\Gamma^{ij} =\frac{2\mu_0\,\omega_0^2}{\hbar}\,\db^*\cdot\text{Im}\,\mathbf{G}(\rb_i,\rb_j,\omega_0)\cdot\db,
\end{equation}
\end{subequations}
where the sign of $J^{ij}$ is taken to be opposite to that of Refs.~\cite{GW96,DKW02,BW07,BI12,HGA16,AHC17}. In the above Hamiltonian, we have neglected Casimir interactions between ground-state atoms (of the form $\hat{\sigma}_{gg}^i\hat{\sigma}_{gg}^j$), as their spatial decay is very fast ($\sim 1/d^6$ in free space, $d$ being the inter-atomic distance) \cite{BI12}.

The dynamics under the master equation can analogously be described in the quantum jump formalism of open systems \cite{MS07}. In this formalism, the atomic wave function evolves deterministically under an effective non-Hermitian Hamiltonian that reads $\mathcal{H}=\hbar\omega_0\sum_{i=1}^N\hat{\sigma}_{ee}^i+\mathcal{H}_{\rm eff}$, with
\begin{equation}\label{heff}
\mathcal{H}_{\rm eff}=-\mu_0\omega_0^2\sum_{i,j=1}^N \,\db^\ast\cdot\mathbf{G}(\rb_i,\rb_j,\omega_0)\cdot\db\,\,\hat{\sigma}_{eg}^i\hat{\sigma}_{ge}^j,
\end{equation}
along with stochastically applied ``quantum jump" operators to account for the population recycling term ($\hge^{j}\hat{\rho}_\text{A}\heg^{i}$) of Eq.~\eqref{lind}. While $\mathcal{H}_{\rm eff}$ just describes the interaction of atoms through emission and re-absorption of photons, one can directly add other terms to the Hamiltonian to account for external driving fields.

To conclude, we point out that although the full formalism above has only been rigorously and generally developed in recent years, many aspects have long been used within atomic physics and quantum optics. For example, for a single atom or other quantum emitter, the spin model becomes trivial and just yields the total spontaneous emission rate. Thus, the calculation of enhancement of spontaneous emission near dielectric structures is standardly reduced to the calculation of the Green's function \cite{CPS1978,LVN04,EFW05}. Alternatively, such equations are often used to model the optical response of dense three-dimensional atomic gases \cite{SSH16,JBS16,JSG16,BZB16,GK17}. 

\section{Free space: Subradiant states}
\label{SecIII}
We now apply the spin model we describe in the previous section to investigate the properties of subradiant states associated with ordered atomic arrays in free space. Recently, the peculiar linear optical properties of periodic atomic arrays have started to attract interest \cite{BGA15,BGA16,BGA16b,SWL17,KSP16,SR16,ZR11,POR15,HFS16,JR12,FJR16}. This includes the identification of guided modes supported by infinite arrays \cite{ZR10,SWL17,SR16}, and states with very long lifetimes in finite arrays \cite{BGA15,BGA16,BGA16b,KSP16,SR16,ZR11,POR15,JR12,HFS16,FJR16}. Here, we provide a clear and intuitive connection between the existence of guided modes in infinite arrays and subradiant states in a finite system. We provide conditions for the lattice constants in 1D and 2D that enable single-excitation guided Bloch modes with zero decay rate to emerge, which are decoupled from free-space radiation due to wave vector mismatch. We then analyze a single excitation in a finite lattice, and show how the guided modes acquire a non-zero decay rate due to scattering into electromagnetic radiation at the system boundaries. We also analyze the scaling of the decay rates with system size and elucidate the spatial structure of subradiant states. Finally, we go beyond previous studies of single-excitation subradiance (where the atoms can equivalently be treated as classical dipoles) to the rich physics of the multi-excitation case. In particular, in one dimension, we show that multi-excitation subradiant states exist for any low density of excitations, and that their wave functions have fermionic character.

The atoms are assumed to be tightly trapped, so that we can treat the positions of the particles as classical points rather than dynamical variables. In this situation, we substitute in Eqs.(\ref{fielddef}-\ref{heff}) the free-space Green's tensor $\GG(\rb_i,\rb_j,\omega_0)=\GG_{0}(\rb_{ij},\omega_0)$, with $\rb_{ij}=\rb_i-\rb_j$. Here $\GG_{0}(\rb,\omega_0)$ is the solution to Eq.(\ref{gf}) when setting $\epsilon(\rb,\omega)=1$, and can be written as:
\begin{align}
\GG_0 (\rb,\omega_0) = \frac{e^{\ii k_0 r}} 
{4\pi k_0^2 r^3} \left[(k_0^2 r^2+\ii k_0 r -1) \mathbb{1} \right.+ \notag \\
\qquad{}\left. + (-k_0^2 r^2 -3\ii k_0 r + 3) \frac{\rb \otimes \rb}{r^2}  \right], \label{Greens_def}
\end{align}
where $r=|\rb|$ and $k_0=2\pi/\lambda_0=\omega_0/c$ is the wave number corresponding to the atomic transition energy. For a single atom, evaluating Eq.~\eqref{dissiprate} simply reproduces the well-known vacuum emission rate $\Gamma^{ii}=\Gamma_0$, where $\Gamma_0=\omega_0^3|\db|^2/3\pi\hbar\epsilon_0 c^3$. The single-atom energy shift $J^{ii}$ in Eq.\eqref{shiftrate} arising from $\GG_0$ formally yields a divergence and will be set to zero in what follows, as it should be incorporated into a re-normalized resonance frequency $\omega_0$. In Sec. IV, the Green's function of a nanofiber will be decomposed into a free-space and a scattered component, $\GG=\GG_0+\GG_\text{sc}$, where $\GG_\text{sc}$ does produce a finite, observable contribution to $J^{ii}$.

For concreteness, we will restrict ourselves to the following lattice geometries: one-dimensional (1D) linear chains and closed circular rings, two-dimensional (2D) square and three-dimensional (3D) cubic lattices. However, it should become clear that the underlying principles should be general to other lattice structures as well. In the following, the number of atoms and lattice constant are denoted by $N$ and $d$, respectively.

\subsection{Infinite Lattice (Single excitation)}
\label{SecIIIA}
\begin{figure}[h!]
\centerline{\includegraphics[width=\linewidth]{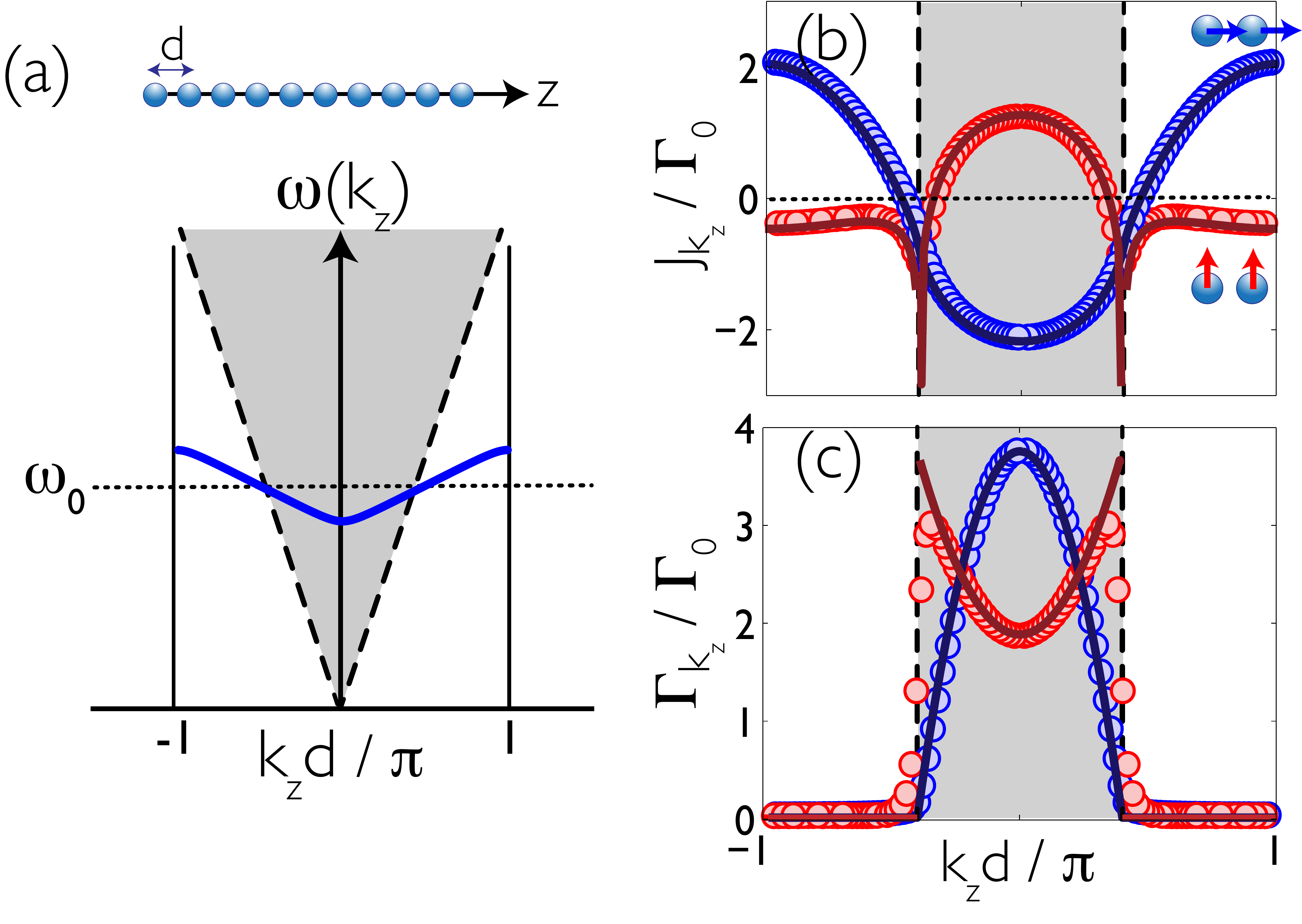}}
\caption{{\bf (a)} Generic dispersion relation of frequency $\omega(k_z)$ versus Bloch wave vector $k_z$ for single-excitation modes of an infinite, one-dimensional chain. The Bloch vector $k_z$ is only uniquely defined within the first Brillouin zone $|k_z|\leq \pi/d$. The dashed black line is the light line, and corresponds to the dispersion relation of light in vacuum propagating along the $\hat{z}$ direction, \textit{i.e.}, $\omega = c |k_z|$. Atomic modes in the region enclosed within the light line (shaded) are generally unguided and radiate into free space. Outside the light line ($|k_z| > \omega /c$) the modes are guided and subradiant, as the electromagnetic field is evanescent in the directions transverse to the chain. The dispersion relation is generally expected to be rather flat and centered around the bare atomic resonance frequency $\omega_0$. {\bf (b)} Collective frequency shifts and {\bf (c)} decay rates for an atomic chain along $\hat{z}$ with lattice constant $d/\lambda_0 = 0.2$, for parallel (blue) and transverse (red) atomic polarization. Circles correspond to the results for a finite system with $N=50$ atoms. The analytical expressions for the infinite chain are denoted by solid lines and approximate well the finite chain results, except for a small region close to the light line. In the infinite lattice case the modes with $|k_z|>\omega_0/c$ are perfectly guided and the decay rate $\Gamma_{k_z}$ is exactly zero. The light line (black dashed) appears vertical over the very narrow frequency ranges plotted here.}
\label{fsFig1}
\end{figure}

Let us consider first a perfectly ordered infinite array of atoms. Despite the infinite lattice being unrealistic, it provides insight into the problem thanks to its mathematical simplicity. In this case, the system is perfectly translationally invariant by any lattice vector displacement, and thus, both atomic and electromagnetic eigenmodes must obey Bloch's theorem. 

For a single excitation stored in the system, the eigenstates of the effective atomic Hamiltonian of Eq.(\ref{heff}) are spin waves, with well-defined quasi-momentum $\kb$, which can always be chosen to be within the first Brillouin zone. For such states, whose creation operators can be written as $S^\dagger_{\kb}= N^{-1/2} \sum_j  e^{\ii \kb \cdot \rb_j} \heg^j$, the single excitation is delocalized and shared in a coherent way among all the atoms. Classically, these states are analogous to oscillating dipoles where the phase of dipole $j$ is given by $e^{\ii \kb \cdot \rb_j}$.

As the Bloch modes are eigenstates of the effective Hamiltonian, they satisfy $\mathcal{H}_{\rm eff} \,S^\dagger_\kb \ket{g}^{\otimes N}=\hbar(J_\kb - \ii \Gamma_\kb/2) \,S_\kb^\dagger \ket{g}^{\otimes N}$. Here $J_\kb$ and $\Gamma_\kb$ are real quantities and can be identified as the frequency shift of mode $\kb$ (relative to the bare atomic frequency $\omega_0$) and the decay rate, respectively. One can readily show that, in terms of the single-atom spontaneous emission rate $\Gamma_0$, they are given by:  
\begin{subequations}
\begin{align}
\frac{J_\kb}{\Gamma_0} &=- \frac{3\pi}{k_0} \dbu^* \cdot \text{Re}  \,\tilde{\GG}_{0} (\kb) \cdot \dbu, 
\end{align}
\begin{align}
\frac{\Gamma_\kb}{\Gamma_0} &= \frac{6\pi}{k_0} \dbu^*\cdot \text{Im} \,\tilde{\GG}_{0} (\kb) \cdot  \dbu, \label{Eq:Couplings}
\end{align}
\end{subequations}
where $\tilde{\GG}_{0} (\kb) = \sum_{j} e^{-\ii \kb \cdot \rb_j} \GG_{0} (\rb_j)$ is the discrete Fourier transform of the free-space Green's tensor.

We will now show that, when the atoms are placed at close enough distances, dipole-dipole interactions can dramatically modify the decay rates of collective states. As the simplest case, let us consider an infinite one-dimensional chain of atoms first, oriented along the $\hat{z}$ direction. In that case, the wave vector $k_z$ constitutes an index for the modes, and one can consider the dispersion relation of frequency $\omega(k_z)=\omega_0+J_{k_z}$ versus $k_z$. For a periodic structure, regardless of the system details, one expects the dispersion relation to exhibit general characteristics [see Fig.~\ref{fsFig1}(a)]. First, and as mentioned before, $k_z$ is only uniquely defined within the first Brillouin zone ($|k_z|\leq \pi/d$) and thus, it suffices to plot the dispersion relation in that region. Second, it is helpful to draw the ``light line'', \textit{i.e.}, the dispersion relation $\omega=c|k_z|$ corresponding to light propagating in free space along the $\hat{z}$ direction [dashed line of Fig.~\ref{fsFig1}(a)]. Physically, the light line is significant because it separates states of very different character, as we now describe. 

To see this, let's consider the field generated by a spin-wave excitation, which is given by Eq.(\ref{fielddef}) under the replacement $\hge^{j}\rightarrow e^{\ii k_z z_j}$ (it is sufficient to consider the limit of classical dipoles for this argument). One can always  expand the field $\Eb(\rb)$ in terms of plane wave components, $\Eb(\rb)=\sum_{q_z,{\bf q_\perp}} \Eb_{q_z,{\bf q_\perp}} e^{\ii q_z z + \ii{\bf q_\perp} \cdot{\bf r_\perp}}$. The state is clearly of Bloch's form, and thus, only a discrete set of wave vectors $q_z = k_z+g_z$ ($g_z$ being any reciprocal lattice vector) will contribute. At the same time, the wave equation requires that the axial and perpendicular components of the wave vector satisfy $(q_z)^2+{\bf q}_\perp \cdot {\bf q}_\perp = (\omega/c)^2$. Thus, one can readily verify that a spin wave outside the light line ($|k_z|>\omega/c$) has an associated electromagnetic field composed of axial wave vectors $|q_z|>\omega/c$. This in turn implies that ${\bf q_\perp}$ is imaginary, and the field is guided and decays evanescently away from the structure. Therefore, these guided modes are decoupled from all optical modes propagating in free space, and their inability to radiate away energy leads to perfect subradiance (exactly zero decay rate). Conversely, modes within the light line are generally unguided and can radiate energy out to infinity.

The concepts outlined above regarding the separation of the dispersion relation into guided and radiative regions are actually quite general, and well-known in the context of periodically modulated dielectric waveguides (``photonic crystals'' \cite{JMW95}). An atomic chain might appear quite different physically, but mathematically the same set of principles apply. Furthermore, while it is difficult to prove independent of lattice geometry and atomic level configuration, one would generally expect that for atoms any guided modes would occur within a narrow bandwidth (on the order of the atomic transition linewidth $\Gamma_0/2\pi \lesssim 10 {\rm~MHz}$) around the resonance frequency ($\omega_0/2\pi \sim 300 {\rm ~THz}$), where the atoms have a significant optical response. Thus, in Fig.~\ref{fsFig1}(a) the band structure will appear rather flat. Then, a sufficient condition for guided modes to exist in an atomic chain is essentially that the light line intersects the edge of the Brillouin zone $k_z = \pi/d$ at a frequency $\omega(k_z)$ greater than the atomic resonance. This condition can be rewritten as a condition on the lattice constant $d<\lambda_0/2$ required to support guided modes.\\

Equipped with this general intuition, we now quantitatively investigate the dispersion relation for the 1D infinite chain of two-level atoms with polarization parallel or transverse to the array. The collective frequency shifts are derived in greater detail in Appendix~\ref{AppA}, and read:
\begin{subequations}
\begin{align}
\frac{J_{k_z}^{||}}{\Gamma_0} &=-\frac{3}{2 k_0^3 d^3}\textrm{Re} \left[ \textrm{Li}_{3}(e^{\ii (k_0+k_z)d})+ \textrm{Li}_{3}(e^{\ii (k_0-k_z)d}) \right. \notag\\
&\quad{}\left. -\ii k_0 d   \textrm{Li}_{2}(e^{\ii (k_0+k_z)d})-\ii k_0 d \textrm{Li}_{2}(e^{\ii (k_0-k_z)d}) \right], \label{Eq:ShiftAnal1}
\end{align}
\begin{align}
\frac{J_{k_z}^{\perp}}{\Gamma_0} &=\frac{3}{4 k_0^3 d^3} \textrm{Re} \left[\textrm{Li}_{3}(e^{\ii (k_0+k_z)d})+ \textrm{Li}_{3}(e^{\ii (k_0-k_z)d}) \right.\notag\\
&\quad{}-\ii k_0 d  \textrm{Li}_{2}(e^{\ii (k_0+k_z)d})- \ii k_0 d  \textrm{Li}_{2}(e^{\ii (k_0-k_z)d})  \notag\\
&\quad{} \left. +k_0^2 d^2 \textrm{Ln}(1-e^{\ii(k_0+k_z)d})  +k_0^2 d^2 \textrm{Ln}(1-e^{\ii(k_0-k_z)d})  \right], \label{Eq:ShiftAnal2}
\end{align}
\end{subequations}
where $\textrm{Li}_{n}(x)$ is the PolyLogarithm of order $n$. These expressions are plotted in Fig.~\ref{fsFig1}(b), for the particular value of $d/\lambda_0=0.2$. Here, the light line is indicated as before by a dashed line, but since $\Gamma_0/ \omega_0 \sim 10^{-8}$, it appears essentially as a vertical line.

As anticipated, we can see in this figure that the bands occupy only a narrow bandwidth around the resonance frequency, except close to the light line for transverse polarization. The exact shape of the bands depends on the value of $d/\lambda_0$ and the polarization direction, and for instance, the effective mass at the zone edge ($|k_z|=\pi/d$) is negative (positive) for parallel (transverse) polarization. Exactly at the light line the expression for $J_{k_z}^{\perp}$ ($J_{k_z}^{||}$) becomes non-analytic and diverges (has a derivative that diverges).

\begin{figure}
\centerline{\includegraphics[width=\linewidth]{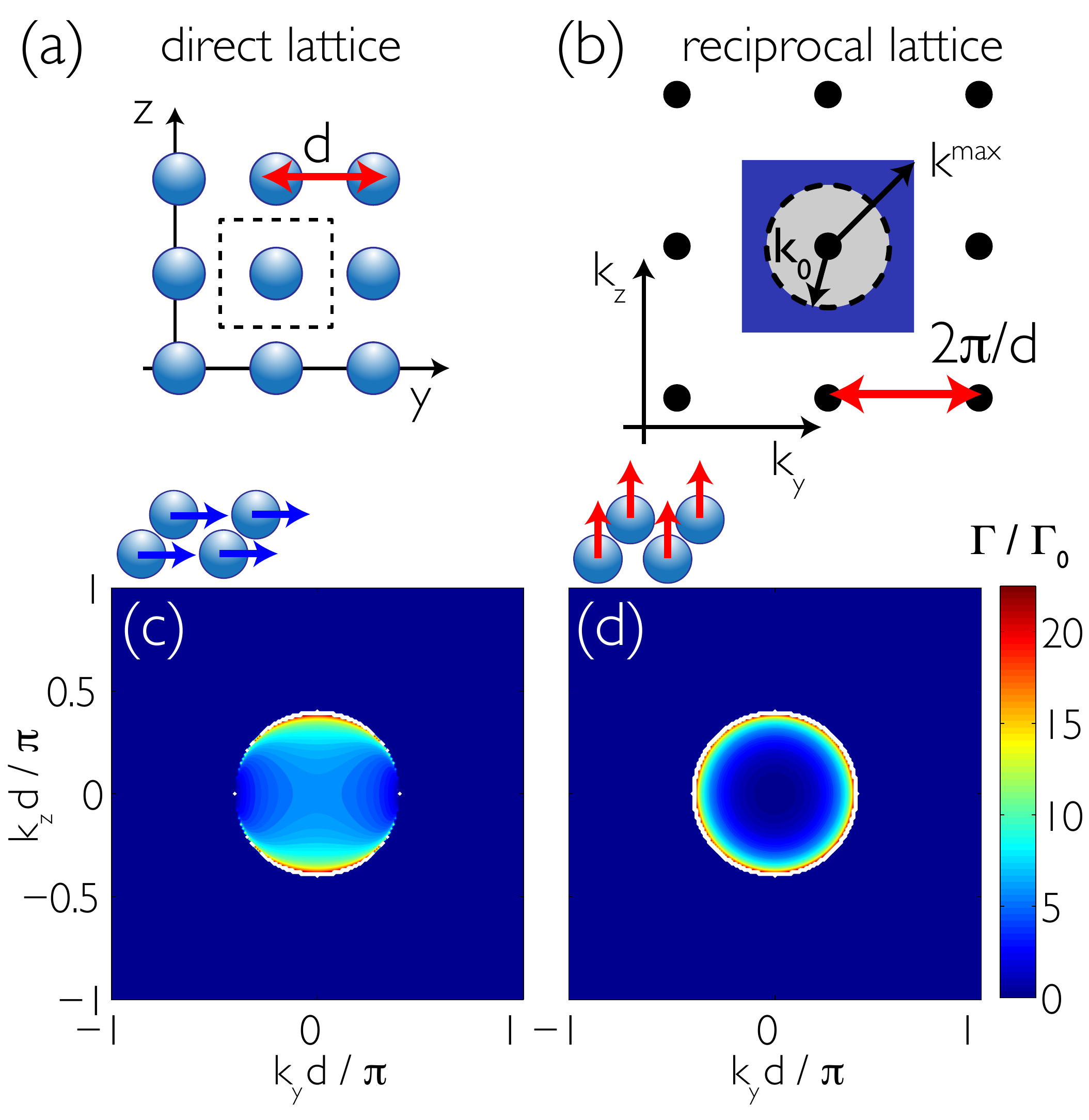}}
\caption{{\bf (a)} Illustration of a square lattice of atoms in the $\hat{y}$-$\hat{z}$ plane, with lattice constant $d$. {\bf (b)} Corresponding reciprocal lattice in 2D, with lattice constant $2\pi/d$. The collective modes have well defined quasi-momentum  ${\kb}=(k_y,k_z)$ within the first Brillouin zone, which is indicated by the blue square. A circle of radius $| \kb |=k_0$ defines the set of propagating electromagnetic modes in vacuum in the $\hat{y}$-$\hat{z}$ plane at the atomic frequency. Collective spin waves outside of this circle will be guided, with a decay rate $\Gamma_{\kb}=0$. For $k_0>k^\textrm{max} = \sqrt{2} \pi / d$ (or equivalently, $d/\lambda_0> 1/\sqrt{2}$) all collective eigenstates lie inside of the circle. {\bf (c)} and {\bf (d)} Collective decay rates for the infinite square lattice for parallel and transverse atomic linear polarization, respectively. The lattice constant is set to $d/\lambda_0=0.2$.  For transverse polarization, $\Gamma_{\kb}$ also vanishes for $\kb \sim (0,0)$, provided that $d/\lambda_0<1$.} \label{fsFig2}
\end{figure}

The collective decay rates can also be analytically derived (see Appendix~\ref{AppA}): 
\begin{subequations}
\begin{align}
\frac{\Gamma_{k_z}^{||}}{\Gamma_0} &=\frac{3\pi}{2 k_0 d} \sum_{\substack{g_z \\ |k_z+g_z|\leq k_0}}\left(1-\frac{(k_z+g_z)^2}{k_0^2} \right),
\end{align}
\begin{align}
\frac{\Gamma_{k_z}^{\perp}}{\Gamma_0} &=\frac{3\pi}{4 k_0 d}  \sum_{\substack{g_z \\ |k_z+g_z|\leq k_0}} \left(1+\frac{(k_z+g_z)^2}{k_0^2} \right).
\end{align}
\end{subequations}
These summations run over reciprocal lattice vectors that satisfy $|g_z +k_z| \leq k_0$. That is, only the diffracted waves enclosed  within the light line will contribute to the decay rate. When $|k_z|>k_0$, there are no values of $g_z$ satisfying the above condition. Thus the decay rates are zero and we mathematically recover the result previously anticipated -- modes beyond the light line are perfectly guided without radiative losses. The decay rates are plotted in Fig.~\ref{fsFig1}(c). As we can see from the expressions above, at the light line the state can be subradiant or radiant depending on the polarization direction. This results in a discontinuity at the light line for transverse polarization.\\

A similar set of results can be obtained for a 2D array. Considering a square lattice in the $\hat{y}$-$\hat{z}$ plane [Fig.~\ref{fsFig2}(a)], the  corresponding first Brillouin zone for Bloch wave vectors extends over the region $|k_y|,|k_z| \leq \pi/d$, as shown in Fig.~\ref{fsFig2}(b). The set of electromagnetic fields propagating in the plane at the atomic frequency $\omega_0$ have a wave vector of magnitude $k_0$ which defines a circle centered around the origin in $\kb$-space, as illustrated in Fig.~\ref{fsFig2}(b). Similar to 1D, a sufficient condition for spin-wave excitations to be guided is that the wave vector lies outside of this circle. It should be noted that the longest ``distance'' in the first Brillouin zone from the origin extends along the diagonal, and has magnitude $k^\text{max} = \sqrt{2}\pi/ d $. Thus, in 2D, guided modes exist as long as $k_0<k^\text{max}$, which translates into a maximum allowed lattice constant $d/\lambda_0 = 1/\sqrt{2}$.

Analogous to the 1D case, we can obtain closed mathematical expressions for the decay rates in the 2D lattice. They are given by:  
\begin{subequations}\label{rate2D}
\begin{align}
\frac{\Gamma_{\kb}^{||}}{\Gamma_0} &=\frac{3\pi}{ k_0^3 d^2}  \sum_{\substack{\gb \\ |\kb+\gb| \leq k_0}} \frac{k_0^2 - |(\kb+ \gb)\cdot \dbu|^2}{ \sqrt{k_0^2 -|\kb + \gb|^2}}, \label{Eq:Decay2D_parallel}
\end{align}
\begin{align}
\frac{\Gamma_{\kb}^{\perp}}{\Gamma_0} &=\frac{3\pi}{k_0^3 d^2}  \sum_{\substack{\gb \\ |\kb+\gb| \leq k_0}}  \frac{|\kb + \gb|^2}{ \sqrt{k_0^2 -|\kb + \gb|^2}}, \label{Eq:Decay2D_trans}
\end{align}
\end{subequations}
from which we recover again the important result that Bloch states with $|\kb|>k_0$ do not radiate out to infinity. For these states, the electromagnetic field is now confined within the plane, and evanescently decays away from the lattice in the transverse direction. The decay rates are plotted in Fig.~\ref{fsFig2}(c)-(d) for atomic polarizations along $\hat{z}$ and $\hat{x}$ directions, for the particular value of $d/\lambda_0=0.2$, which defines the light line as the circle $k_0=0.4\pi/d$, beyond which the decay rate is exactly zero.  

We would like to remark that the previous considerations are valid regardless of the specific atomic structure, provided that the atom in question only contains a single ground state (see Sec.~V for a discussion of the subtleties associated with a ground-state manifold).

The previous analysis for the 2D lattice provides another interesting result. As Fig.~\ref{fsFig2}(d) shows, for transverse atomic polarization in a 2D square lattice, subradiance can emerge not only outside the light line, but also at the center of the Brillouin zone, that is, for Bloch states with quasi-momentum $\kb \sim (0,0)$. Physically, the origin of this effect can be understood as follows.  On one hand, and as previously discussed, for $d/\lambda_0 < 1$ the field created by such a state is generally evanescent at all diffraction orders except for the component $\gb = 0$ [c.f. Eq.(\ref{Eq:Decay2D_trans})], which corresponds to a plane wave propagating perpendicularly to the atomic plane. On the other hand, the state $\kb = (0,0)$ corresponds to an array of dipoles that are in phase. However, dipoles oscillating in phase and perpendicularly to the atomic plane are forbidden to radiate energy in the perpendicular direction, and thus, the state must be subradiant. In contrast, as soon as $d/\lambda_0 > 1$, there will  be other $\gb$ components that are not evanescent, yielding a radiative state. We note that, although only the state $\kb = (0,0)$ has exactly zero decay rate, other modes around this point will also show a strong suppression in the emission rate relative to $\Gamma_0$.\\

\begin{figure*}
\centerline{\includegraphics[width=\linewidth]{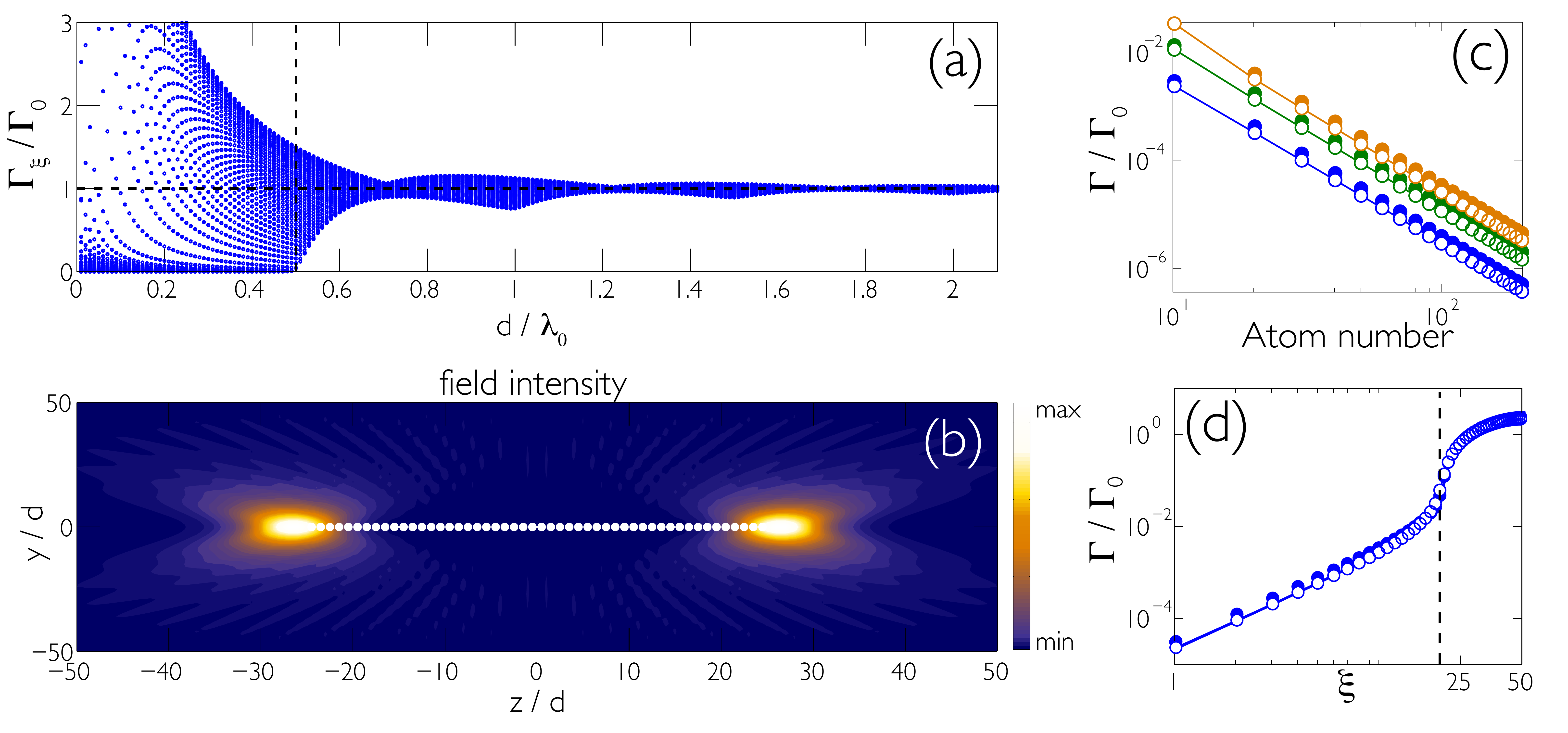}}
\caption{Single-excitation collective modes in a finite 1D chain of two-level atoms with polarization along the chain. {\bf (a)} Decay rates of the $N$ modes at different lattice constants $d/\lambda_0$ for a finite chain of $N=50$ atoms. Subradiant modes only arise if  $d/\lambda_0\leq 1/2$. A vertical cut at the fixed value of $d/\lambda_0 = 0.2$ corresponds to the blue circles depicted in Fig. \ref{fsFig1}(c). {\bf (b)} Field intensity (arbitrary units) in the $\hat{y}$-$\hat{z}$ plane ($x=5d)$ created by the most subradiant mode in a chain of $N=50$ atoms. The field is largely evanescent transverse to the bulk of the chain, while most of the energy is radiated out through scattering at the ends of the chain. White circles denote  atomic positions. {\bf (c)} Scaling with atom number of decay rates for the three most subradiant modes. A fit for large $N$ yields $\Gamma \sim N^{-3}$. {\bf (d)}  Scaling with mode index $\xi$ of decay rates at fixed $N=50$. Here $\xi$ is used to label the magnitude of the decay rates in increasing order ($\xi=1$ is the most subradiant state, while $\xi=N$ has the largest decay rate). A fit for small $\xi$ yields $\Gamma \sim \xi^2$.  Open and solid symbols denote the results obtained by exact diagonalization and from the ansatz of Eq.(\ref{Eq:SingleParticleStates}), respectively. The black dashed line corresponds to the eigenstate whose dominant wave vector $k$ crosses the light line ($k=k_0$). (b), (c) and (d) are for $d/\lambda_0=0.3$.} \label{fsFig3}
\end{figure*}

\subsection{Finite Lattice (Single excitation)}
\label{SecIIIB}
In this section, we analyze the decay rates and spatial properties of single-excitation eigenstates, for a lattice of finite size. We show that all eigenstates now acquire a non-zero decay rate, and subradiant states can be identified as those for which the rate is suppressed with increasing system size. The small value of the decay can be interpreted as arising from the finite system boundaries, which scatter a mode propagating in the bulk into free space.\\

{\bf{1D Linear Chain.}}\\
In the following, we consider a finite chain of atoms along $\hat{z}$, with a linear polarization along the chain (unless otherwise stated). However, a similar set of conclusions is obtained for the transverse polarization case.

{\bf \textit{Scaling of the most subradiant decay rates with system size.--}} The effective atomic Hamiltonian of Eq.(\ref{heff}) conserves the excitation number in the system, and thus, it can be diagonalized in blocks with fixed excitation number. Before proceeding futher, we discuss a technical but important point. Since the effective Hamiltonian is non-Hermitian, in general the eigenstates will not be orthogonal in the standard quantum mechanical sense (\textit{i.e.}, two eigenstates $\ket{\psi_i}$ and $\ket{\psi_j}$ will not satisfy $\Braket{\psi_i}{\psi_j}=\delta_{ij}$) \cite{AHC17}. The infinite lattice case presented an exception, as Bloch's theorem is still enforced. While this implies that general quantum mechanical rules, such as for eigenstate decompositions of states and observables, do not apply, we will nonetheless investigate the properties of the eigenstates further. This is physically motivated as they still represent non-evolving states under the Hamiltonian (aside from an overall phase and amplitude); thus, for example, they might be expected to shed light on how a general state behaves at long times.

We consider the case of a 1D chain of $N$ atoms with lattice constant $d$, for which numerical diagonalization of $\mathcal{H}_\textrm{eff}$  in the one-excitation manifold produces $N$ eigenstates (denoted by $\ket{\psi_\xi}$, $1\leq \xi \leq N$) and complex eigenvalues. As in the infinite case, the eigenvalues can be written in the form $J_{\xi}-i\Gamma_\xi /2$, with $J_{\xi}$ and $\Gamma_{\xi}$ representing the frequency shift relative to $\omega_0$ and decay rate, respectively. For concreteness, the eigenstates will be ordered in increasing decay rate, such that $\xi=1$ represents the most subradiant state and $\xi=N$ the most radiant one. To start understanding the properties of this system we fix the atomic number $N=50$ and change the lattice constant $d$. For each value of $d$, we diagonalize $\mathcal{H}_\textrm{eff}$, and obtain the $N$ different values for the decay rates and frequency shifts associated with each of the collective modes. Figure \ref{fsFig3}(a) shows the resulting single-excitation decay rates $\Gamma_\xi$ for each collective mode, normalized by the free-space single-atom emission rate $\Gamma_0$, in the case of atomic polarization parallel to the chain. In this plot, a vertical cut at a fixed value of $d/\lambda_0$ contains the $N$ different values of $\Gamma_\xi$. As expected, for large interparticle distances, the collective decay rates tend to the spontaneous emission rate of a single atom. As the distance decreases, they are periodically modulated, showing for $d/\lambda_0 < 1/2$ a qualitatively distinct behavior. In this region, the decay rates of some of the modes are dramatically suppressed ($\Gamma_\xi/\Gamma_0 \ll 1$), in accordance with the condition for the emergence of modes with zero decay rate derived in the infinite lattice case.

The subradiant modes in the finite chain are closely related to those derived in the infinite chain. First, having established that subradiant states in the infinite chain correspond to guided modes, the nonzero decay rates in the finite chain can be interpreted as emerging from scattering of these guided modes from the ends of the system. This can be seen in Fig.~\ref{fsFig3}(b), where we have plotted the field intensity in the plane $x=5d$ generated by the most subradiant state when the atomic polarization is parallel to the chain (we choose a distance $x$ offset from the $x=0$ plane containing the atomic chain in order to avoid seeing the divergent near-fields associated with each atom). Clearly, this figure shows that the field vanishes when moving away from the chain transversally, while it is very intense at the tips of the chain, where the spin wave scatters into an outgoing photon. The field intensity is computed from Eq.(\ref{fielddef}), by taking $\bra{\psi_1} \hat{\Eb}^- (\rb) \hat{\Eb}^+ (\rb) \ket{\psi_1}$. As the input field is vacuum, the intensity only involves calculating two-body correlations $\hat{\sigma}_{eg}^i \hat{\sigma}_{ge}^j$ of the eigenstate.

While the wave vector $k_z$ is strictly a good index for the modes only in the case of the infinite chain, in practice one can also unambiguously associate a distinct, dominant wave vector $k$ with each of the modes $\xi$ in the finite case. Specifically, the discrete Fourier transform of the coefficients that define each mode is peaked around a different value $k_\xi$, which can be used to label the state.  In particular, let us consider a general single-excitation state, which can be written as $\ket{\psi_\xi} =\sum_j c_{\xi}^j \ket{e_j}$, where $\ket{e_j}\equiv \heg^j \ket{g}^{\otimes{N}}$ is defined as the state where atom $j$ is excited while all others are in their ground states. Then, we define the discrete Fourier transform of the associated coefficients $\tilde{c}_{\xi}^k \equiv N^{-1/2} \sum_j e^{\ii k j d} c_{\xi}^j$, for discrete values of $k =2\pi m /N d$ ($1 \leq m \leq N$). For each value of $\xi$, the function $\tilde{c}_{\xi}^{k}$ shows a well defined peak at a distinct value of $k=k_{\xi}$. In Figs.~\ref{fsFig1}(b),(c) (circles) we plot the decay rates $\Gamma_{\xi}$ and energy shifts $J_{\xi}$ of each mode as indexed by the dominant wave vector, for $N=50$ atoms and both transverse and parallel polarizations, overlaid with the infinite lattice result. There is good agreement between them. For the decay rates of the finite chain, the points also correspond to those along a vertical cut in Fig.~\ref{fsFig3}(a), at the fixed value of $d/\lambda_0=0.2$. 

The exact behavior of $\Gamma_\xi$ depends on the microscopic details, such as the polarization of the atoms. For instance, for two-level atoms, the smallest decay rate decreases monotonically as $d/\lambda_0\rightarrow 0$, while for transverse polarization it oscillates. Regardless of these details, however, the scaling with $N$ of the few lowest decay rates seems to show a universal behavior, going like $\Gamma_{\xi}/\Gamma_0 \sim \xi^2/N^{3}$. In Fig.~\ref{fsFig3}(c), we show the $1/N^3$ scaling for the three lowest eigenstates as a function of $N$, while in Fig.~\ref{fsFig3}(d) we show the $\xi^2$ scaling for fixed $N=20$. The scaling with $\xi$ is satisfied for all $\Gamma_{\xi}/\Gamma_0 \ll 1$. For transverse polarization, there is a particular value of lattice constant (that tends to $d/\lambda_0 \sim 0.25$ as the atom number increases), for which the decay rates do not follow exactly the scaling with $\xi$. We believe that this is related to the fact that for transverse polarization and $d/\lambda_0=0.25$ the band structure becomes flat at the edge of the Brillouin zone. Nevertheless, and as we discuss in Appendix~\ref{AppB}, the scaling $\Gamma_{\xi}\sim \xi^2/N^{3}$ seems to appear rather generically for finite-size, one-dimensional photonic crystal structures.\\

\begin{figure}
\centerline{\includegraphics[width=\linewidth]{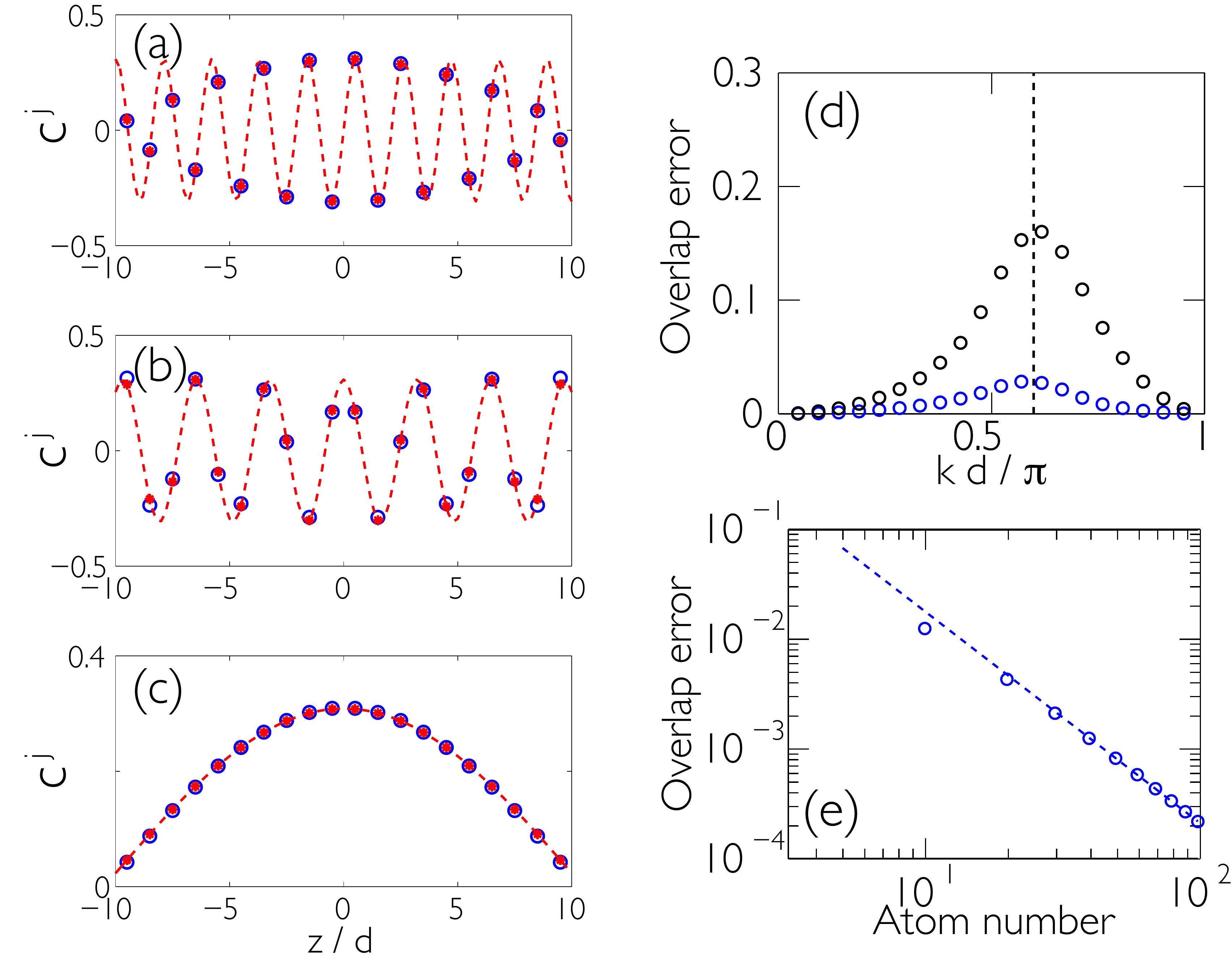}}
\caption{Comparison between ansatz of Eq.(\ref{Eq:SingleParticleStates}) and exact single-excitation eigenstates in an atomic 1D chain. Here, we identify a selected number of modes based upon their dominant wave vector k, and compare the spatial wave-function coefficients $c^j$ with an ansatz built from the same wave vector: {\bf (a)} $k d=\pi N /(N+1)$ (most subradiant state), {\bf (b)} $k \sim k_0$ (close to light line) and {\bf (c)} $k d=\pi/(N+1)$ (most radiant state), for parallel atomic polarization. Blue and red circles denote the coefficients of the exact state and ansatz respectively, while the dashed red line indicates the function $\cos(k_n z)$ or $\sin(k_n z)$ associated with each mode. {\bf (d)} Error in overlap $\varepsilon$ between exact state and ansatz as a function of $k$, for parallel (blue) and transverse (black) atomic polarization. The error decreases far from the light line (denoted by black dashed line). {\bf (e)} Scaling of $\varepsilon$ with particle number $N$ for the most subradiant state. The lattice constant is set to $d/\lambda_0=0.3$ and [except in (e)] $N=20$.}
\label{fsFig4}
\end{figure}

\begin{figure*}
\centerline{\includegraphics[width=\linewidth]{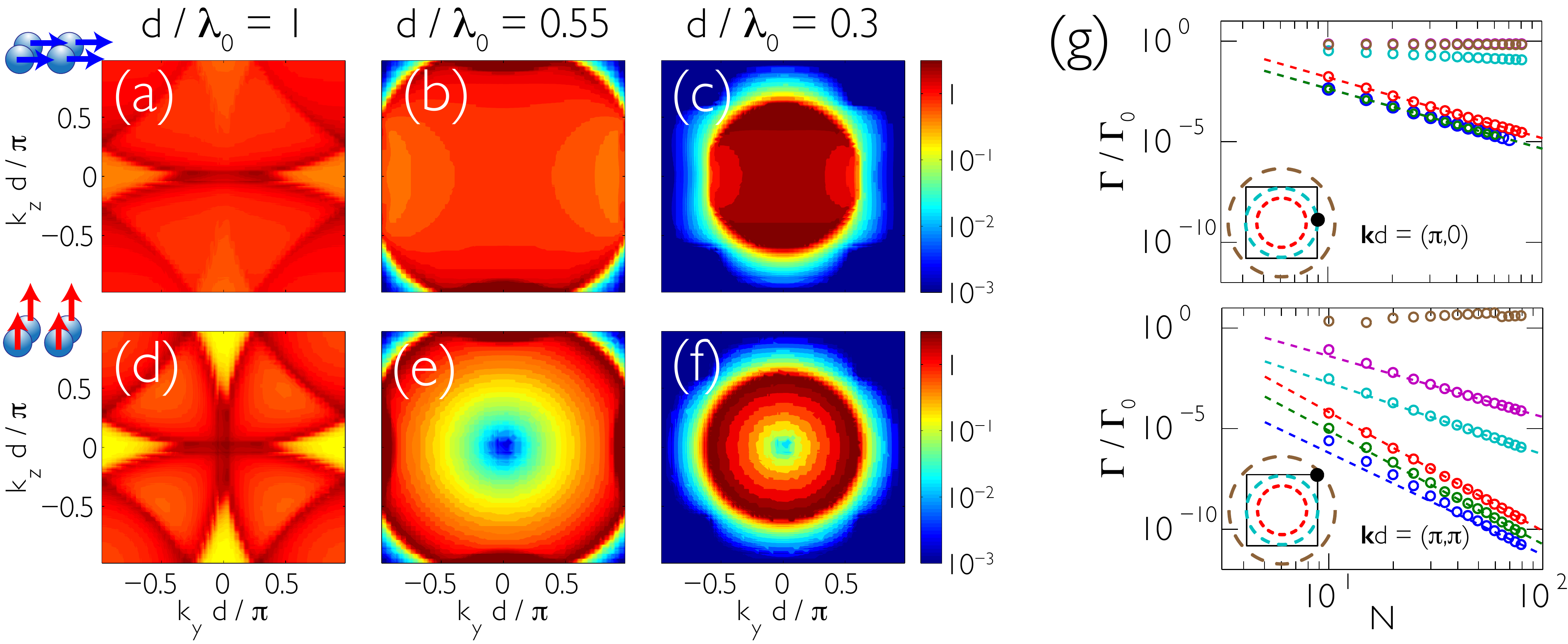}}
\caption{Single-excitation collective modes in 2D square array of $50 \times 50$ two-level atoms. Collective decay rates as a function of predominant wave vector ($k_y$,$k_z$) associated with the mode, for different values of $d/\lambda_0$: {\bf (a)}, {\bf (b)} and {\bf (c)} are for parallel atomic polarization along $\hat{y}$ axis; {\bf (d)}, {\bf (e)} and {\bf (f)} are for transverse polarization. Subradiant and guided modes arise outside the circle defined by the light line ($|\kb| = k_0$), if $d/\lambda_0 < 1/\sqrt{2}$. For $d/\lambda_0 < 1$ and transverse atomic polarization, a different class of subradiant states emerge at $\kb=(0,0)$. {\bf (g)} Scaling of decay rates with $N$ ($N^2$ being the number of atoms) for two particular modes $(k_y,k_z) = (\pi/d,0)$ (top) and $(k_y,k_z) = (\pi/d,\pi/d)$ (bottom). Different colors are for different values of $d/\lambda_0 = 0.25,0.3,0.4,0.5,0.6,1/\sqrt{2}$ (blue, green, red, cyan, purple and brown, respectively). The dashed lines are guides to the eye for $N^{-3}$ and $N^{-6}$ scalings.}
\label{fsFig5}
\end{figure*}

{\bf \textit{Ansatz for single-excitation collective modes.--}}
If the chain is finite the single-excitation collective modes are not spin waves with pure wave vector $k_z$, and contrary to the infinite lattice case, they are not orthonormal in general. Nevertheless, as discussed earlier, the eigenstates can be characterized by a dominant wave vector $k$ that connects well with the infinite case. Furthermore, we find that the states far from the light line (including the most subradiant modes as well as those where $\kb \sim 0$) are almost orthonormal, and display a relatively simple spatial structure. This motivates us to find an orthonormal set of functions that approximates well these modes. For an even number of sites $N$, the wave-function coefficients $c_\xi^ j$ of the exact collective modes are close to the orthonormal set of functions defined by wave vector $k_n$:
\begin{align}
c_{{\rm ans,}k_n}^{j} &= \sqrt{2/(N+1)}\cos(k_n x_j) &\qquad{} \textrm{if $n$ odd}\notag\\
c_{{\rm ans,}k_n}^{j}  &= \sqrt{2/(N+1))} \sin(k_n x_j) &\qquad{} \textrm{if $n$ even}.
\label{Eq:SingleParticleStates}
\end{align}
Here $k_n d = \pi n/(N+1)$, $n=1,2,...,N$ and the atomic positions $x_j  = j d -x_0$ ($1\leq j \leq N$) and  $x_0=(N+1) d/2$. Figure \ref{fsFig4}(a)-(c) shows the exact coefficients $c_{\xi}^j$ for the most subradiant state ($\xi=1$), a state with dominant wave vector near the light line, and the most radiant state ($\xi=N$), for $N=20$ atoms, together with the corresponding ansatz coefficients. The error between the exact wave function and ansatz can be quantified by considering the mismatch in overlap between the two states, $\varepsilon=1-|\Braket{\psi_\text{ans}}{\psi_\xi}|^2$. In Fig.~\ref{fsFig4}(d) this is plotted as a function of the wave-number $k$ associated to each of the modes. Generally, the error is negligible, except for states close to the light line. In Fig.~\ref{fsFig4}(e) we show that for the most subradiant state this error vanishes with chain length $N$ as $\varepsilon \sim N^{-2}$.

Moreover, this ansatz not only approaches the spatial pattern of the wave function, but its decay rate defined as $\Gamma_{\rm ans} = -(2/\hbar) \textrm{Im}\bra{\psi_\text{ans}} \mathcal{H}_\textrm{eff} \ket{\psi_\text{ans}}$ captures the same scaling with the index $\xi$ and $N$: $\Gamma_\textrm{ans} \propto \xi^2 / N^3$. The overall proportionality constant varies depending on the microscopic details (such as the polarization or the value of $d/\lambda_0$). For instance, for $d/\lambda_0=0.3$, $\Gamma_\text{ans}/\Gamma_{\xi=1} \approx 3/2$ (parallel polarization) and $\Gamma_\text{ans}/\Gamma_{\xi =1} \approx 8$ (transverse polarization). The comparison between $\Gamma_\xi$ and $\Gamma_\text{ans}$ is shown in Fig.~\ref{fsFig3}(c)-(d) (solid circles correspond to the ansatz).\\

{\bf{2D Square Array.}}\\
The previous results are not specific to the one-dimensional chain and can be extended to other lattice geometries. As an example, let us consider a finite square array of $N \times N$ atoms spanning the $\hat{y}$-$\hat{z}$ plane. Just like in the linear chain, we can diagonalize the block Hamiltonian with a single excitation and find the decay rates associated with the $N^2$ collective modes. We can also define an ansatz wave function with coefficients $c_{{\rm ans,}\kb}^\textrm{\bf j} \propto c_{{\rm ans,}k_y}^ {j_y}  c_{{\rm ans,} k_z}^ {j_z}$, where $c_{{\rm ans,}k}^ {j}$ are the coefficients for the one-dimensional ansatz Eq.~(\ref{Eq:SingleParticleStates}). We can then associate with each of the collective modes a pair of values $\left( k_y, k_z \right)$, which lies  within the first Brillouin zone, and for which the corresponding ansatz produces the highest overlap with the exact state. 

In Fig.~\ref{fsFig5} we have plotted the decay rates as a function of $\left( k_y,k_z \right)$ after diagonalizing the Hamiltonian for an array of $50 \times 50$ atoms, for different values of $d/ \lambda_0 = \left\{1, 0.55, 0.3\right\}$. Figures \ref{fsFig5}(a)-(c) depict the case of polarization parallel to the array, and show the emergence of subradiant states (corresponding to wave vectors beyond the light line) for $d/\lambda_0 < 1/\sqrt{2}$. As it can be seen in this figure, the most subradiant modes correspond to those at the edges of the Brillouin zone, \textit{i.e.}, $\left( |k_y|,|k_z| \right) \sim (\pi/d,\pi/d)$. The wave-function amplitude of this mode is a generalization of the one shown in Fig.˜\ref{fsFig4}(a) for the 1D chain, where now the alternating plus and minus sign in the amplitude exhibits a checkerboard pattern. Figures \ref{fsFig5}(d)-(f) depict the decay rates for the case of transverse polarization. Here, one sees a set of subradiant states emerges beyond the light line for $d/\lambda_0 < 1/\sqrt{2}$ as before, and also a set of subradiant modes around $(k_x,k_y)=(0,0)$ for $d/\lambda_0 <1$, in agreement with the infinite lattice analysis.

While we can expect that in general the decay rate of the most subradiant modes will be suppressed with the system size, the scaling is more complex than for the linear chain. Nevertheless, we have numerically verified that for collective modes at the edge of the Brilllouin zone, and if $d/\lambda_0$ is small enough, the decay rate will scale as $\Gamma \sim N^{-\alpha}$. In particular, we find that for $\left( k_y,k_z \right) = (\pi/d,\pi/d)$ (most subradiant state), $\alpha = 6$ for $d/\lambda_0 <1/2$ and $\alpha = 3$ for $1/2 \leq d/\lambda_0 < 1/\sqrt{2}$, while for $\left( k_y,k_z \right) = (\pi/d,0)$, $\alpha = 3$ for $d/\lambda_0 < 1/2$. For ranges of $d/\lambda_0$ not included above, the decay rates are not suppressed with increasing system size, since in that case the wave vector $\left(k_x,k_y\right)$ of the two states lies within the light line. These scalings are shown for the two states in Fig.~\ref{fsFig5}(g).\\

\begin{figure}[t]
\centerline{\includegraphics[width=0.9\linewidth]{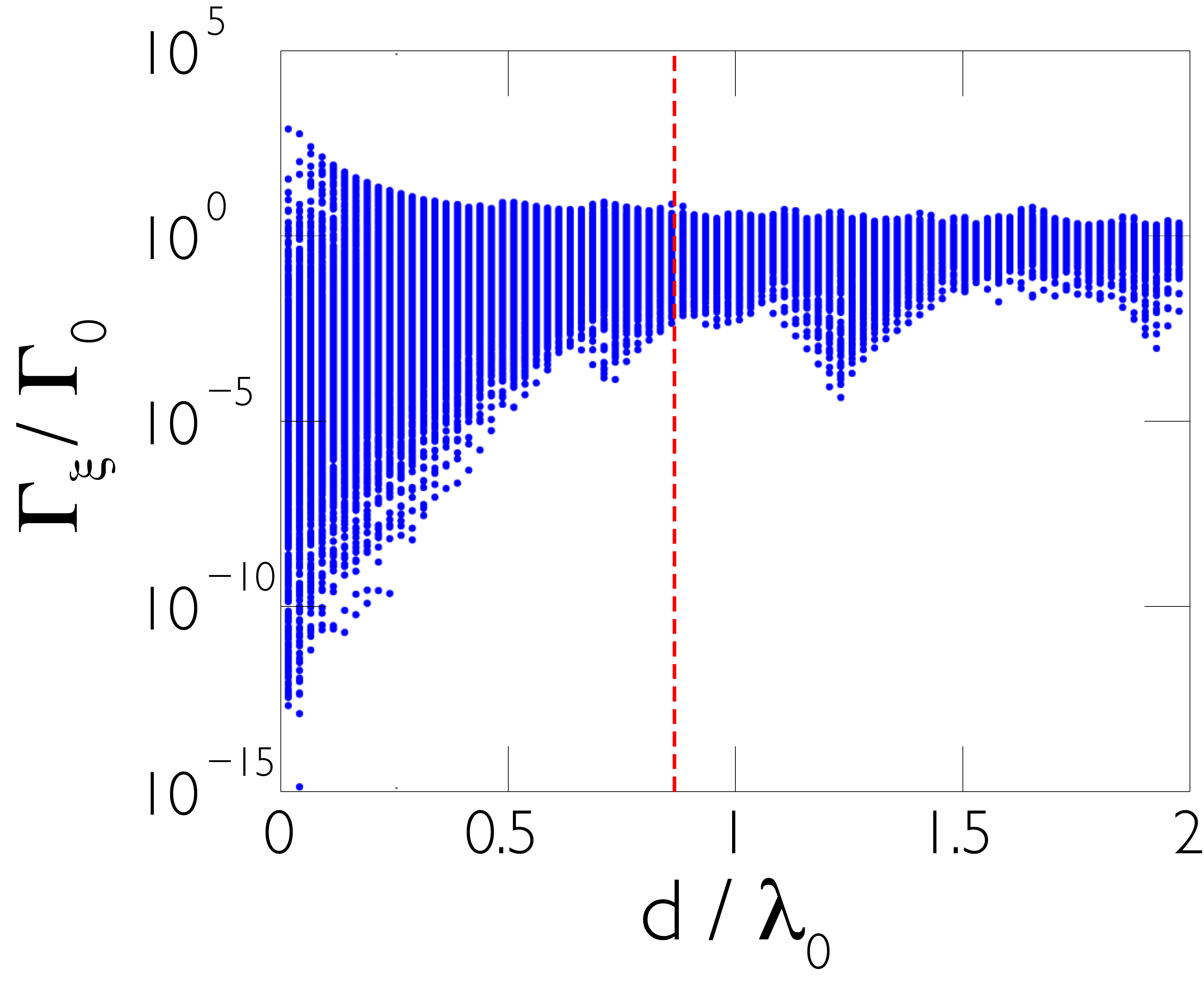}}
\caption{Decay rates of the $N^3$ modes at different lattice constants $d/\lambda_0$ for a cubic lattice of $10 \times 10 \times 10$ atoms with linear polarization along one of the lattice axes. The dashed red line corresponds to the particular value of $d/\lambda_0=\sqrt{3}/2$, where the light line touches the edge of the first Brillouin zone in 3D. In 3D, subradiant states can exist even beyond this value.} \label{fsFig6}
\end{figure}

{\bf{3D Cubic Array.}}\\
While the extension from 1D chains to 2D arrays is conceptually straightforward, it appears that three-dimensional lattices are governed by different physics. In particular, in infinite 1D and 2D arrays, Bloch modes diagonalize the system, and subradiant modes can be characterized as ``guided'' as the associated electromagnetic fields are evanescent in the spatial directions transverse to the array. In contrast, while Bloch modes still diagonalize the system in 3D, the associated fields are necessarily extended over space in all directions. Thus, for a finite-size system, it does not appear that subradiant states can be identified based on the infinite-system results, as was possible in 1D and 2D.

Nonetheless, for completeness, we can still numerically investigate the decay rates for a single excitation in a 3D finite-size lattice. In Fig.~\ref{fsFig6}, we plot the decay rates $\Gamma_{\xi}$  for the $N=10^3$ eigenstates associated with a $10 \times 10 \times 10$ lattice of two-level atoms, in the case that the polarization of the transition is aligned with one of the cubic axes. It can be seen that while decay rates are still most prominently suppressed for lattice constants $d \ll \lambda_0$, the effect of subradiance can survive even for lattice constants $d>\lambda_0$. It would be interesting to further explore the nature of subradiance in 3D systems in future work, and identify conceptual similarities it has to arrays in lower dimensions, if any.\\

\begin{figure}[t]
\centerline{\includegraphics[width=0.85\linewidth]{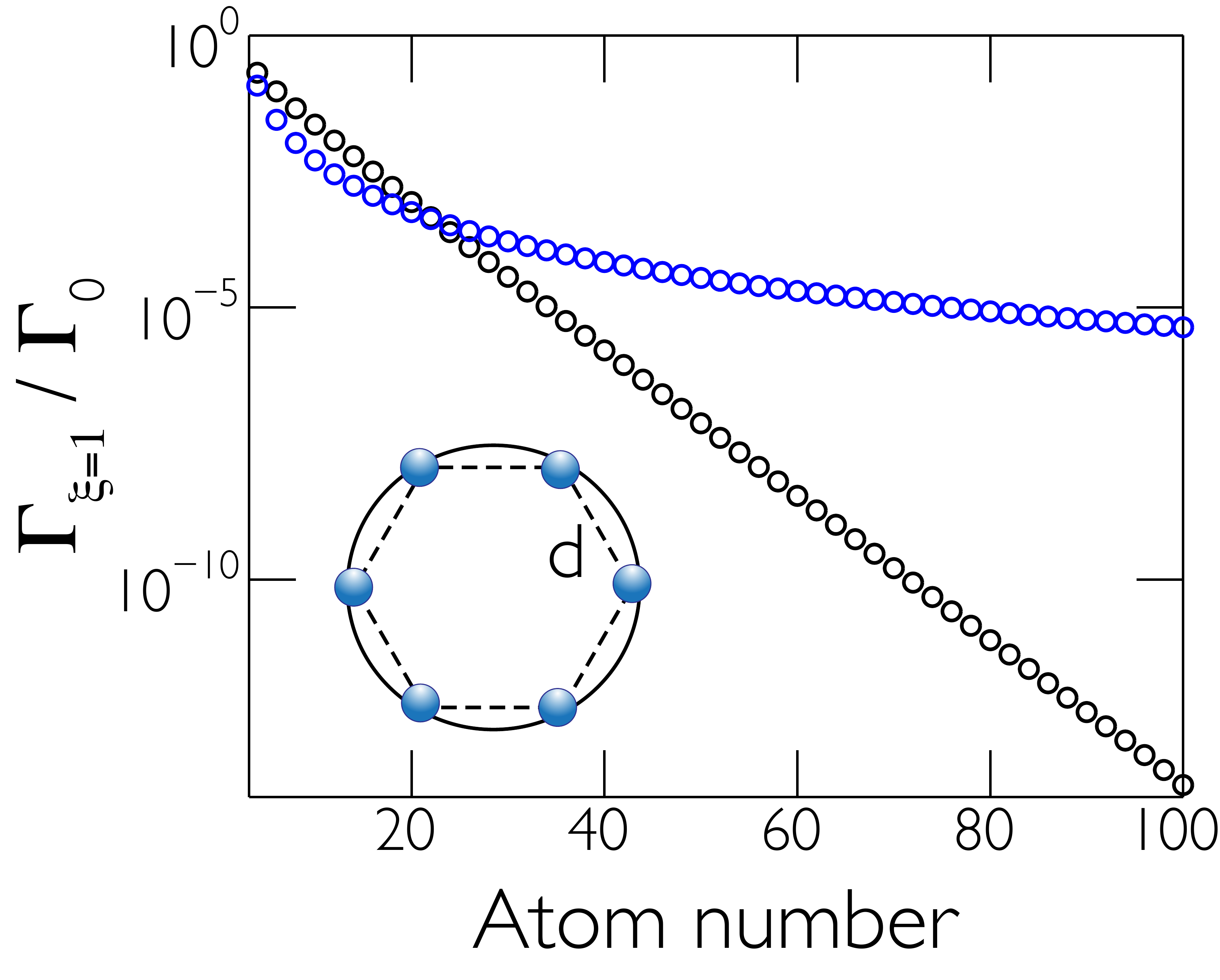}}
\caption{Decay rate of the most subradiant mode as a function of atom number $N$ in a circular ring of 2-level atoms (black circles). Two consecutive atoms are separated a distance $d$ as shown in the sketch (here $d/\lambda_0=0.3$), and the atomic polarization is transverse to the plane defined by the ring. For this geometry, there is an exponential suppression of the most subradiant decay rate with $N$. For comparison, the decay rate of the most subradiant mode of a linear chain with lattice constant $d$ is shown (blue circles), with only polynomial suppression with $N$.} \label{fsFig7}
\end{figure}

{\bf{Atoms in a ring configuration.}}\\
The result that we have found for 1D linear chains, indicating that subradiant modes are guided and that radiation leakage is primarily from the system ends, motivates us to study the decay rates when the atoms form a closed configuration, since this might lead to a stronger suppression in the decay. In particular, we consider now that the atoms are sitting on a circular ring separated by an equal distance $d$ (see sketch in Fig.~\ref{fsFig7}) and with linear polarization transverse to the plane of the ring.

In Fig.~\ref{fsFig7}, for a distance of $d/\lambda_0=0.3$, we numerically diagonalize the single-excitation block Hamiltonian, and plot the decay rate $\Gamma_{\xi=1}$ of the most subradiant state versus atom number $N$. It can be seen that an exponential suppression emerges, $\Gamma_{\xi=1}\sim \exp(-N)$. For the chosen parameters, the minimum decay rate for a ring drops below that of an open chain for $N\gtrsim 20$ atoms. The subradiant modes of the ring can be interpreted as ``whispering gallery modes'', which weakly radiate into free space only via the finite radius of curvature. The exponential suppression with ring radius is analogous to the scaling of radiation losses in a conventional whispering gallery resonator \cite{BK03}.\\

\begin{figure*}[t]
\centerline{\includegraphics[width=\linewidth]{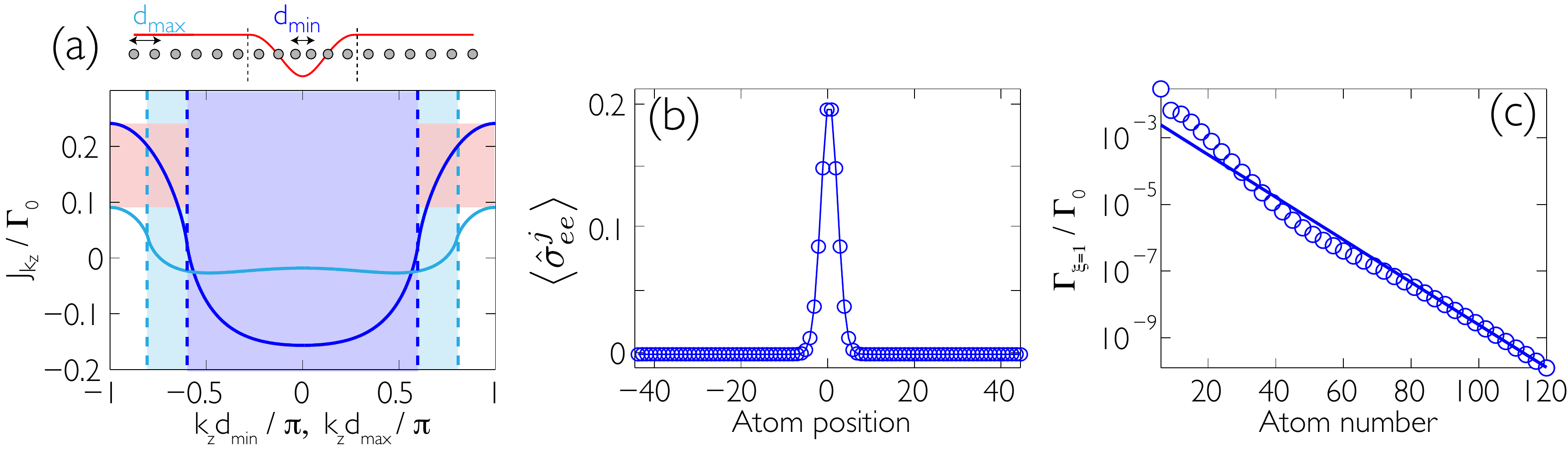}}
\caption{Cavity in an atomic chain with slowly varying lattice constant. The chain along $\hat{z}$ is divided into three regions: in the left and right regions the lattice constant $d$ is uniform and equal to $d_{\rm max}$, while in the middle it changes slowly (from $d_{\rm max}$ to $d_{\rm min}=0.75 d_{\rm max}$). \textbf{(a)} Dispersion relation for two infinite lattices with constant $d_{\rm max}$ (cyan) and $d_{\rm min}$ (dark blue) vs wave vector along the chain (in units of the corresponding lattice constant). The light line is indicated by the vertical dashed lines with the same color. A bandwidth of frequencies (shaded pink) emerges wherein propagating modes exist for lattice constant $d_\text{min}$, but not for $d_{\rm max}$. This allows localized resonances to form within the non-uniform chain. An illustration of the chain is shown (top), where the red line represents how the lattice separation changes along the chain. \textbf{(b)} Excited state population vs atom position corresponding to the fundamental mode in the cavity, illustrated for $N=90$ atoms. \textbf{(c)} Decay rate of the fundamental mode vs atom number, showing an exponential suppression. We have chosen $d_{\rm max}/\lambda_0 = 0.4$ and atomic polarization along the chain axis.} \label{fsFig10}
\end{figure*}
{\bf{Localized resonance in an atomic chain.}}\\
Here, we also show how to achieve a spatially confined mode in a linear 1D chain of atoms, which also exhibits an exponential suppression of decay rate with atom number. This can be achieved by introducing a smooth, local variation in the lattice constant, in analogy to the principles that govern the design of a conventional photonic crystal cavity \cite{JMW95}.

To illustrate this, we consider the geometry schematically depicted in Fig.~\ref{fsFig10}(a). The atomic chain along $\hat{z}$ has been divided into three regions: in the left and right regions the lattice constant  $d$ is uniform and equal to $d_{\rm max}$, while in the middle it changes slowly (from $d_{\rm max}$ to $d_{\rm min}=0.75 d_{\rm max}$) following the red line and creating a defect. The lattice constant in the middle is chosen to follow a sinusoidal variation, $d(z_i) = d_{\rm max}+(d_{\rm min}-d_{\rm max}) \sin^2(3\pi z_i/N)$, where $z_i$ ($i=N/3,...,2N/3$) denotes the atom position. In the same figure, the band structures for an infinite lattice with constant $d_{\rm max}=0.4\lambda_0$ and $d_\text{min}=0.3\lambda_0$ are plotted for the case of atomic polarization parallel to the chain. It can be seen that the smaller lattice constant $d_\text{min}$ supports propagating modes over a range of frequencies (pink shaded region) that lies within the band gap of lattice $d_{\rm max}$. Thus, for the system with slowly varying lattice constant, a set of localized resonances can appear in the middle region, with frequencies situated in the gap of lattice $d_{\rm max}$, and unable to propagate into the left and right regions.

The atomic excited state population associated with the fundamental localized mode (\textit{i.e.}, the mode with no nodes in the population) is illustrated in Fig.~\ref{fsFig10}(b), for a representative case of $N=90$ atoms. For a smooth variation in the lattice constant (here occurring over a region of size $N/3$), one expects that the fundamental mode will have a Fourier transform with an exponentially small weight of wave vectors lying within the light line. Likewise, the leakage of this mode through the left and right regions to the ends of the chain will be exponentially suppressed, leading to an overall exponential suppression in decay rate with increasing atom number $N$. The numerically calculated decay rate of this mode is plotted in Fig.~\ref{fsFig10}(c) as a function of $N$, and clearly confirms the expected behavior.

\begin{figure}[h!]
\centerline{\includegraphics[width=1\linewidth]{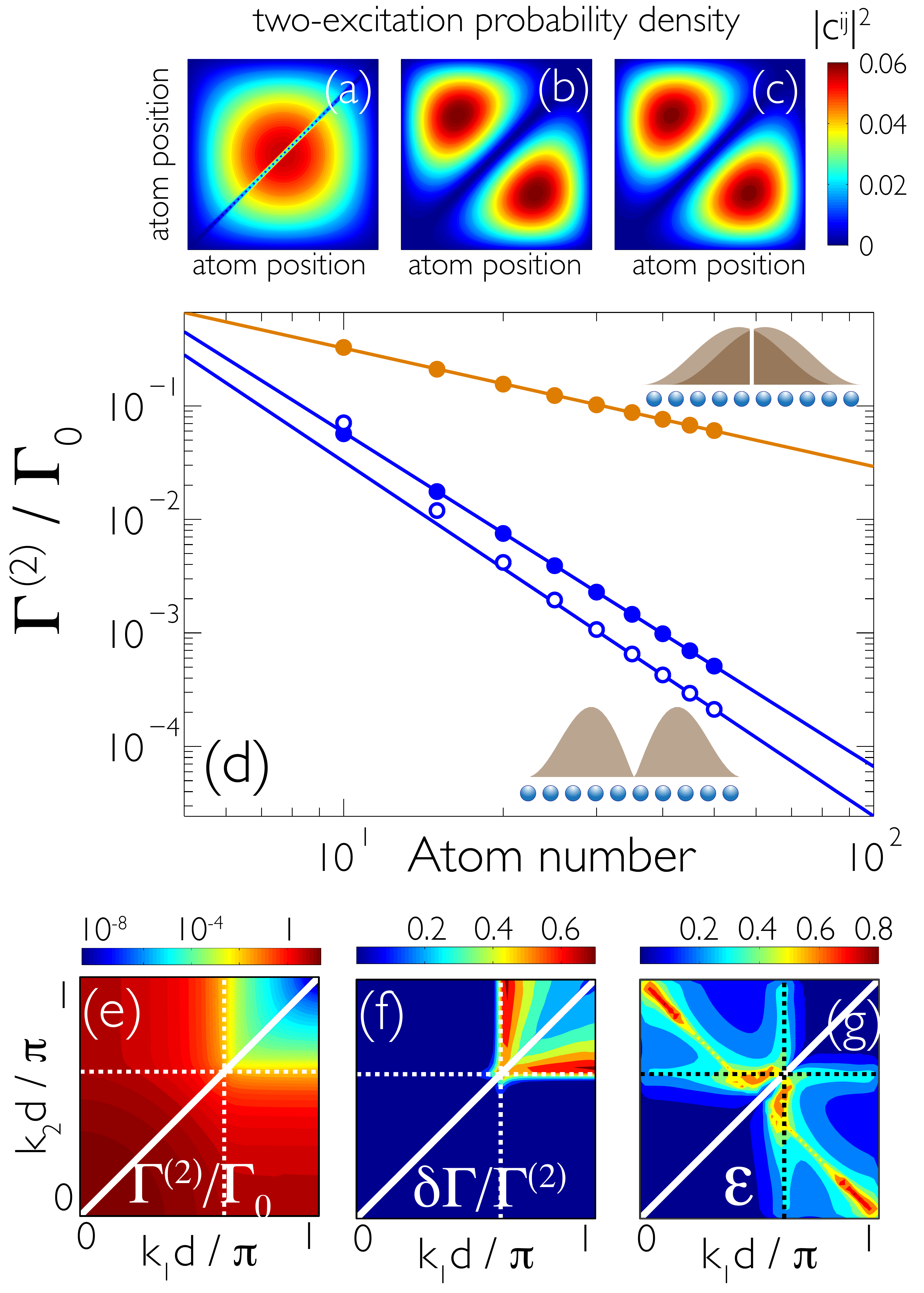}}
\caption{Two-excitation collective modes in a 1D chain of two-level atoms with polarization parallel to the chain. Probability $|c^{ij}|^2$ of atoms $i$ and $j$ to be excited (with $c^{ii}=0$) for: {\bf (a)} state resulting from occupying the most subradiant single-excitation mode twice, \textit{i.e.}, $(S_{\xi=1}^{\dagger} )^{2} \ket{g}^{\otimes N}$; {\bf (b)} most subradiant mode $c^{ij}_{\xi=1}$ (obtained by exact diagonalization); {\bf (c)} fermionic ansatz $c^{ij}_\textrm{ans}$. The two axes denote the excited atoms position ($i$ and $j$). {\bf (d)} Scaling of the collective decay $\Gamma^{(2)}$ with atom number corresponding to the states (a) (orange circles), (b) (blue open circles) and (c) (blue solid circles). The lines are polynomial fits (to the last five points), close to $\sim 1/N$ (a) and $\sim 1/N^3$ (b),(c). {\bf (e)} Decay rates as a function of the associated pair of quasi-momentum values $(k_1,k_2)$ of the two excitations (by construction $k_1\neq k_2$). Subradiant states arise when both $k_1,k_2>k_0$, where $k_0$ corresponds to the light line (dotted lines). {\bf (f)} Relative error $\delta \Gamma / \Gamma^{(2)} = |\Gamma^{(2)}_{\rm sum}-\Gamma^{(2)}| / \Gamma^{(2)}$ between the numerically exact decay rate, and the decay rate estimated from the sum of single-excitation decay rates $\Gamma^{(2)}_{\rm sum}$. {\bf (g)} Error in overlap between exact and antisymmetrized ansatz. In all plots $N$ is fixed to $50$ atoms [except (d)] and $d/\lambda_0 = 0.3$.}\label{fsFig8}
\end{figure}

\subsection{Multi-excitation modes}
\label{SecIIIC}
We now turn to the problem of multiple excitations stored in an atomic 1D lattice. As in the previous sub-section, we consider that the atoms are linearly polarized along the chain axis. However, similar results are found for the transverse polarization case. First, we note that if the Hamiltonian of Eq.(\ref{heff}) were composed of bosonic particles instead of spins (\textit{i.e.}, $\hge,\heg \rightarrow a,a^{\dagger}$) the multiple excitation case would be trivial. In particular, the resulting Hamiltonian would be quadratic in the creation and annihilation operators, and a Fock state of $n$ excitations in a given mode would simply have a decay rate $n$ times that of a single excitation. The fact that we are dealing with spins, where a single spin cannot be excited twice ($\heg^2=0$), leads to highly non-trivial properties of multiply excited states. In this section, we will analyze in detail the spatial properties and scaling of decay rates of multi-excitation subradiant states. While we will not explicitly utilize these states in later sections, these findings might help to provide some initial insight into how many-body physics can be encoded into subradiant manifolds.\\

Let us first consider the two-excitation manifold. A general state within this manifold can be written as $\ket{\psi^{(2)}} = \sum_{i<j} c^{ij} \ket{e_i e_j}$ where now $\ket{e_i e_j}=  \heg^{i} \heg^{j}\ket{g}^{\otimes N}$ corresponds to the state whith atoms $i$ and $j$ excited while the rest remain in the ground state. Although it is necessary only to specify $c^{ij}$ for $i<j$ to define the wave function, in the following plots and for visual appeal we also assign values to $c^{ij}$ for $j<i$, by simply defining $c^{ij}=c^{ji}$. To illustrate that the spin system behaves differently than a bosonic system, we begin by considering the two-excitation state formed by occupying the same single-excitation mode twice. In particular, we construct the two-excitation state given by $\ket{ \psi^{(2)}_{\rm b}} \propto (S_{\xi=1}^{\dagger} )^2 \ket{g}^{\otimes N}$ (properly normalized). Here, $S_{\xi=1}^\dagger = N^{-1/2} \sum_j c^j_{\xi=1} \heg^j$ is the collective operator that creates the most subradiant single excitation in a chain of $N$ atoms, when applied to the ground state. In Fig.~\ref{fsFig8}(a) we plot the corresponding probability density $|c_{\rm b}^{ ij}|^2$ for the case $N=50$. The two-excitation wave function appears relatively smooth, except for a sharp cut $c_{\rm b}^{ ii }=0$ along the diagonal, owing to the fact that a single spin cannot be excited twice. As a result of this feature, the Fourier transform of this state will be relatively broad, and in particular, will contain many components that lie within the light line and can subsequently radiate. Its decay rate, defined as $\Gamma^{(2)}_\text{b} = -(2/\hbar)\textrm{Im} \bra{\psi^{(2)}_\textrm{b}} \mathcal{H}_\textrm{eff} \ket{\psi^{(2)}_\textrm{b}}$ is only suppressed with the length of the chain as $\Gamma^{(2)}_\text{b}  \sim N^{-1}$, in stark contrast to the single-excitation case. This scaling is shown in Fig.~\ref{fsFig8}(d) (orange circles).

Numerically, we now exactly diagonalize the Hamiltonian in the two-excitation manifold, and identify the most subradiant state. The scaling of its decay rate with $N$ is plotted in Fig.~\ref{fsFig8}(d) (blue open circles), and is seen to preserve the scaling $\Gamma \sim N^{-3}$ present in the single-excitation manifold. The probability density $|c^{ij}_{\xi=1}|^2$ for the case of $N=50$ is plotted in Fig.~\ref{fsFig8}(b). The wave function appears distinctly different than that of Fig.~\ref{fsFig8}(a), and in particular, it appears that the two excitations are smoothly repelled from one another.

From the previous considerations, it is apparent that the most subradiant states should simultaneously satisfy that they are composed predominantly of wave vectors beyond the light line, and that the excitations are smoothly repelled from one another in order to avoid sharp kinks in the wave function. This inspires us to try an antisymmetric (or fermionized) ansatz $\ket{\psi_{\rm ans}^{(2)}}$ for the wave function of the form $c_{{\rm ans,}k_1 k_2}^{ij} \propto c_{\text{ans},k_1}^i c_{\text{ans},k_2}^j - c_{\text{ans},k_2}^i c_{\text{ans},k_1}^j$ (properly normalized). Here $c_{\text{ans,} k_n}^i$ denote the coefficients of the single-excitation orthonormal ansatz Eq.(\ref{Eq:SingleParticleStates}) associated with the wave vector $k_n$. Such an ansatz naturally constructs a state that incorporates ``Pauli exclusion'', and a smooth separation of excitations in space. Taking $k_1d = \pi N/(N+1)$ and $k_2 d=\pi (N-1)/(N+1)$, \textit{i.e.}, building a two-excitation state from the two most subradiant single-excitation states, yields a wave function $c_{\rm{ans}}^{ij}$ that agrees well with the exact one, as seen in Fig.~\ref{fsFig8}(c) where the probability density is plotted. Moreover, the decay rate associated with this state scales again with the particle number as $\Gamma^{(2)}_\textrm{ans} \sim N^{-3}$, as it is shown in Fig.~\ref{fsFig8}(d) (blue solid circles). 

We can then associate with each of the exact two-excitation collective states a pair of quasi-momentum values $\left\{ k_1, k_2 \right\}$ for which the wave-function overlap with the ansatz is maximum. The exact decay rates as a function of these values are plotted in Fig.~\ref{fsFig8}(e). This figure shows that when both $k_1, k_2 > k_0$ the decay rates are strongly suppressed, and we can identify this region as the one containing the subradiant states. In fact, the sum of decay rates of the single-excitation modes used to construct the ansatz, \textit{i.e.}, $\Gamma^{(2)}_\textrm{sum} = \Gamma_{\textrm{ans},k_1}+\Gamma_{\textrm{ans},k_2}$, is not far from the exact value. This is quantified by the relative error $\delta \Gamma \equiv |\Gamma^{(2)}-\Gamma^{(2)}_\textrm{sum}|/\Gamma^{(2)}$ and it is plotted in Fig.~\ref{fsFig8}(f). For completeness, we also show in Fig.~\ref{fsFig8}(g) the error in overlap between each two-excitation eigenstate and the best-matched ansatz state, $\varepsilon=1-|\Braket{\psi}{\psi^{(2)}_\text{ans}} |^2$. This error is very small in the subradiant region.

\begin{figure}[b]
\centerline{\includegraphics[width=0.9\linewidth]{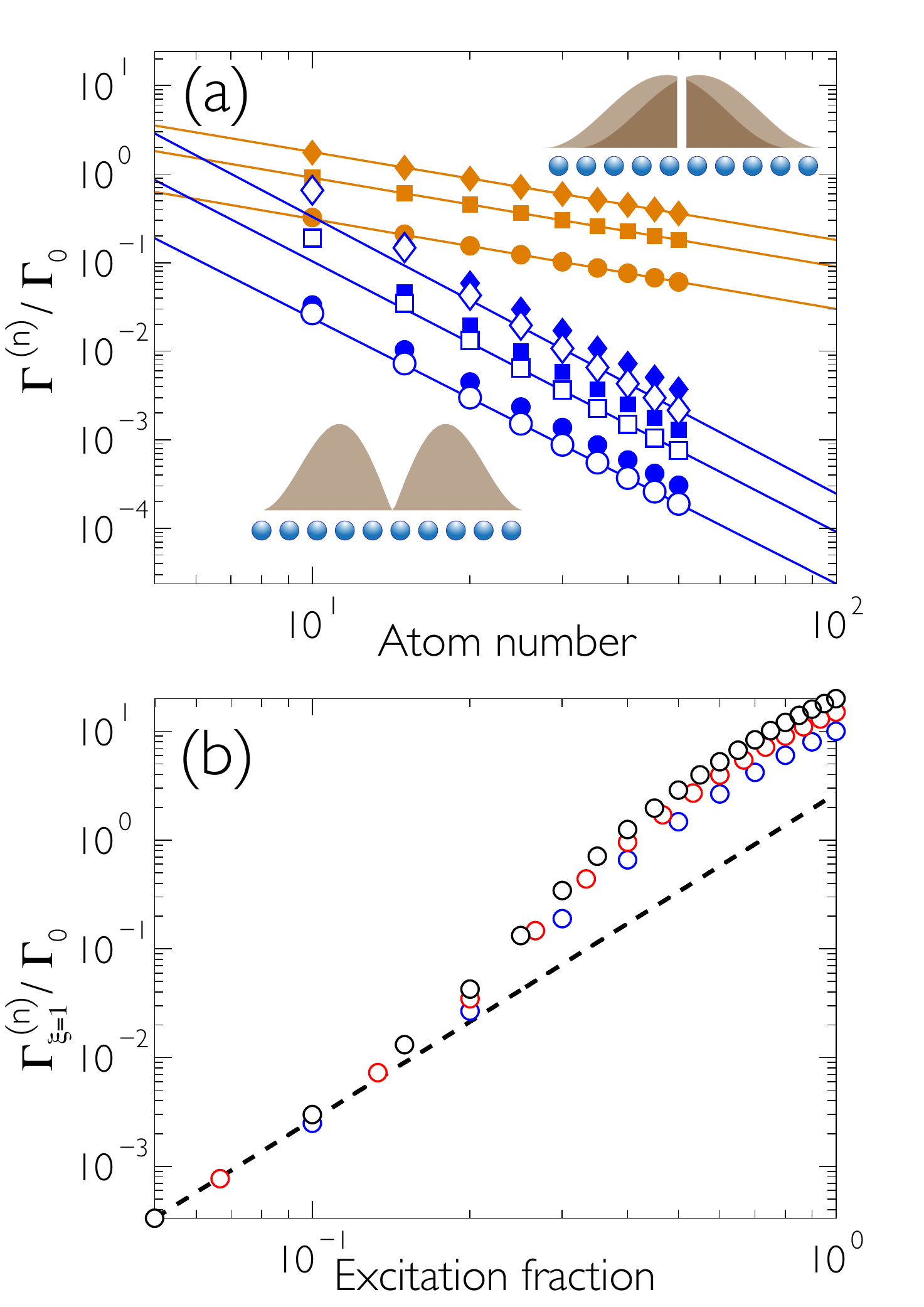}}
\caption{Multi-excitation states in a 1D chain of two-level atoms with polarization parallel to the chain. {\bf (a)} Scaling of the decay rate $\Gamma^{(n)}_{\xi = 1} / \Gamma_0$ of the most subradiant $n$-excitation state with the atom number $N$ (blue open symbols). For comparison, the decay rates of the Fock state constructed with $n$ excitations in the most subradiant single-excitation mode $\Gamma^{(n)}_{\rm b}$ (orange solid symbols), and for the fermionic ansatz $\Gamma^{(n)}_{\rm ans}$ (blue solid symbols) are shown ($n=2,3,4$ are denoted by circles, squares and diamonds). The lines are polynomial fits close to $\Gamma^{(n)}_{\xi=1} \sim N^{-3}$ and $\Gamma^{(n)}_{\rm b} \sim N^{-1}$, respectively. The sketches on top of the atomic chain represent the two-excitation density profile. {\bf (b)} Scaling of the decay rate $\Gamma^{(n)}_{\xi=1} / \Gamma_0$ of the most subradiant state with the excitation density $n/N$. The dashed line corresponds to the predicted scaling $\sim (n/N)^3$ valid at low excitation density ($n/N \ll 1$). Blue, red and black are for $N=10$, $N=15$ and $N=20$, respectively. All plots are for $d/\lambda_0=0.3$ and atomic polarization parallel to the chain.}\label{fsFig9}
\end{figure}

In the more general case of $n$ excitations, the most subradiant mode and its decay rate $\Gamma^{(n)}_{\xi = 1}$ can be found as in the previous cases by exactly diagonalizing the corresponding block Hamiltonian. For a low density of excitations, the scaling of the decay rate with the chain length is still as in the single and two-excitation manifolds, \textit{i.e.}, $\Gamma^{(n)}_{\xi = 1} \sim N^{-3}$, as shown in Fig.~\ref{fsFig9}(a) (open blue symbols). For comparison, we also show in the same figure (orange symbols) the decay rate of the state with $n$ excitations in the most subradiant mode, \textit{i.e.}, $\ket{ \psi^{(n)}_{\rm b}} \propto (S_{\xi=1}^{\dagger} )^n \ket{g}^{\otimes N}$, which scales as in the two-excitation case, $\Gamma_{\rm b}^{(n)} \sim N^{-1}$. 

One can also numerically evaluate the error $\varepsilon$ in overlap between the most subradiant state and an ansatz state, $\ket{\psi_{\rm ans}^{(n)}} = \sum_{i_1<i_2<...<i_N} c_{\textrm{ans},k_1 k_2 ... k_N}^{i_1 i_2... i_N}$. Here the wave-function amplitudes $c_{ \textrm{ans},k_1 k_2...k_N}^{i_1 i_2... i_N}$ are generalized from the two-excitation case, and constructed from the Slater determinant of $n$ single-excitation wave-function ansatz coefficients  $c_{\textrm{ans},k_n}^{i_n}$. For the most subradiant mode these correspond to the $n$ most subradiant single-excitation modes and in the large atom number limit the error is found to scale like $\varepsilon \sim N^{-2}$. 

If the ansatz holds, then for $n$ excitations one expects that the decay rate for the most subradiant state scales like $\Gamma^{(n)}_{\xi =1}/\Gamma_0 \sim \sum_{m=1}^{n} m^2/N^3 \sim (n/N)^3$. In Fig.~\ref{fsFig9}(b), we compare this predicted scaling with numerically calculated values of $\Gamma_{\xi = 1}^{(n)}$, and find qualitatively good agreement for low excitation density $n/N\ll 1$.  We note that the prediction of the ansatz also seems physically reasonable in that it can be extended to the thermodynamic limit, as it predicts a decay rate that only depends on the density $n/N$ of excitations.

\section{Atoms coupled to a nanofiber: selectively-radiant states}
In the previous section we elucidated the nature of subradiant states in atomic arrays, whose long-lived nature arises from weak coupling to all propagating electromagnetic modes. Subradiant manifolds themselves might be useful for many purposes, for example, to accumulate interactions without dissipation in order to realize strongly correlated states. However, to realize an efficient atom-light interface, one would instead prefer to utilize a set of atomic states that strongly radiate into a desired electromagnetic mode (or set of modes) through constructive interference, while destructive interference simultaneously suppresses the emission rate into all undesired modes. We term states that satisfy this property to be ``selectively radiant," as the overall emission rate might not be small, but the branching ratio into desired versus undesired channels could be extremely high. It should be noted that such a definition of ``selectively radiant" is somewhat arbitrary -- for example, even a single isolated atom emitting into a dipole radiation pattern is selectively radiant, if the preferred optical mode is defined to be the dipole pattern itself. In practice, however, the collection efficiency of a dipole pattern with realistic optics is quite small \cite{EK00, EK01,DJD05,KCA08,WGH08,HSH11}, and a functionally useful definition should involve a mode (\textit{e.g.}, focused Gaussian beam or guided mode of a dielectric structure) that is generally accepted to be efficient to match to.

Here, we show that one natural way to realize and utilize selectively radiant states is to couple one-dimensional atomic chains to the guided modes of an optical nano-structure (such as an nanofiber). Qualitatively, for sufficiently small lattice constants $d<\lambda_0/2$, a set of spin-wave excitations with associated wave vector $|k_z|>k_0$ emerge, which inefficiently radiate into free space as the wave vector lies beyond the light line. However, as an optical mode guided by a high-index dielectric itself has a wave vector $|k_z|>k_0$, we show that it is possible that a set of spin-wave excitations simultaneously experiences an enhanced emission rate into the guided modes while being subradiant to free space. We will provide an explicit construction of a protocol where selectively radiant states are exploited, involving a quantum memory or photon storage. We find in particular that these states enable an exponential improvement in the error probability versus atom number, over previously known bounds.

This section is organized as follows. Section IV A describes the nanofiber and provides the Hamiltonian that governs the interactions between the atoms located in the vicinity of the nanostructure. We introduce the ``collective emission" model, which accounts for atom-atom interactions both through the guided and non-guided modes of the fiber. We also present the ``independent emission" model, in which atoms interact through the guided modes but coupling via free-space modes is neglected. This thus represents the ``standard model" of atom-light interactions specifically applied to nanofibers. In particular, it reproduces previously accepted bounds for fidelities of photon storage, against which the ``collective emission" model can be compared. Section IV B describes linear optical processes (\textit{i.e.}, single-photon transmission and reflection) for two-level atoms coupled to the fiber. We show how the conventional figure of merit, the optical depth, is not sufficient to characterize optical transport through the array when the collective emission into non-guided modes is taken into account. In Section IV C we study how selective radiance influences electromagnetically induced transparency \cite{HFI1990,HFI1991,FL00,L03,FIM05}, a phenomenon that is commonly used in photon storage protocols. In particular, we show that the bandwidth-delay product, which quantifies the number of photons that can be stored in the atomic medium, scales linearly with the number of atoms. The linear scaling is characteristic of ideal, non-lossy systems, and is in contrast to the independent emission model, which predicts a scaling that goes with the square root of the optical depth. In this sub-section, we also provide the first glimpse of improvement in photon storage beyond traditional bounds. Finally, in Section IV D we demonstrate how to achieve an exponential suppression with the atom number on the infidelity of a quantum memory. 

\subsection{Description of the nanofiber}
\begin{figure}
\centerline{\includegraphics[width=\linewidth]{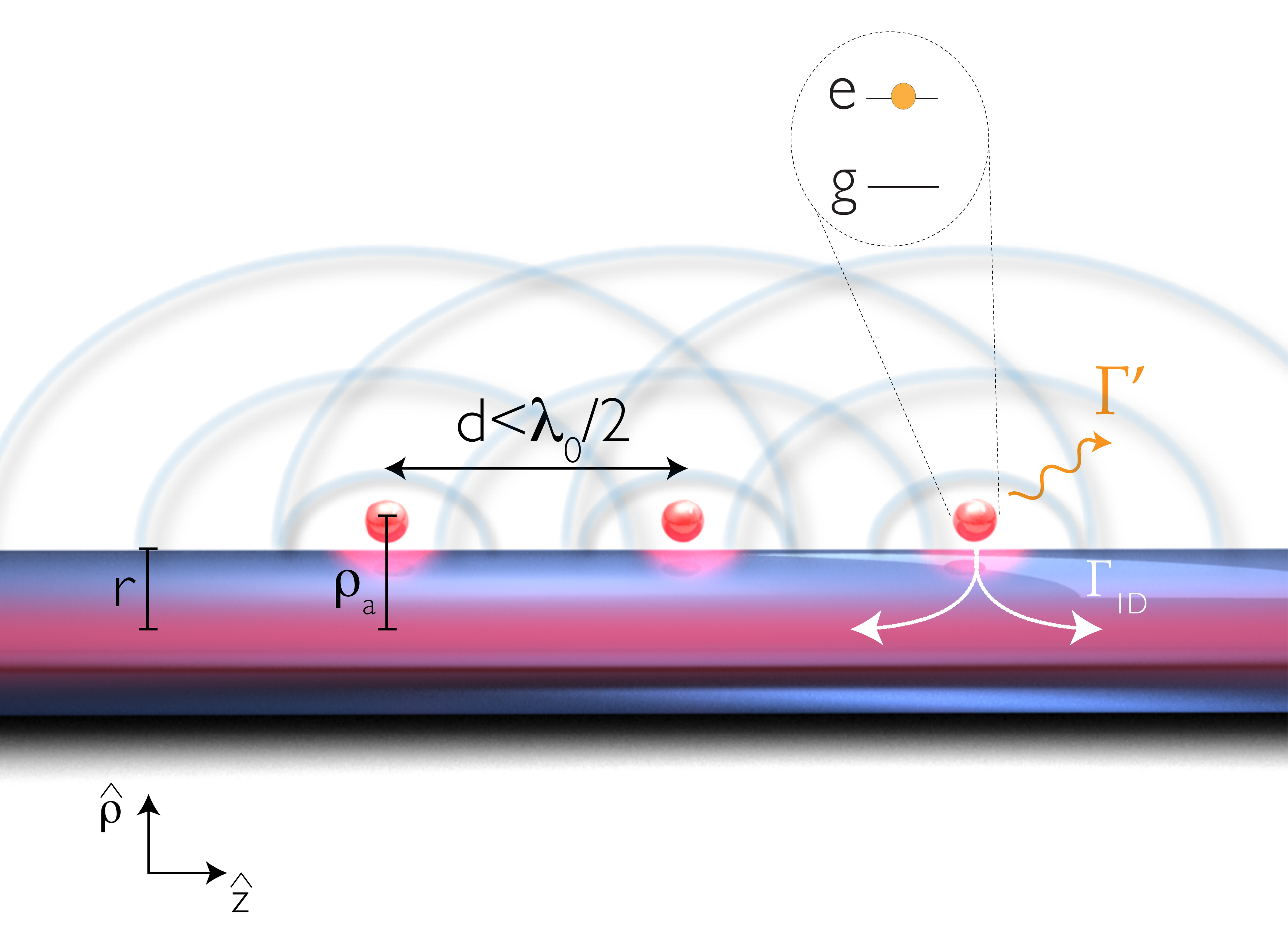}}
\caption{Schematic of the setup under consideration: $N$ two-level atoms are located in the vicinity of a dielectric nanofiber of dielectric constant $\epsilon$ and radius $r$, at a distance $\rho_\text{a}$ from the center of the fiber, and at a constant distance $d$ from each other. For the calculations in this manuscript, we take $k_0r=1.2$, $\rho_\text{a}=1.5r$, and $\epsilon=4$. The atoms interact with each other not only through the guided mode, but also through non-guided photons. The single-atom emission rates into the fiber and into free space are $\ga$ and $\Gamma'$, respectively.} \label{nFig1}
\end{figure}
The possibility of enhancing atom-light interactions through selective radiance should exist for any nanophotonic structure where atoms can be periodically trapped, including in nanofibers \cite{VRS10,GCA12,polzik,CGC16} and 1D and 2D photonic crystal waveguides \cite{DHH15,DHC15,GHH15,HGA16}. For complicated structures, however, the Green's function cannot be obtained analytically. Furthermore, while the Green's function can be calculated numerically \cite{S00}, to do so with sufficient accuracy appears quite challenging (in particular, it must be calculated with enough accuracy so that diagonalization correctly captures subradiant emission rates that scale like large inverse powers of $N$). Motivated by this observation, here we focus on a special geometry where the Green's function can be exactly obtained, which consists of a chain of atoms coupled to guided modes of an infinite, cylindrical nanofiber.

We consider that the chain of atoms lies parallel to the axis of a dielectric nanofiber oriented along $\hat{z}$, with radius $r$ and relative permittivity $\epsilon$ (or corresponding refractive index $n_\text{fiber}=\sqrt{\epsilon}$). As shown in Fig.~\ref{nFig1}, the distance between the atoms and the center of the nanofiber is $\rho_\text{a}$, and the orientation of their dipole transition is directed along $\hat{\rho}$, perpendicular to the axis of the nanofiber (\textit{i.e.}, $\db=\wp\hat{\rho}$). The Green's function for such a nanofiber can be found analytically \cite{S1941,YS08_2,KGB05,DSF10,LR14,QBH16,LR17}. In particular, we follow the work of Klimov and Ducloy ~\cite{KD04}. In the following, we provide a qualitative description of the derivation, while details are given in Appendix~\ref{AppC}. The first step in the derivation is to separate the Green's function into two terms, \textit{i.e.}, $\mathbf{G}(\rb_i,\rb_j,\omega_0)=\mathbf{G}_0(\rb_i,\rb_j,\omega_0)+\mathbf{G}_\text{sc}(\rb_i,\rb_j,\omega_0)$. Here $\mathbf{G}_0$ is the already-known vacuum Green's function given by Eq.~\eqref{Greens_def}, which corresponds to the field emitted by a dipole in free space, and $\mathbf{G}_\text{sc}$ is a general solution to the sourceless wave equation, which will physically correspond to the (thus far unknown) field scattered by the nanofiber. Exploiting the cylindrical symmetry of the problem, one can employ separation of variables and expand the vacuum and scattered Green's functions using a set of functions $f_{m, k_\parallel}(\rho)e^{ik_\parallel z+im\phi}$. Here $k_\parallel$ is the longitudinal wave vector and $m$ denotes angular momentum. The coefficients in the expansion of  $\mathbf{G}_\text{sc}$ associated with each value of $k_\parallel$ and $m$ inside and outside the fiber are a priori unknown, but can be solved for through equations that enforce electromagnetic field continuity relations at the surface of the fiber.

The fiber supports a set of guided modes, \textit{i.e.}, electromagnetic modes that propagate along the nanostructure and are confined in the transversal direction. These modes are denoted by their angular momentum $m$, and their associated  wave vectors  $k_{m} (\omega_0)$ always satisfy $|k_m(\omega_0)|>\omega_0/c$, as the guiding mechanism is by total internal reflection [here we have dropped the ``$\parallel$" subscript associated with the guided mode wave vector $k_m(\omega_0)$, for notational simplicity]. In other words, these modes are evanescent, and their dispersion relations are situated beyond the light line.  The number of guided modes is determined by the fiber radius and dielectric constant. We will restrict ourselves to a single-mode fiber (with $m=\pm 1$), which occurs for a sufficiently small fiber radius. Instead of working with $\mathbf{G}_0$ and $\mathbf{G}_\text{sc}$, for our purposes it is convenient to isolate the guided mode contribution and separate the Green's function $\mathbf{G}(\rb_i,\rb_j,\omega_0)=\mathbf{G}_\text{1D}(\rb_i,\rb_j,\omega_0)+\mathbf{G}'(\rb_i,\rb_j,\omega_0)$ into two terms: one that characterizes the excitation of the guided mode of the fiber, $\mathbf{G}_\text{1D}(\rb_i,\rb_j,\omega_0)$, and another that describes the non-guided electromagnetic modes, $\mathbf{G}'(\rb_i,\rb_j,\omega_0)$. In particular, the guided Green's function takes the form $\mathbf{G}_\text{1D}(\rb_i,\rb_j,\omega_0)=\mathbf{g}(\boldsymbol{\rho}_\text{a})\,e^{\ii k_\text{1D}|z_i-z_j|}$, where $\mathbf{g}(\boldsymbol{\rho}_\text{a})$ is a tensor that only depends on the radial and azimuthal position of the atoms (assumed to be identical), and $\kg=|k_{\pm 1}(\omega_0)|$.

The dynamics of the atoms is governed by the non-Hermitian Hamiltonian $\mathcal{H}_{\rm eff}$ of Eq.~(\ref{heff}), which can be similarly split, \textit{i.e.}, $\mathcal{H}_\text{eff}=\mathcal{H}_{\rm 1D}+\mathcal{H}'$. From the form of $\mathbf{G}_\text{1D}$ given above, it follows that
\begin{align}\label{hguided}
\mathcal{H}_{\rm 1D}=-\ii\frac{\hbar\ga}{2} \sum_{i,j=1}^Ne^{\ii\kg |z_i-z_j|} \heg^{i}\hge^j,
\end{align} 
where $\Gamma_{\rm 1D}=(2\mu_0\,\omega_0^2 \wp^2/\hbar) \,\text{Im}\,G^{\rm 1D}_{\rho \rho}(\rb_i,\rb_i,\omega_0)$ is the spontaneous emission rate of a single atom into the fiber guided mode. The plane-wave dependence reflects the fact that the guided photon propagates without diffraction between two atoms and thus produces an infinite-range interaction.

The non-guided term 
\begin{align}\label{hprime}
\mathcal{H}'=-\frac{3\pi\hbar\Gamma_0}{k_0}\sum_{i,j=1}^N G'_{\rho\rho}(\rb_i,\rb_j,\omega_0)\,\hat{\sigma}_{eg}^i\hat{\sigma}_{ge}^j
\end{align} 
accounts for the interaction through the remaining non-guided electromagnetic modes. Already for just a single atom, the self term of the non-guided Green's function $G'_{\rho \rho}(\rb_i,\rb_i,\omega_0)$ gives rise to both a frequency shift and a decay rate that we will denote as $J'-\ii\Gamma'/2=-(\mu_0\,\omega_0^2 \wp^2/\hbar)  \,G'_{\rho \rho}(\rb_i,\rb_i,\omega_0)$. This self-term reflects the fact that the modification of electromagnetic modes by the nanofiber causes a single atom to have a resonance frequency $\omega_0+J'$ shifted from its vacuum value, and a decay rate into radiative modes $\Gamma'$ different than $\Gamma_0$. For many atoms, the above Hamiltonian accounts for collective emission into non-guided modes, as it takes into account atom-atom interactions that are not mediated by the guided mode. Unlike $\mathbf{G}_\text{1D}$, $\mathbf{G}'$ does not admit a simple form, and in what follows it will be evaluated numerically using the prescription detailed in Appendix~\ref{AppC}. Throughout this manuscript, we will refer to the dynamics generated by $\mathcal{H}_\text{1D}+\mathcal{H}'$ as the ``collective emission" model.

Whether in free space or a nanofiber (or other guided structures), exact collective effects involving modes that are not directly of interest (such as those captured by $\mathcal{H}'$) are typically difficult to treat in the context of applications of atomic ensembles. It is usually heuristically argued that photon-mediated interactions through these modes are not relevant, particularly for disordered or dilute atomic gases, and the ``standard'' model within quantum optics is to ignore such terms \cite{HSP10, GAF07,CJG12,QBH16}. Specifically, the terms of $\mathbf{G}'(\rb_i,\rb_j,\omega_0)$ involving two different atoms ($\rb_i\neq\rb_j$) are assumed to be zero, and the Hamiltonian accounting for emission into non-guided modes reduces to 
\begin{align}\label{hindep}
\mathcal{H}'_\text{indep}=\hbar(J'-\ii\Gamma'/2)\sum_{j=1}^N\hat{\sigma}_{ee}^j.
\end{align}
In this approximation, the non-guided modes of the fiber introduce a modified Lamb shift due to the presence of the fiber surface, and more importantly, provide independent baths for each atom to emit into, at a rate $\Gamma'$. We will refer to the dynamics generated by $\mathcal{H}_\text{1D}+\mathcal{H}'_\text{indep}$ as the ``independent emission" model.

Before proceeding further, we digress to clarify the different usages of the terms super/sub-radiance in literature. Within the independent emission model, the concept of superradiance and subradiance has also been discussed, since $\mathcal{H}_\text{1D}$ alone yields a set of collective atomic states that radiate strongly or weakly into the waveguide \cite{CJG12,LFL13,AHC17}. Similar effects have also been pointed out in cavities (with collective states emitting strongly or weakly into the cavity mode) \cite{BCW12,RAK15,CFB15}. Protocols for photon generation and storage and other quantum information tasks have been built around the manipulation of these states \cite{CJG12,GVC15}. However, as these models still assume independent emission into free space, these protocols do not surpass conventional error bounds.

Throughout this manuscript, the nanofiber radius is taken to be $k_0r=1.2$, the distance between the atoms and the center of the nanofiber is $\rho_\text{a}=1.5r$, and the dielectric constant of the fiber is $\epsilon=4$ (as that of silicon nitride). As an illustration, for the D$_1$ line of Cesium (of resonance frequency $\omega_0=2\pi\times 335.1$ THz), the radius of the fiber would be $r\simeq 170$~nm, and the distance between the atoms and the fiber surface would be approximately $85$~nm. The wave vector of the photonic guided mode is found to be $k_{\rm 1D}\simeq 1.3 k_0$, larger than any wave vector within the light line, as the guided mode is confined. The single-atom decay rates are calculated to be $\Gamma_{\rm 1D}\simeq 0.4\Gamma_0$ to the guided mode, and $\Gamma'\simeq 1.3\Gamma_0$ to the non-guided modes. The modified Lamb shift due to the fiber is $J'\simeq-0.5\Gamma_0.$

In the following sub-sections, we will utilize the formalism introduced above to identify novel phenomena that emerge when collective emission is exactly accounted for, which cannot be predicted from the independent emission approximation, and will show how collective emission enables exponential improvement in performance for quantum memories of light.

\subsection{Linear optics for two-level atoms}

\begin{figure*}
\centerline{\includegraphics[width=\linewidth]{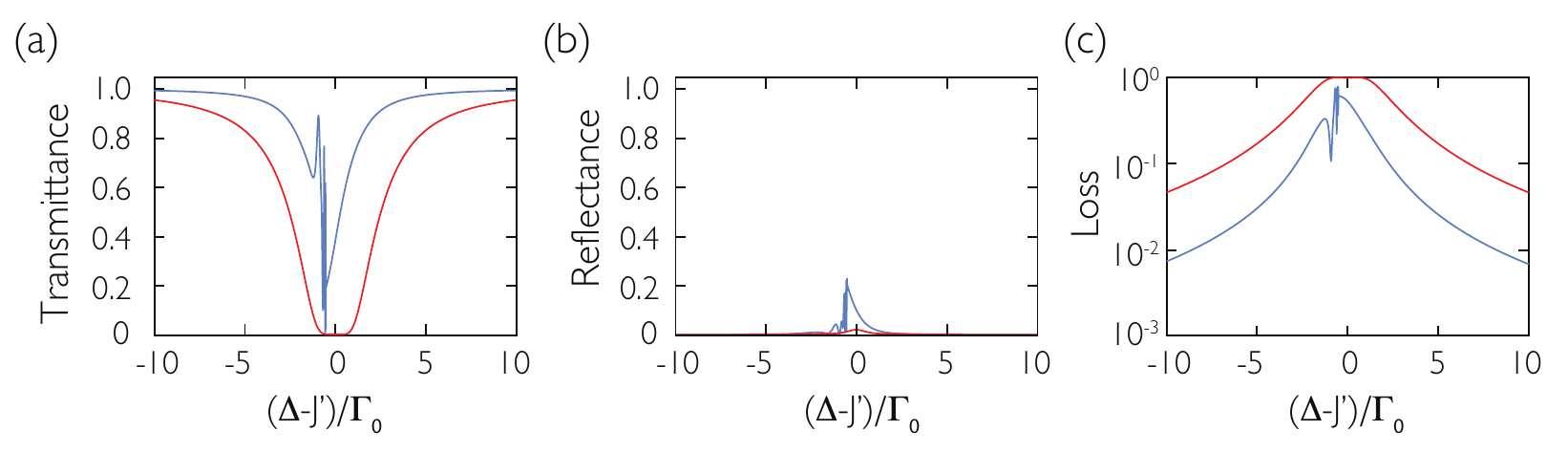}}
\caption{Linear optics for a chain of $N=20$ atoms coupled to a nanofiber, for $d=\lambda_\text{1D}/4$ ($k_\text{1D}d=\pi/2$). \textbf{(a)} Transmittance, \textbf{(b)} reflectance, and \textbf{(c)} loss probability as a function of the atom-probe detuning. The blue curves are obtained by including the collective emission into free-space [see Eq.~\eqref{hprime}], and the red lines are produced within the ``independent emission" model, where it is assumed that free-space emission is a single-atom effect [see Eq.~\eqref{hindep}]. The parameters characterizing the nanofiber are given in Fig.~\ref{nFig1}.} \label{nFig2}
\end{figure*}

We begin by studying the transmission and reflection properties of a chain of atoms coupled to the fiber within the ``independent emission" model. The effective Hamiltonian that describes the atomic dynamics under a coherent-state guided-mode probe field of frequency $\omega_\text{p}$ reads $\mathcal{H}_\text{tot}=\mathcal{H}_\text{drive}+\mathcal{H}'_\text{indep}+\mathcal{H}_\text{1D}$. The Hamiltonians $\mathcal{H}_\text{1D}$ and $\mathcal{H}'_\text{indep}$ are defined in Eqs.~\eqref{hguided}, and \eqref{hindep}, respectively, and the driving term is given by
\begin{align}\label{hdrive}
\mathcal{H}_\text{drive}=-\hbar\Delta\sum_{i=1}^N\hat{\sigma}_{ee}^i-\hbar\sum_{i=1}^N\left(\Omega\, e^{\ii\kg z_i}\,\hat{\sigma}_{eg}^i+\text{h.c.}\right).
\end{align}
In the above expression, $\Delta=\omega_{\rm p}-\omega_0$ is the detuning between the probe field frequency and the atomic resonance frequency. We have also defined the Rabi frequency of the guided-mode probe field as  $\Omega=\db^*\cdot\Eb_{\rm p}^+(\rho_\text{a})/\hbar$, where $\Eb_{\rm p}^+(\rb)\equiv\braket{\hat{\Eb}_{\rm p}^+(\rho_\text{a})}$ is the amplitude of a coherent probe field that implicitly contains the radial position $\rho_\text{a}$ of the atoms. For the remainder of this subsection, we consider that the probe field is weak and does not saturate the atoms. Therefore, all the calculations can be performed in the single-excitation manifold, the realm of classical linear optics.

In the single excitation subspace, the wave function of the atomic ensemble is written as the superposition $\ket{\psi(t)}= c_g (t)\ket{g}^{\otimes N}+ \sum_{j=1}^N c_e^{\, j} (t)\ket{e_j}$, where $\ket{e_j}=\heg^j \ket{g}^{\otimes N}$. In the low saturation regime, with $c_g\simeq1$, the evolution equations for the amplitude of the $\ket{e}$ states are found to be
\begin{align}\label{sigmas}
\dot{c}_e^{\,j}(t)=&\ii\left(\Delta_{\rm}-J'+\ii\frac{\Gamma'}{2}\right)c_e^{\,j}(t)+\ii \,\Omega\, e^{\ii\kg z_j}\\\nonumber
-&\frac{\ga}{2}\sum_{i=1}^N e^{\ii\kg |z_i-z_j|}\,c_e^{\,i}(t).
\end{align}
The generalized input-output expression of Eq.~\eqref{fielddef} allows us to calculate the guided-mode field at any point of the fiber, which reads
\begin{align}\label{iosimple}
\hat{\Eb}^+(\rb)&=\hat{\Eb}_{\rm p}^+(\rb)+\mu_0 \omega_0^2\sum_{j=1}^N \GG_\text{1D}(\rb,\rb_j,\omega_0)\cdot\db\,\,\hge^j.
\end{align}
It is important to notice that the Green's function appearing in the field equation is not the total one, but just that of the guided mode, as it describes the propagation of the photonic guided field along the nanostructure. Due to the cylindrical symmetry of the fiber, the guided modes with angular momenta $m=\pm 1$ are degenerate. One can alternatively take superpositions of these to obtain quasilinearly-polarized $H$ and $V$ modes \cite{LR14,QBH16}. The polarization basis of the fiber can always be set so that the $H$ mode at the atomic positions has polarization components along the $\hat{\rho}$ and $\hat{z}$ directions. We will consider the case where the probe field is $H$-polarized, in which case the atoms scatter solely back into $H$, and the $V$-polarized mode de-couples from the problem.

We can thus project the input-output equation into 1D equations for the $H$-modes, and further separate the guided fields into left- and right-propagating components. The resulting equations are given by
\begin{subequations}\label{leftright}
\begin{align}\label{rightf}
\hat{E}_\text{1D,R}^+(z)&=\hat{E}_\text{in,R}^+(z)+\Omega e^{\ii \kg z}\\\nonumber
&+\ii\frac{\ga}{2} \sum_{j=1}^N \,e^{\ii \kg(z-z_j)}\Theta(z-z_j)\hge^j,
\end{align}
\begin{align}\label{leftf}
\hat{E}_\text{1D,L}^+(z)&=\hat{E}_\text{in,L}^+(z)\\\nonumber
&+\ii\frac{\ga}{2} \sum_{j=1}^N \,e^{\ii \kg(z_j-z)}\Theta(z_j-z)\hge^j,
\end{align}
\end{subequations}
where $\hat{E}_\text{in,R(L)}^+(z)$ are the right(left)-going vacuum fluctuation fields, and $\Theta$ is the Heaviside function. The vacuum fluctuations do not contribute to any of our observables of interest. For convenience, we have re-scaled the fields so that the atomic parameters $\Omega$ and $\ga$ directly appear.

In the quasistatic limit ($\dot{c}_e^{\,j}=0$), the solutions of Eq.~\eqref{sigmas} for the $\ket{e}$-state amplitudes are directly proportional to the probe field Rabi frequency $\Omega$. Together with Eqs.~(\ref{rightf},\ref{leftf}), this allows us to find the reflection and transmission coefficients for the guided field. For example, the transmittance is found by evaluating $T=\braket{\psi|\hat{E}_\text{1D,R}^-(z)\hat{E}_\text{1D,R}^+(z)|\psi}/\Omega^2$, where $z$ is a position immediately after the right-end of the atomic chain, and $\hat{E}_\text{1D,R}^-(z)$ is the Hermitian conjugate of $\hat{E}_\text{1D,R}^+(z)$. A similar expression can be found for the reflectance $R$. One can also calculate the loss probability due to scattering into free space, which is given by $\kappa=1-T-R$.

We choose the distance between the atoms to be $d=\lambda_\text{1D}/4$, with $\lambda_\text{1D}=2\pi/k_\text{1D}$ being the guided-mode wavelength. Any other separation except for the so-called mirror configuration, \textit{i.e.}, $d=\lambda_\text{1D}/2$ or integer multiples thereof \cite{CJG12}, would display qualitatively similar optical properties. In Appendix~\ref{AppD}, we analyze the linear optics of such a special configuration, which has been theoretically known and experimentally observed to become a very reflective mirror \cite{DSR95,SZC11,CJG12}, around which powerful protocols for quantum information processing can be built \cite{CJG12,GVC15,PKG16}. In the mirror configuration and within the independent emission model, there is only one atomic collective state that couples to the guided mode of the fiber, decaying superradiantly into it at a rate $N\ga$. 

In contrast, for any other separation, every atomic collective state is excited by the probe field, and contributes to light transmission and reflection. Therefore, the behavior of the atoms cannot be attributed to a single ``super-atom" of enhanced decay rate, and the transmission spectrum -- depicted by the red line of Fig.~\ref{nFig2}(a) -- differs significantly from a Lorentzian \cite{AHC17}. For large enough number of atoms, and for low single-atom coupling efficiency into the waveguide ($\ga\lesssim \Gamma'$), the transmittance approximately follows the expression 
\begin{align}\label{random}
T_\text{indep}\simeq \text{exp}\left[\frac{-D}{1+4(\Delta-J')^2/\Gamma'^2}\right],
\end{align}
in accordance with the result obtained for a free-space atomic gas \cite{AHC17}. On resonance (when $\Delta-J'=0$), the figure of merit that determines how much light is transmitted is the optical depth, $D=2N\ga/\Gamma'$. For a chain of $N=20$ atoms, the expression of Eq.~\eqref{random} nicely reproduces the transmittance spectrum shown in Fig.~\ref{nFig2}(a). The corresponding reflectance spectrum is displayed by the red curve of Fig.~\ref{nFig2}(b), which shows a very small bump, as the distance $d=\lambda_\text{1D}/4$ minimizes reflection due to destructive interference \cite{CMS15}. As both transmission and reflection are very small close to resonance, the dominant process is photon loss due to atom-mediated scattering into free space. The loss probability $\kappa$ is shown in Fig.~\ref{nFig2}(c). 

If the atoms are closely packed, the above calculations are no longer valid due to the atomic interactions mediated by non-guided modes. Nevertheless, the previous techniques can be straightforwardly modified to calculate the new transmission and reflection coefficients. Within the ``collective emission" model, the atoms evolve under the Hamiltonian $\mathcal{H}_\text{tot}=\mathcal{H}_\text{drive}+\mathcal{H}'+\mathcal{H}_\text{1D}$, where $\mathcal{H}'$ replaces $\mathcal{H}'_\text{indep}$. In the low saturation limit, the evolution equations for the $\ket{e}$ state amplitudes now read
\begin{align}
\dot{c}_e^{\,j}(t)=&\ii\Delta_{\rm}c_e^{\,j}(t)-\frac{\ga}{2}\sum_{i=1}^N e^{\ii\kg |z_i-z_j|}\,c_e^{\,i}(t)\\\nonumber
&+\ii \,\Omega\, e^{\ii\kg z_j}+\ii\frac{3\pi\Gamma_0}{k_0}\sum_{i=1}^N G_{\rho\rho}'(\rb_i,\rb_j,\omega_\text{p})\,c_e^{\,i}(t).
\end{align}
Once again, we evaluate Eqs.~(\ref{rightf},\ref{leftf}) using the steady state solution for the atomic wave function, in order to reconstruct the electromagnetic field along the nanofiber. We overlay our results for the transmission, reflection, and loss probability spectra in Figures~\ref{nFig2}(a-c). The transmittance spectrum displays many sharp peaks that are not observed within the independent emission model, similar to what was found in Ref.~\cite{KSP16}. These peaks correspond to the interference between different collective atomic modes, whose response can be observed due to the diminished photon loss. Close to resonance, reflectance is significantly larger than that obtained within the independent emission model. As a matter of fact, accounting for cooperative emission into non-guided modes lowers significantly the probability $\kappa$ of photon scattering into free space, as can be observed in Fig.~\ref{nFig2}(c). However, this decrease in loss is not uniform for all detunings, and close to resonance this spectrum also showcases sharp peaks. Globally, the behavior is much more complex than that of a ``standard" atomic ensemble, such as given by Eq.~\eqref{random}.

One interesting question is how the loss scales with the number of atoms. We find that, far from resonance, $\kappa/\kappa_\text{indep}\sim N^{-1}$, where $\kappa_\text{indep}$ is the photon loss probability when the collective emission into non-guided modes is neglected. However, in Section III we have found that for a chain of atoms in free space the decay rate of the most subradiant mode scales as $N^{-3}$. Taken together, Figs.~\ref{nFig2}(a-c) suggest a simple reason for this apparent discrepancy. Based on previous arguments of Sec. IIIA, both the infinite atomic chain and the fiber have sets of perfectly guided modes, which experience zero radiation into free space. However, the dispersion relation of the effective medium formed by the fiber and the atomic chain is different from that of the bare fiber alone (\textit{i.e.}, for a given guided-mode frequency, there is a different wave vector). This impedance mismatch leads to large scattering loss at the interface between the two different systems (bare fiber versus fiber with atoms), in close analogy to what occurs between different conventional waveguides \cite{MDD1997,APL03}. 

\subsection{Electromagnetically induced transparency}
\begin{figure}
\centerline{\includegraphics[width=\linewidth]{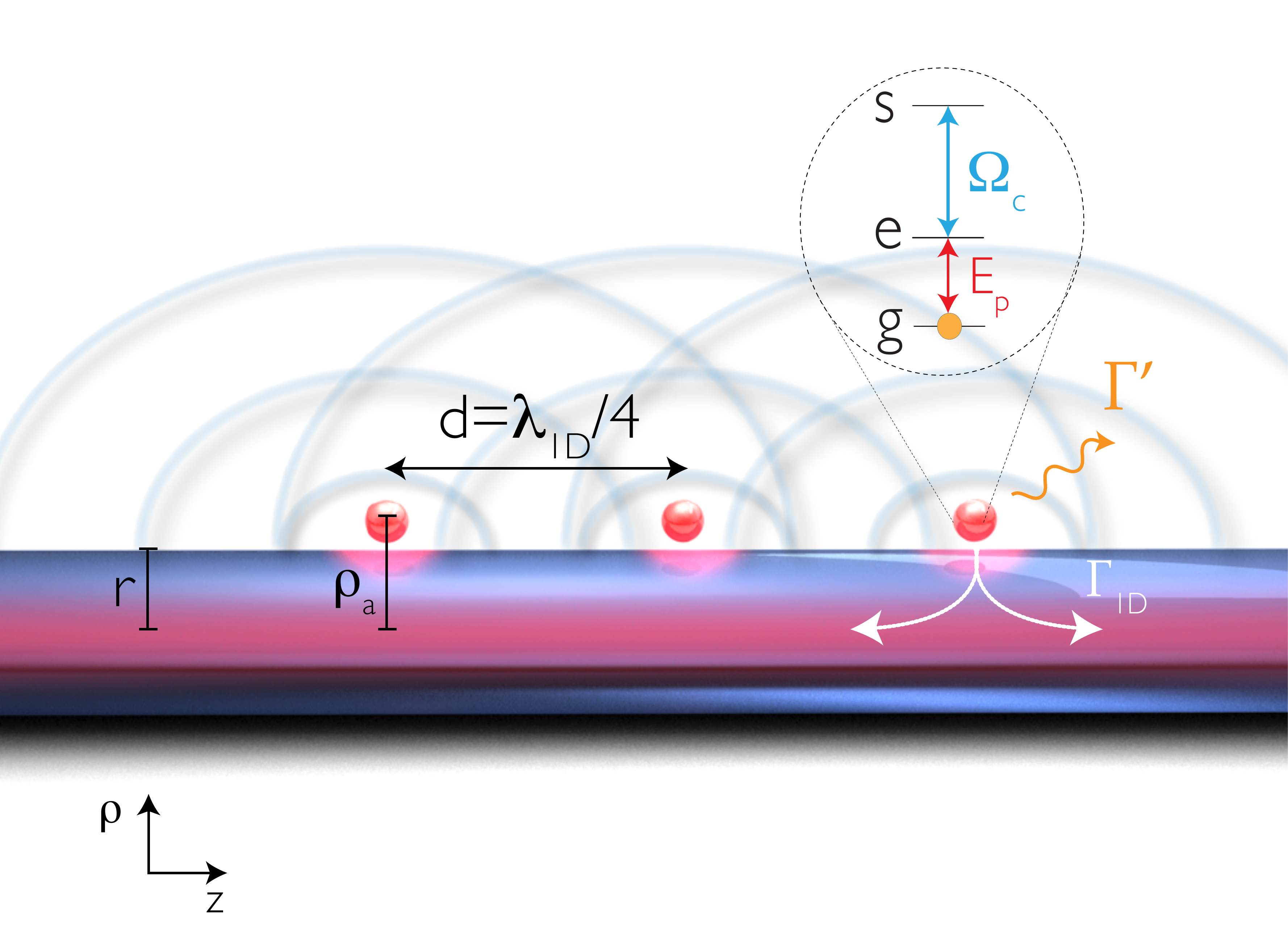}}
\caption{Electromagnetically-induced transparency scheme. The $\ket{g}$ to $\ket{e}$ transition is coupled to the guided mode, and the $\ket{s}$ to $\ket{e}$ transition is driven by an external, classical control field of Rabi frequency $\Omega_\text{c}$. The distance between the atoms is a quarter of the guided mode wavelength, $d=\lambda_\text{1D}/4$.} \label{nFig3}
\end{figure}

\begin{figure*}
\centerline{\includegraphics[width=\linewidth]{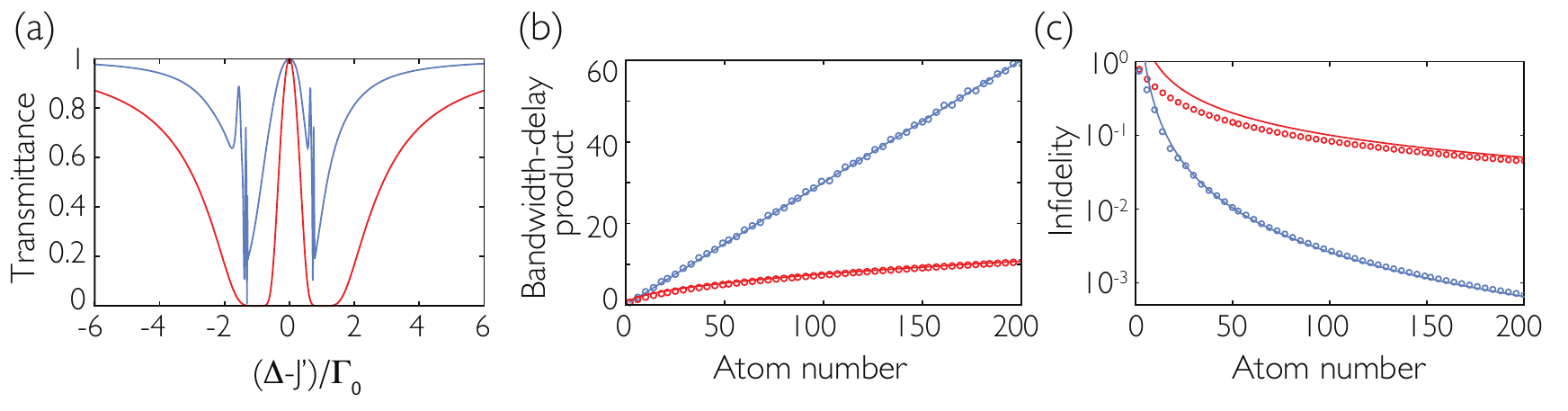}}
\caption{Electromagnetically induced transparency and photon storage efficiency, within the independent emission (red) and collective emission (blue) models. \textbf{(a)} Transmittance spectrum for a chain of $N=20$ atoms. The control field Rabi frequency is set to $\Omega_\text{c}=\Gamma_0$, while other system parameters can be found in the main text. \textbf{(b)} Scaling of the bandwidth-delay product with the number of atoms $N$. The circles  represent the numerical calculation. The red line shows the theoretically predicted answer within the independent emission model, $\mathcal{P}=\sqrt{D}\simeq 0.76\sqrt{N}$, and the blue curve represents the best fit of the numerical data to a linear scaling, $\mathcal{P}\simeq 0.30 N$. The control field intensity is the same as in (a). \textbf{(c)} Infidelity in the retrieval of the spin excitation given by Eq.~\eqref{spinwave}. The circles show the numerics. The red line represents the expected scaling derived in Ref.~\cite{GAL07}, within the independent emission model, $\varepsilon=5.8/D\simeq 10/N$, and the blue line is the best fit of the numerical results to a $\propto N^{-2}$ scaling, and shows $\varepsilon\simeq26 /N^2$ (the range for the fit is $N\in[30,200]$).} \label{nFig4}
\end{figure*}

Having posited that scattering at the interface between the bare fiber and the atomic chain dominates the losses observed in the two-level case, we now attempt to reduce these losses by better matching the dispersion relations of the two regions, using three-level atoms under conditions of electromagnetically induced transparency (EIT) \cite{HFI1990,FL00,L03,FIM05}.

The system under consideration is illustrated in Fig.~\ref{nFig3}. In addition to the  $\ket{g}$ to $\ket{e}$ transition studied earlier, a third metastable level $\ket{s}$ (of frequency $\omega_s$) is added. We assume that the $\ket{e}$ to $\ket{s}$ transition does not couple to the optical fiber (\textit{e.g.}, due to its dipole matrix element being orthogonal to the guided mode polarization), but can be addressed by an external classical control field of Rabi frequency $\Omega_c$ that propagates through free space. Through a two-photon interference effect mediated by the control field, a near-resonant guided photon interacting with an atom originally in state $\ket{g}$ can be coherently mapped to state $\ket{s}$, with minimal excitation of state $\ket{e}$. The lack of population in $\ket{e}$ and associated photon scattering causes the otherwise optically opaque medium to become transparent, and thus EIT nominally preserves the effective refractive index of the guided mode.

Again, we consider the case where atoms are separated by a distance $d=\lambda_\text{1D}/4$ ($\kg d=\pi/2$), to guarantee minimal reflection. Any other distance of the form $d=n\lambda_\text{1D}/4$ (with $n$ being odd) also strongly suppresses reflection and should suffice, as long as $d$ fulfils the subradiance condition. In particular, as atoms nominally do not alter the effective index under EIT, the guided wave vector $\kg$ itself should lie outside the light line. Without atoms this is clearly always true, as the fiber mode is guided. With atoms, however, one must ensure that $\kg$ lies outside the light line when folded back into the first Brillouin zone. If we set the distance between the atoms to be such that $\kg d=\pi/2$, then $\kg$ lies within the first Brillouin zone and automatically satisfies this constraint, $k_0<\kg\leq \pi/d$. However, if $\kg d=3\pi/2$, the condition on the guided mode wave vector, $\kg>3k_0$, becomes much more stringent. In fact, for the radius and dielectric constant of the fiber here considered, the subradiance condition is not met for $d=3\lambda_\text{1D}/4$.

We begin by solving for the characteristics of EIT under the independent emission model, which we find to reproduce previously derived and well-known results in free-space atomic ensembles. In particular, we consider the system evolving under the effective Hamiltonian $\mathcal{H}_\text{tot}=\mathcal{H}_\text{1D}+\mathcal{H}'_\text{indep}+\mathcal{H}_\text{drive}+\mathcal{H}_\text{c}$. The Hamiltonians $\mathcal{H}_\text{1D}$, $\mathcal{H}'_\text{indep}$, and $\mathcal{H}_\text{drive}$ are defined in Eqs.~\eqref{hguided}, \eqref{hindep}, and \eqref{hdrive}, respectively. $\mathcal{H}_\text{c}$ captures the interaction of the atoms with the control field, and is given by 
\begin{align}
\mathcal{H}_{\rm c}=-\hbar\sum^N_{i=1}\Delta_{\rm s}\hat{\sigma}_{ss}^i-\hbar\sum^N_{i=1}\Omega_c (z_i)\left(\hat{\sigma}_{es}^i+\hat{\sigma}_{se}^i\right),
\end{align}
where $\Delta_{\rm s}=\omega_{\rm p}+\omega_c-\omega_{\rm s}$ is the two-photon detuning. We take the control field Rabi frequency $\Omega_c$ to be real, and allow for a possible spatial dependence. We have also assumed that $\ket{e}$ has a negligible decay rate into $\ket{s}$, as in the case of a dipole-forbidden transition or ladder system. For EIT within the independent emission model, this assumption is not necessary, as the emission rate from $\ket{e}$ to $\ket{s}$ can be incorporated into $\Gamma'$ and simply leads to a moderate decrease of optical depth $D$. Such a condition, however, becomes important when considering the collective emission case (see more detailed discussion about multi-level structure in Sec. V).

Within the single excitation manifold, the wave function of the atomic ensemble is $\ket{\psi(t)}=c_g (t)\ket{g}^{\otimes N}+ \sum_{j=1}^N c_e^{\, j} (t)\ket{e_j}+ \sum_{j=1}^N c_s^{\, j} (t)\ket{s_j}$, with $\ket{s_j}=\hat{\sigma}_{sg}^j \ket{g}^{\otimes N}$. For a uniform control field [$\Omega_\text{c}(z_i)\equiv\Omega_\text{c}$] and in the low saturation limit [$c_g(t)\simeq 1$], the equations for the evolution of the amplitudes of the $\ket{e}$ and $\ket{s}$ states read
\begin{subequations}\label{eqEIT}
\begin{align}
\dot{c}_e^{\,i}(t)&=\ii\left(\Delta-J'+\ii\frac{\Gamma'}{2}\right) c_e^{\,i}(t)+\ii \,\Omega\, e^{\ii\kg z_i}\\\nonumber
&+\ii\Omega_c c_s^{\,i}(t)-\frac{\ga}{2}\sum_{j=1}^N e^{\ii\kg |z_i-z_j|}\,c_e^{\,j}(t),\\
\dot{c}_s^{\,i}(t)&=\ii\Delta_{\rm s}c_s^{\,i}(t)+\ii\Omega_c\,c_e^{\,i}(t).
\end{align}
\end{subequations}
We solve these equations in the steady state and readily find $c_s^{\,i}=-(\Omega_c/\Delta_s)\,c_e^{\,i}$, and 
\begin{align}\label{eqEITstat}
\left(\Delta-J'-\frac{\Omega_c^2}{\Delta-J'}+\ii\frac{\Gamma'}{2}\right)&c_e^{\,i}+\Omega_i\,e^{\ii\kg z_i}\\\nonumber
+&\ii\frac{\ga}{2}\sum_{j=1}^N e^{\ii\kg |z_i-z_j|}\,c_e^{\,j}=0.
\end{align}
Here, we have chosen $\Delta_{\rm s}=\Delta-J'$, which assures total transparency when the probe field is resonant with the (shifted) $\ket{e}-\ket{g}$ transition ($\Delta-J'=0$). Having found the steady state solution of the spin wave function, it is now possible to calculate the transmitted guided-mode field by means of the input-output expression of Eq.~\eqref{iosimple}, and, therefore, the transmission coefficient of the array under EIT, $t_{\rm EIT}$. 

The transmission coefficient gives us enough information to calculate two key quantities that describe the EIT medium: the group velocity of the polariton, and the bandwidth-delay product, a parameter that quantifies how many spatially separate photons can be stored in the atomic ensemble \cite{CSH11}. After propagating along the atomic chain, the guided mode field acquires a phase $t_{\rm EIT}\equiv e^{\ii k_{\rm eff} Nd}$, where $k_{\rm eff}$ is a complex effective wave vector that encodes both light absorption and dispersion. Up to second order in the atom-probe detuning, the effective wave vector reads \cite{CSH11,AHC17}
\begin{align}\label{keff}
k_{\rm eff}&=k_\text{1D}+\frac{\ga}{2d\Omega_c^2}(\Delta-J')\\\nonumber
&+\ii\frac{\ga(\Gamma'+\eta\ga/2N)}{4d\Omega_c^4}(\Delta-J')^2,
\end{align}
where $\eta=0 (1)$ for an even (odd) number of atoms. It can be seen that when $\Delta=J'$, the effective wave vector perfectly matches that of the bare fiber, $k_\text{eff}=k_\text{1D}$, as originally desired. From the above expression, the group velocity at the center of the transparency window is found to be $v_\text{g}=\left(\partial k_{\rm eff}/\partial \Delta\right)^{-1}=2\Omega_c^2d/\ga$. The delay time, \textit{i.e.}, the time it takes for this slow polariton to traverse the medium is $\tau=Nd/v_\text{g}=N\ga/2\Omega_c^2$. The bandwidth of the transparency window, which dictates how spectrally narrow a photon has to be to propagate with high transparency, is defined as 
\begin{align}\label{deltaeit}
\Delta_\text{EIT}=2\delta=2\Omega_c^2\sqrt{2/N\ga(\Gamma'+\eta\ga/2N)},
\end{align}
where $\delta$ is the detuning for which $|t_\text{EIT}|^2=1/e$. Therefore, the bandwidth-delay product, $\mathcal{P}=\tau\Delta_\text{EIT}=\sqrt{2N\ga/(\Gamma'+\eta\ga/2N)}\simeq\sqrt{D}$, scales with the square root of the optical depth $D=2N\ga/\Gamma'$, for realistic values of $\Gamma'$. This is the same scaling that is predicted in free-space atomic ensembles, when atoms are assumed to emit independently \cite{L03,FIM05}. In contrast, for some idealized system without loss (\textit{i.e.}, $\Gamma'=0$), the bandwidth-delay product scales simply with the number of atoms, $\mathcal{P}\sim N$ \cite{CSH11,SPP05}. This result does not follow from the perturbative expansion of Eq.~\eqref{keff}. Rather, one can perform an exact calculation of the optical band structure and the bandwidth, $\Delta_\text{EIT}\sim\Omega_c^2/\ga$, all of which is usable in the absence of loss~\cite{CSH11}. 

Figure~\ref{nFig4}(a) depicts a representative transmittance spectrum for a chain of $N=20$ atoms (red curve). 
For $\Delta-J'=0$, \textit{i.e.}, when the probe field is in resonance with the (shifted) $\ket{e}-\ket{g}$ transition, the transmittance is perfect. However, total transparency is only exactly achieved at this precise frequency, decreasing with the detuning from resonance. The medium can be considered roughly transparent within a small window of bandwidth $\sim\Delta_\text{EIT}$, for which the transmittance $T>1/e$. The scaling of the  bandwidth-delay product with the number of atoms is shown in Figure~\ref{nFig4}(b). The numerical results (red dots) are obtained by solving Eq.~\eqref{eqEITstat}, then calculating the transmission as a function of the atom-probe detuning, and finally numerically finding the values of $\Delta$ where the transmittance drops to $1/e$. The calculations follow perfectly the simple scaling $\mathcal{P}=\sqrt{D}$ derived above (continuous red line). We should note that the usual definition of $D$ --in terms of exponential reduction of transmittance on resonance--  does not apply any more to EIT. However, we maintain the definition $D=2N\ga/\Gamma'$, as it represents a physical resource.

As the final step in our summary, we turn our attention to the problem of the efficiency of an EIT-based quantum memory. Qualitatively, the large bandwidth-delay product associated with an optically dense ensemble enables an incident pulse to become spatially compressed and localized completely within the ensemble, while propagating with a reduced group velocity $v_\text{g} \ll c$. The slow group velocity is associated with the photon mixing strongly with a collective spin excitation $\hat{\sigma}_{sg}$, to form a ``dark-state'' polariton. Once the pulse is completely inside, the pump field can be adiabatically decreased to zero ($\Omega_c=0$), in which case $v_\text{g}\rightarrow 0$ and the pulse becomes stored, while simultaneously the polariton becomes a pure spin excitation \cite{L03,FIM05}. This process of photon mapping can be reversed by ramping up the control field intensity at a later time, which allows for an ``on demand" retrieval of the stored photon. Gorshkov and co-workers demonstrated that, due to time reversal symmetry, the optimal efficiency of photon storage is identical to that of photon retrieval \cite{GAF07}. Therefore, our discussion will focus on the latter. Neglecting collective emission into non-guided modes, Gorshkov \textit{et al.} predicted that any smooth spin wave fitting inside the atomic medium should be retrieved with error $\varepsilon\sim1/D$ \cite{GAF07,GAL07}. The reason for such scaling is that the optical depth sets the branching ratio between emitting the photon into the desired channel (the guided mode of the fiber) and into the undesired reservoir (free space). 

\begin{figure}
\centerline{\includegraphics[width=\linewidth]{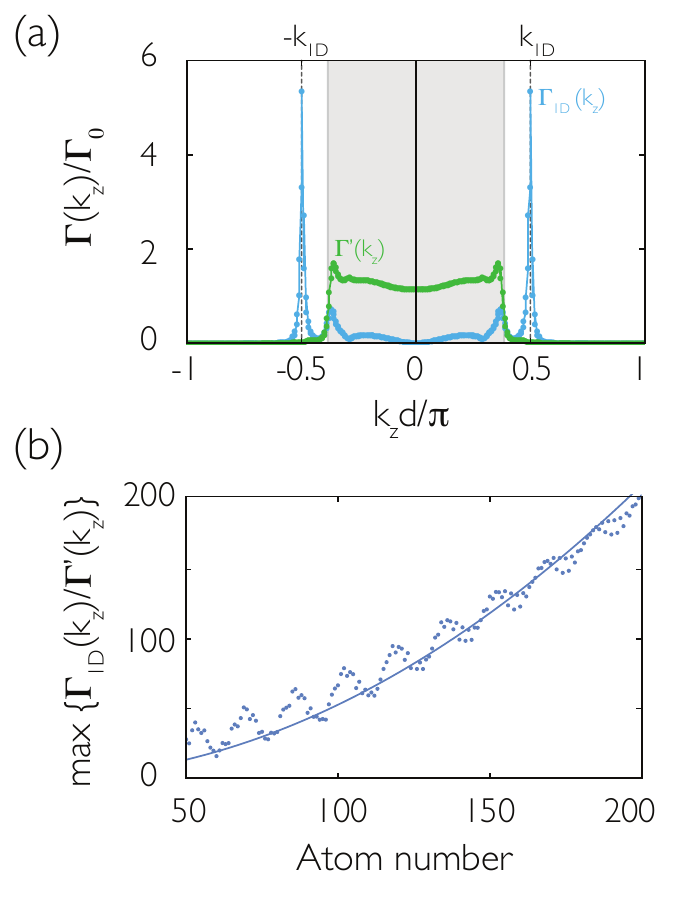}}
\caption{Selectively radiant states of the $s$-branch. \textbf{(a)} Guided [$\ga(k_z)$, light blue] and non-guided [$\Gamma'(k_z)$, green] decay rates of the single-excitation eigenstates of $\mathcal{H}_\text{1D}+\mathcal{H}'+\mathcal{H}_\text{c}$ vs the dominant wave vector $k_z$ of each eigenstate, for a chain of $N=200$ atoms coupled to the fiber. The control field Rabi frequency is $\Omega_c=4\Gamma_0$, and the plot is restricted to eigenstates that consist mostly of population in the $s$ states (the s-branch). The gray shaded area represents the region within the light line, the dashed lines show the guided mode wave vector $\pm\kg$, and the color lines are guides to the eye. \textbf{(b)} Scaling of the ratio $\ga(k_z)/\Gamma'(k_z)$ with the number of atoms, at the wave vector $k_z$ where it is maximum. The dots are numerical results, and the curve represents the best quadratic fit, $\text{max}\{\ga(k_z)/\Gamma'(k_z)\}\simeq 0.0053 N^2$.}
\label{nFig5}
\end{figure}

In order to demonstrate that our calculations match the previously known results, we initialize a single-excitation spin wave of the form 
\begin{align}\label{spinwave}
\ket{\psi(t=0)}=\mathcal{N}\sum_{j=1}^N \,j\, \,e^{\ii \kg z_j} \ket{s_j},
\end{align}
where $\mathcal{N}$ is a normalization constant, and the phase $e^{\ii \kg z_j}$ guarantees retrieval of this excitation as a photon in the right-propagating guided mode, Eq.~\eqref{rightf}. This peculiar-shaped spin wave (in particular, the relative population of atom $j$ grows like $j^2$) presents a balance between the pulse being sufficiently smooth (such that its wave vector components fit within the transparency window), and the majority of the pulse sitting at the forward end of the medium, such that it does not accumulate propagation losses over a large distance. In the limit of large optical depth, such a polariton is predicted to be of the optimal shape to yield maximal retrieval efficiency (in particular, $\varepsilon\simeq 5.8/D$ \cite{GAL07}). At $t=0$ we switch on the control field and let the atomic wave function evolve under the effective Hamiltonian $\mathcal{H}_\text{1D}+\mathcal{H}'_\text{indep}+\mathcal{H}_\text{c}$ until no excitation is left in the atomic chain (having been emitted into the waveguide or free space). We calculate the infidelity in the photon retrieval in two different manners, which yield identical results. The first method consists in integrating over time the radiative emission into non-guided photonic modes. The error is thus $\varepsilon=\int_0^{\infty}\, dt\,\kappa'_\text{indep}(t)$, where $\kappa'_\text{indep}(t)=-(2/\hbar)\,\text{Im}\braket{\mathcal{H}'_\text{indep}}=\Gamma'\braket{\sum_{j=1}^N\hat{\sigma}_{ee}^j}$. The second way to calculate the infidelity is to realize that a successful retrieval occurs whenever the photon is emitted to the guided mode of the fiber. Then, the error is $\varepsilon=1-\int_0^{\infty}\, dt\,\kappa_\text{1D}(t)$, where the time-dependent decay rate into the guided mode is $\kappa_\text{1D}(t)=-(2/\hbar)\,\text{Im}\braket{\mathcal{H}_\text{1D}}$. Technically, the efficiency should be calculated only accounting for emission into the preferred (right-going) direction of the fiber, using Eq.~\eqref{rightf}. We have checked that this gives nearly an identical answer, as emission in the left-going direction is negligible. The scaling of the retrieval infidelity with the number of atoms is shown by red circles in Fig.~\ref{nFig4}(c). The numerical results agree very well with the expected scaling ($\varepsilon\simeq 5.8/D$ \cite{GAL07} \footnote{In Ref. \cite{GAL07} the scaling is $\varepsilon\simeq 2.9/D$, due to a factor of 2 difference in their definition of $D$.}, red line) for large number of atoms. In principle, the shape of the outgoing photon can be further tailored via a time-dependent control field, but we will not treat that case here.

Now that we have reviewed the basic parameters characterizing an EIT medium as well as the fidelity of a quantum memory, we analyze how collective emission into non-guided modes modifies the relevant figures of merit. In this case, the system evolves under the effective Hamiltonian $\mathcal{H}_\text{tot}=\mathcal{H}_\text{1D}+\mathcal{H}'+\mathcal{H}_\text{drive}+\mathcal{H}_\text{c}$, where now collective emission is taken into account through the $\mathcal{H}'$ term (instead of the previous $\mathcal{H}'_\text{indep}$). Before proceeding to the calculation of the optical properties, we would like to discuss the decay rates of the eigenstates of the system without guided-mode driving, \textit{i.e.}, of $\mathcal{H}_\text{1D}+\mathcal{H}'+\mathcal{H}_\text{c}$. Due to the presence of the $s$-states, the number of eigenstates in the single excitation subspace is $2N$. If the population of the $s$-states of a given eigenstate is larger than that of the $e$-states, we say that this eigenstate belongs to the ``$s$-branch", and vice versa. For any finite control field, there is  mixing between the $e$- and $s$-branches, meaning that the eigenstates do not purely consist of $\ket{e}$ or $\ket{s}$ states. 

In Fig.~\ref{nFig5}(a) we show the guided [$\ga(k_z)=-(2/\hbar)\,\text{Im}\braket{\mathcal{H}_\text{1D}}$] and non-guided [$\Gamma'(k_z)=-(2/\hbar)\,\text{Im}\braket{\mathcal{H}'}$] decay rates of the numerically calculated eigenstates that belong to the $s$-branch, for a fixed number of atoms $N=200$. As in Section III A, we have performed a finite Fourier transform to associate an effective wave vector $k_z$ to each of the atomic spin eigenstates. As expected, the non-guided decay rates are negligible when the dominant wave vector $k_z$ lies beyond the light line. On the contrary, the guided decay rates peak strongly outside the light line, at $k_z=\pm \kg$. It can also be seen that these same states experience a decay rate into free space of $\Gamma'(k_z)/\Gamma_0 \ll 1$, and are thus the ``selectively radiant" states that we previously anticipated. Some of the eigenstates with $|k_z|<k_0$ have a non-zero $\ga(k_z)$ decay rate into the guided mode. This occurs because the eigenstates are not purely Bloch waves with a perfectly-determined $k_z$, but instead can have some finite contributions from all $k_z$. As a technical note, we remark that only when the chain of atoms is infinite can $\mathcal{H}_\text{1D}$ and $\mathcal{H}'$ be simultaneously diagonalized. For any finite number of atoms, the eigenstates of $\mathcal{H}_\text{1D}+\mathcal{H}'+\mathcal{H}_c$ are not simultaneously eigenstates of its guided and non-guided parts.

One can also consider the behavior of the selectively radiant states, as a function of atom number. In particular, of interest is the maximum possible branching ratio $\ga (k_z)/\Gamma'(k_z)$ of all the eigenstates, as a function of $N$. We plot this quantity in Fig.~\ref{nFig5}(b), where we find an approximate scaling of max$\{\ga(k_z)/\Gamma'(k_z)\}\propto N^2$. We find that this scaling is in fact independent of the magnitude of the control field, and is in contrast to the $\sim N$ scaling in the case of the independent emission model. We later show that this same scaling manifests itself in the photon storage/retrieval error probabilities.

Let's now calculate EIT transmittance spectra. Under the same conditions as before (low saturation, uniform control field), the evolution equations for the state amplitudes are found to be
\begin{subequations}\label{eqEIT2}
\begin{align}\nonumber
\dot{c}_e^{\,i}(t)&=\ii\Delta c_e^{\,i}(t)+\ii \Omega_i e^{\ii\kg z_i}-\frac{\ga}{2}\sum_{j=1}^N e^{\ii\kg |z_i-z_j|}\,c_e^{\,j}(t)\\
&+\ii\Omega_cc_s^{\,i}(t)+\ii\frac{3\pi\Gamma_0}{k_0}\sum_{j=1}^N G_{\rho\rho}'(\rb_i,\rb_j,\omega_\text{p})\,c_e^{\,j}(t),\\
\dot{c}_s^{\,i}(t)&=\ii\Delta_{\rm s}c_s^{\,i}(t)+\ii\Omega_c\,c_e^{\,i}(t).
\end{align}
\end{subequations}

While analytical approximations are not as readily obtained, the numerical procedures follow exactly as presented for the case of independent emission. The blue curve in Fig.~\ref{nFig4}(a) shows how the transmittance spectrum is modified by selective radiance. The first noticeable consequence of collective suppression of the emission into non-guided modes is that the transparency window becomes wider, as expected if the loss becomes smaller. This is further confirmed in Fig.~\ref{nFig4}(b), which displays a linear scaling of the bandwidth-delay product with the atom number, in contrast to the conventional square root dependence. As mentioned before, such a scaling is characteristic of a system without losses \cite{SPP05}. This scaling, along with the conclusion in Sec. III that suppression of emission into free space can occur for low densities of excitations, suggests that it might be possible to store a number of photons in an atomic medium that scales linearly with atom number (in contrast to the $\sim\sqrt{D}$ scaling within the independent emission model).

Finally, Fig.~\ref{nFig4}(c) shows the improvement in the infidelity of retrieval of the spin wave given by Eq.~\eqref{spinwave}. The error is now calculated including collective emission as $\varepsilon=\int_0^{\infty}\, dt\,\kappa'(t)$, where 
\begin{align}\label{kappa}
\kappa'(t)=-\frac{2}{\hbar}\,\text{Im}\braket{\mathcal{H}'}.
\end{align}
Again, this error matches the one calculated by taking into account the component of the photon that is released into the guided mode, \textit{i.e.}, $\varepsilon=1-\int_0^{\infty}\, dt\,\kappa_\text{1D}(t)$, with $\kappa_\text{1D}(t)=-(2/\hbar)\,\text{Im}\braket{\mathcal{H}_\text{1D}}$. As before, we have checked that emission in the left-going direction is negligible. We find that by exploiting collective emission, the error decreases with atom number like $\varepsilon\propto 1/N^2$. This result is consistent with the scaling of branching ratios for the most selectively radiant eigenstates, previously plotted in Fig.~\ref{nFig5}(b). Moreover, by varying the radial positions of the atoms over a limited range, thus modifying the ratio $\ga/\Gamma'$, we are able to separate the contributions of the number of atoms and the optical depth to the infidelity. We obtain $\varepsilon\simeq 15/(ND)$, where the numerical prefactor is not necessarily universal, as it probably depends on the fiber properties.
\begin{figure*}
\centerline{\includegraphics[width=\linewidth]{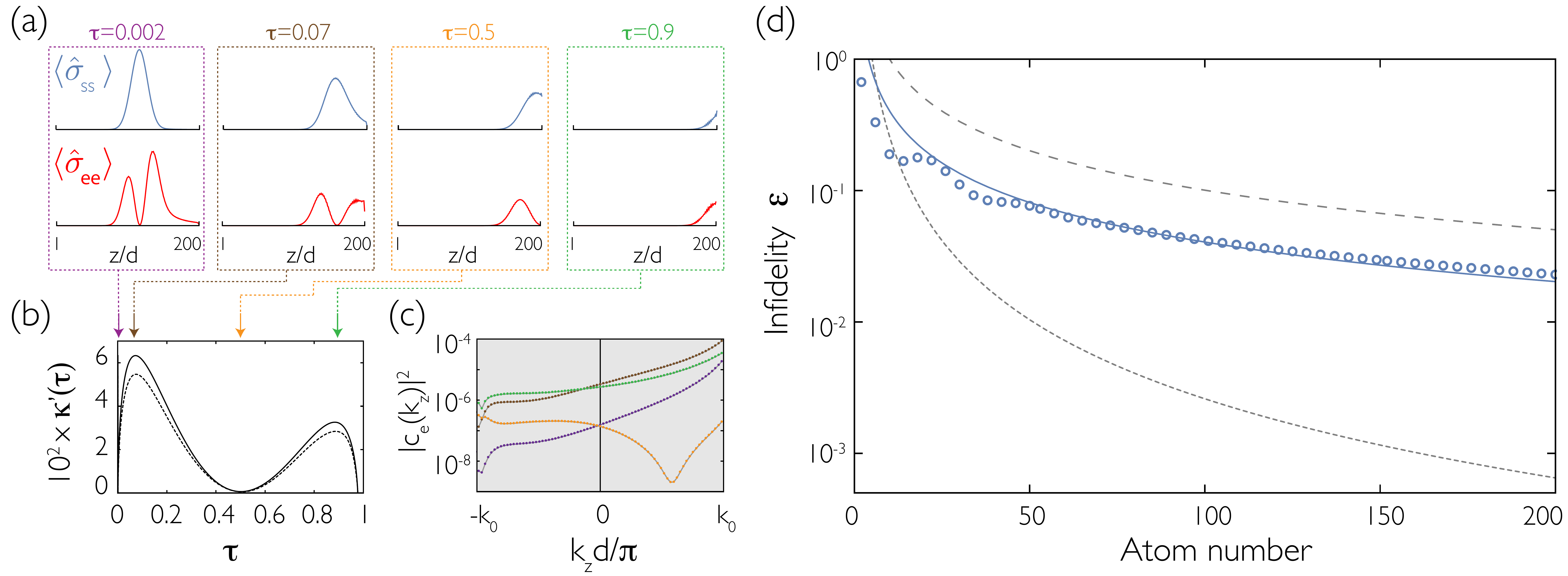}}
\caption{Retrieval of a Gaussian spin wave given by Eq.~\eqref{gaussianwave}, under a spatially uniform control field of Rabi frequency $\Omega_c=0.1\Gamma_0$, and within the collective emission model. \textbf{(a)} Evolution of the population in the $\ket{s}$ states (upper plot) and $\ket{e}$ states (lower plot) as a function of position and at several selected points in time $\tau$, for a chain of $N=200$ atoms, in arbitrary units. The time is normalized such that at $\tau=1$ all the spin population has completely decayed. \textbf{(b)} Instantaneous rate of photon scattering into free space $\kappa'(\tau)$. The solid line is a numerical calculation based upon Eq.~\eqref{kappa}, while the dashed line represents an estimate based upon taking a Fourier decomposition of the spin amplitude $c_e(z_j,t)$ and weighting each component by a wave vector-dependent decay rate. A large instantaneous scattering rate occurs when a large excited-state population is found at the end of the system [for example, at times $\tau=0.07$ and $\tau=0.9$ in figure (a)]. \textbf{(c)} Excited-state population $|c_e(k_z,\tau)|^2$ of the different wave vector components of the spin wave, for different evolution times [corresponding to the snapshots in (a)]. Only the region inside the light line is shown. \textbf{(d)} Scaling of the retrieval loss with the atom number $N$. The blue dots show the numerical calculation, whereas the blue line is the best fit to them and represents $\varepsilon=4.1/N$. The infidelities for the initial spin wave of Eq.~\eqref{spinwave} within the independent and collective emission models are shown by the dashed and dotted lines, respectively.} \label{nFig6}
\end{figure*}

An interesting question is why the error of photon storage improves `only' by a factor of $N$ (from $1/N$ to $1/N^2$). In particular, given that single excitations in a free-space chain can experience a suppression in the emission rate of up to $1/N^3$, one might have expected a greater suppression of errors of up to $1/N^4$ in photon storage. An initial -- but somewhat erroneous -- guess would be to attribute this ``bad scaling" to an unfavorable spatial profile of the initial spin wave. Perhaps surprisingly, although EIT nominally matches the effective guided mode indices of the bare fiber and the composite system of fiber and atomic chain, we show in the next subsection that the slight impedance mismatch away from perfect resonance $\Delta=J'$ is still responsible for the majority of scattering losses into free space. We thus present an improved impedance matching scheme, which allows for exponential improvement with $N$ of the quantum memory fidelity.

\subsection{Quantum memory with exponential fidelity}

The importance of residual impedance mismatch can be seen in a simple example, where one considers an initial Gaussian spin-wave profile,
\begin{align}\label{gaussianwave}
\ket{\psi(t=0)}=\mathcal{N}\sum_{j=1}^N\,e^{\ii \kg z_j} \,e^{-(z_j-z_c)^2/2\sigma^2} \ket{s_j},
\end{align}
and investigates the dynamics of the retrieval process more carefully. In the above expression, $\mathcal{N}$ is a normalization constant, $z_c=(N-1)d/2$ is the center of the atom chain, and $\sigma=\sqrt{N} d$ is the standard deviation of the Gaussian spin wave. Figure~\ref{nFig6}(a) shows the evolution of the  spin excitation at different times $\tau$, for a chain of $N=200$ atoms. Here, to aid in visualization, we have defined a re-scaled dimensionless time $\tau \in [0,1]$, where $\tau$ represents the total amount of atomic population that has decayed (\textit{i.e.}, at $\tau$=1 the spin wave has fully decayed and the photon has been completely released). We have plotted not only the population in the $\ket{s}$-state through the ensemble, which essentially matches population of the dark-state polariton, but also the excited state population, which is ultimately responsible for any emission into free space. 

\begin{figure}
\centerline{\includegraphics[width=\linewidth]{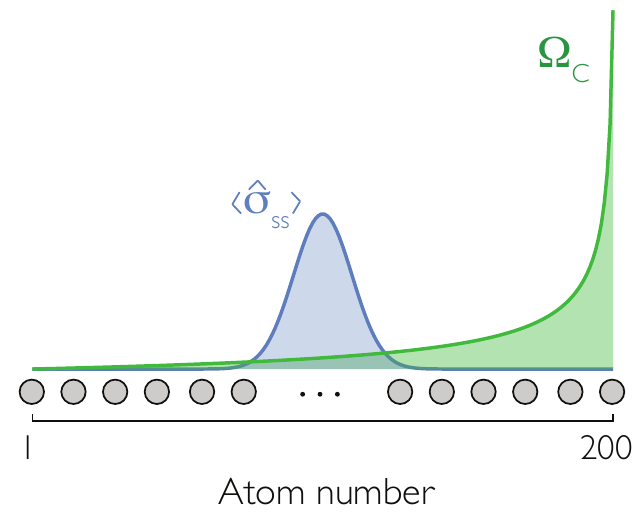}}
\caption{Schematic of the spatial profile of the control field $\Omega_c(z_j)\sim\sqrt{N/(N+1-j)}$ (green curve, arbitrary units) for a chain of $N=200$ atoms. The initial spin wave, given by Eq.~\eqref{gaussianwave}, is overlaid in blue.} \label{nFig6b}
\end{figure}

\begin{figure*}
\centerline{\includegraphics[width=\linewidth]{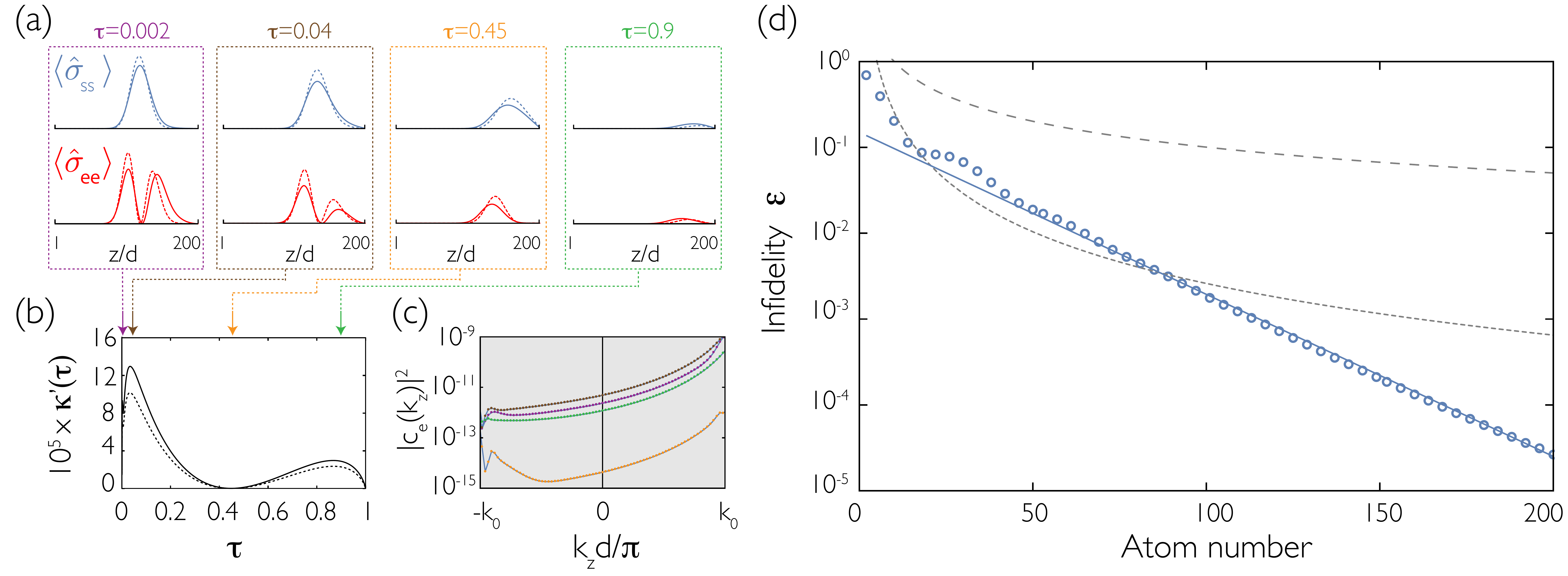}}
\caption{The same as Fig.~\ref{nFig6}, but for a spatially-varying control field of the form $\Omega_c(z_j)=0.005\sqrt{N/(N+1-j)}\Gamma_0$. In (a), the dotted lines show the analytical model. In (d), the blue line is a guide to the eye, and follows $\varepsilon=0.15 e^{-N/23}$.} \label{nFig7}
\end{figure*}

For a spatially uniform control field, and for a theory of EIT within a uniform medium (\textit{i.e.}, where the atomic density is treated as smooth rather than discrete points), it can readily be shown  \cite{FL02} that the excited state population is proportional to $|\partial_z \hat{\sigma}_{gs}(z)|^2$ (also see Appendix~\ref{AppE}), a result that also agrees well with our numerical results. This excited state population is necessarily associated with a pulse of finite extent or bandwidth, and in complementary ways reflects the fact that perfect transparency in EIT only occurs at a single frequency, or that there are non-adiabatic corrections to the formation of a dark-state polariton. At certain times, such as at $\tau=0.07$ or $\tau=0.9$, the excited-state spin wave has a large amplitude at the edge of the atomic chain. Most of the error on retrieval occurs at these times, as can be seen in Fig.~\ref{nFig6}(b), which shows the instantaneous loss $\kappa'(\tau)$. Here, $\kappa'(\tau)$ is re-scaled as well, so that its integral provides the total loss, $\int_{0}^{1} d\tau \kappa'(\tau) = \varepsilon$.

A plausible cause of this behavior is the discontinuity of the excited state population at the system's end, which is associated with a large number of wave vector components $k_z$ that lie within the light line and couple to free space. To further confirm this intuition, we have developed a model for the time-dependent loss based upon the Fourier decomposition of the spatial profile of the excited state amplitude. At every time of the evolution, we calculate $c_e(k_z,\tau)$ by doing a finite Fourier transform of the excited state amplitudes, $c_e(z_j,\tau)\equiv c_e^{\,j}(\tau)$. Then, we find the Fourier-based instantaneous loss as $\tilde{\kappa}'(\tau)=(d/2\pi)\int_{-k_0}^{k_0} dk_z\,\Gamma'(k_z) |c_e(k_z,\tau)|^2$, where $\Gamma'(k_z)$ is obtained from classifying the decay rates of the eigenstates of $\mathcal{H}'$ according to their dominant wave vector. This calculation, represented by the dashed line in Fig.~\ref{nFig6}(b), shows good agreement with the numerics. Figure~\ref{nFig6}(c) depicts the components of $|c_e(k_z)|^2$ inside the light line for different times [corresponding to the snapshots of Fig.~\ref{nFig6}(a)]. For initial times (purple curve), the wave function has minimal population within the light line, suggesting it propagates with little loss down the atomic chain. The population drastically increases as the pulse hits the end of the chain (brown curve). A large population of wave vector components within the light line is correlated with increased instantaneous loss and sharp features in the profile of the excited state population at the system edge. Finally, Fig.~\ref{nFig6}(d) displays the scaling of the infidelity in photon retrieval with the atom number. The scaling is poor ($\varepsilon=4.1/N$), as large losses occur when the polariton hits the ends of the atomic chain. 

We now describe how to smoothly reduce the excited state population at the end of the chain, by introducing a spatially-dependent control field. Heuristically, the idea is to increase the control field at the ends of the chain, as shown in Fig.~\ref{nFig6b}. As the EIT bandwidth is proportional to the control field intensity [see Eq.~\eqref{deltaeit}], the atomic medium becomes more transparent at the edges, where the excited state population is reduced. One can develop an effective continuum wave equation to predict the evolution of the populations in $\ket{s}$ and $\ket{e}$ during the retrieval process, in the presence of a spatially dependent control field (see Appendix~\ref{AppE}) \cite{FL02}. Similar to the case of a uniform control field presented earlier, in principle the scattering loss can then be estimated and minimized from these populations. This optimization process seems quite challenging in practice, however, as it depends on the initial spin wave, the control field profile, and on the integral of momentum components over the entire history of evolution. We will not do such an optimization here, but instead show that rather simple choices can already lead to significant improvements over the infidelity scalings that we found in the previous subsection.

For the initial spin wave of Eq.~\eqref{gaussianwave}, Fig.~\ref{nFig7}(a) shows the evolution of the e- and s-states populations of a chain of $N=200$ atoms for the spatially-dependent control field $\Omega_c(z_j)=\Omega_c^{(0)}\sqrt{N/(N+1-j)}$, which exhibits a rapid increase at the right edge of the chain [see Fig.~\ref{nFig6b}]. Since the control field is switched on suddenly in our simulations, its magnitude is taken to be very small ($\Omega_{c}^{(0)}=0.005\Gamma_0$) to minimize rapid, non-adiabatic accelerations of the spin wave that would artificially increase the losses at initial times. Both the $\ket{e}$- and $\ket{s}$-state populations exhibit smooth profiles at every time of the evolution, and in particular, one sees that the excited state population smoothly vanishes at the edge of the system. The dashed lines in the plots show the results from the analytical model developed in Appendix~\ref{AppE}, which agree well with the numerics. Both the instantaneous loss and the amount of spin-wave population lying within the light line are several orders of magnitude smaller than in the case of a uniform control field, as can be seen in Figs.~\ref{nFig7}(b) and (c), respectively. Moreover, the excited state population within the light line does not significantly increase from its initial values. Finally, Fig.~\ref{nFig7}(d) shows the exponential decrease of the retrieval infidelity $\varepsilon$ as a function of the atom number $N$ for this given profile. We anticipate that optimized initial spin waves and control field profiles will result in a much steeper exponential scaling. Nonetheless, we have demonstrated that a fairly trivial selection of those settings already exponentially improves previously-known bounds for photon storage.

\section{Physical implementations and possible challenges}
Having analyzed the physics of both subradiance and selective radiance, we devote this section to discussing suitable experimental platforms as well as the challenges that might be encountered to observe this physics.

\subsection{Physical implementations}
To potentially observe the physics that we described in Sec. III requires atoms to be regularly trapped, forming an ordered lattice. Moreover, among all the possible collective atomic states, one should be able to access the subradiant manifold. We shall start our discussion with possible physical platforms. As we have demonstrated, the minimal distance at which subradiant states appear depends on the dimensionality of the atomic array ($\lambda_0$ in 2D, and $\lambda_0/2$ in 1D). Standard free-space optical lattices \cite{B05} can achieve lattice constants of $d\sim \lambda_0/2$, and quantum gas microscopes \cite{BGP09} are able to generate single 2D arrays. In such systems, both bosonic \cite{BPT10} and fermionic \cite{GPM16} Mott insulator phases -- where the number of atoms per site can be limited to one -- have been realized.

Very recently, several experimental groups have built almost defect-free 1D \cite{EBK16} and 2D \cite{LLK15,BLL16} lattices in an atom-by-atom manner using optical tweezer arrays. While the inter-atomic distance achieved so far is still larger than the free space wavelength, due to the problem of interference between the tweezers at close distances, it could be possible that further improvements enable sub-wavelength distances to be reached. It might also be possible to employ a transition with a shorter wavelength for the trapping scheme, and use another of longer wavelength to explore subradiant phenomena. Finally, periodically patterned 1D or 2D dielectric structures can readily yield sub-wavelength trapping potentials, with the periodicity of the structure itself \cite{HMC13,DHC15}. While cold atoms can now routinely be trapped near dielectric structures \cite{VRS10,TTL13,GHY14,GHH15,HGA16}, the filling fractions remain quite low and new approaches (such as integration with tweezer arrays) must be developed to achieve near-perfect filling.

Overcoming the second requirement, that of exciting the subradiant manifold efficiently, is not trivial. As subradiant states are characterized by a wave vector that lies beyond the light line, they do not naturally couple to a laser beam that propagates through free space. There are several options to overcome this hurdle. An already-suggested possibility \cite{POR15} is to map superradiant states, which are easy to excite, to subradiant ones via magnetic field gradients. Specifically, a laser can efficiently excite a spin wave $\sim \sum_{j} e^{\ii k_z z_j} \ket{e_j}$ whose wave vector $k_z$ lies within the light line. In the case that the excited state is magnetic-field sensitive, a field gradient would then imprint a spatially dependent phase shift $\ket{e_j}\rightarrow e^{\ii (\beta t)z_j}\ket{e_j}$ in time, which then could allow the wave vector $k_z \rightarrow k_z+\beta t$ to be mapped outside of the light line. In 1D chains in free space, one might exploit the fact that the emission of subradiant states occurs primarily from the ends, and in a pattern that can be collected reasonably well with conventional optics. Note that the question of efficient excitation does not come up with selectively radiant states, as by definition they are well-coupled to a mode of interest.

Finally, we would like to stress that the exploration of both subradiance and selective radiance is not restricted to atoms. Molecules \cite{FTH14} and solid-state emitters should also exhibit these properties, although they pose a different set of challenges. The ability of deterministically placing quantum dots \cite{LSJ08,YTB15,LMS15}, rare earth ions \cite{KXK16,JGW16}, and color centers \cite{SES16} has significantly improved in the past years, thus putting ordered arrays within reach. However, one of the main appeals of atoms is that they are identical to each other and that their decay is purely radiative. An open question is how to translate these features into the domain of solid-state emitters, as it would require high homogeneity among them as well as a large emission into the zero-phonon line (minimizing non-radiative losses).

\subsection{Atomic level structure}
There are a number of potential imperfections that could limit the observation of subradiance and selective radiance, and the performance of protocols that exploit them. A number of these imperfections are conceptually clear (if not necessarily straightforward to analyze), such as disorder in atomic positions, classical and quantum motion, dephasing, and imperfect site filling. A more thorough investigation of these effects will be left to future work.

Here, we would like to discuss a more subtle issue, related to the complications associated with multi-level atomic structure. For most of this manuscript, we have assumed that atoms are two-level systems, with a single ground state and excited state. For atoms with hyperfine structure (and thus a ground state manifold), an effective two-level system is often achieved by exploiting a cycling transition \cite{MV99}, where an excited state of maximum angular momentum can only decay back into a single ground state, also of maximum angular momentum. Such a transition only responds to pure circularly polarized light. In the case of multiple atoms, an important issue is that light scattered from one atom does not display circular polarization globally in space. Thus, for example, the resulting dipole-dipole interactions can potentially drive other atoms to excited states outside of the cycling transition. This can be avoided in the specific case of a 1D chain (where the re-scattered field has the same polarization along the axis of the chain), but not in general.

Another possibility to avoid the full complexity of hyperfine structure is to use atoms without nuclear spin, such as bosonic Ytterbium or Strontium \cite{XLH03,SSK13}. In this case, there is a single ground state but three excited states with orthogonal dipole matrix elements (giving an isotropic optical response to the atoms). Then, one can exploit the fact that in 1D arrays, dipole-dipole interactions involving different excited states decouple from each other, to effectively yield two-level physics (similar to the case of circular polarization described above). In 2D arrays, the transitions involving a dipole matrix element out of the plane decouple, while the two in-plane transitions can hybridize, and calculating the band structure for an infinite system involves diagonalizing a $2\times 2$ matrix associated with the Fourier transform of the in-plane components of the Green's function, $\tilde{G}_{0,\alpha\beta}(\textbf{k})$, with $\{\alpha,\beta\}=\{y,z\}$. This solution qualitatively maintains the same properties as the two-level case analyzed in Sec. III [for example, see the discussion surrounding Eq.~\eqref{rate2D}].

In the presence of hyperfine structure, and excluding the special case of a 1D array described previously, the complication with regard to subradiance can be understood with the following simple example. Suppose that atoms are initialized in a single ground state $\ket{g}$, from which a single-excitation spin wave of the form $\ket{\psi}\sim \sum_j e^{\ii kz_j} \ket{e_j}$ is somehow generated. If the wave vector $k$ is beyond the light line, then as argued in Sec. III, collective dissipative interactions [such as those encoded in the $\heg^{i}\hge^{j}$ term of Eq.~\eqref{ham}] will suppress emission of an excited state back into $\ket{g}$ through destructive interference. However, dipole-dipole interactions will also generally exist between that excited state and any other ground state $\ket{s}$ connected by a dipole-allowed transition, \textit{e.g.}, of the form $\hat{\sigma}_{es}^{i}\hat{\sigma}_{se}^{j}$. Since the initially prepared spin wave $\ket{\psi}$ does not contain any population in $\ket{s}$, there is no interference that prevents decay via this channel, and thus the spin wave would experience a decay rate into $\ket{s}$ equal to that of a single, isolated atom excited to $\ket{e}$. Interesting recent work suggests that it is possible to encode subradiance in a more complex initial state beyond the simple product state $\ket{g}^{\otimes N}$ \cite{HKO17}, which would be worth exploring further. 

Within the context of the enhancement of EIT-based storage protocols through collective emission, studied in Sec. IV, this implies that the state $\ket{s}$ cannot be another state in the ground-state manifold that is directly connected to $\ket{e}$ by a dipole-allowed transition. Various possibilities to implement EIT and retain the desired collective interference effects include the use of a ladder scheme, with the state $\ket{s}$ being a long-lived excited state (\textit{e.g.}, a Rydberg level), or to use a state $\ket{s}$ in the ground-state manifold that is connected only through a two-photon transition \cite{ECZ97,PC08}.

\section{Summary and outlook}

In summary, we have shown that subradiant states acquire an elegant interpretation in 1D and 2D atomic arrays, in terms of optically guided modes whose decay rate is only limited by the system boundaries. We have provided a first glimpse into the nature of subradiance for multiple excitations, and introduced a new concept of selective radiance that should enable the construction of more efficient atom-light interfaces. As a concrete example, we have constructed a protocol for quantum memories for light using selectively radiant states in an optical nanofiber, whose infidelity decreases with atom number at a rate exponentially better than previously known bounds.

Even though memories are a very relevant quantum technology, the improvement in their performance is just an example of the bountiful possibilities spawned by subradiance and selective radiance. We anticipate that exploiting these phenomena could yield new error bounds and protocols for many applications of interest, ranging from nonlinear optics to metrology. At the same time, the nature of subradiance for multiple excitations or internal states could itself constitute a rich new many-body problem. 

\vspace{10pt}
\textbf{Acknowledgments.}-- We are grateful to C. Regal, A. M. Rey, A. Gorshkov, E. Polzik, S. Yelin, M. Lukin, H. Ritsch, E. Shahmoon, and J. Muniz for stimulating discussions. H.J.K. funding is provided by the AFOSR Quantum Many-Body Physics with Photons and QuMPASS MURI, NSF Grant PHY-1205729, the Office of Naval Research (ONR) Award N00014-16-1-2399; the ONR QOMAND MURI; and the IQIM, an NSF Physics Frontiers Center. A.A.-G. was supported by the IQIM Postdoctoral Fellowship and the Global Marie Curie Fellowship LANTERN (655701). D.E.C. acknowledges support from Fundaci\'{o} Privada Cellex, Marie Curie CIG ATOMNANO, Spanish MINECO Severo Ochoa Programme SEV-2015-0522, MINECO Plan Nacional Grant CANS, CERCA Programme/Generalitat de Catalunya, and ERC Starting Grant FOQAL.

\appendix
\section{Analytical expressions for the infinite lattice case}
\label{AppA}
In this appendix we provide a formal derivation for the collective frequency shifts and decay rates of an infinite 1D linear and 2D square lattice of two-level atoms. As discussed in the main text, in an infinite lattice the effective Hamiltonain $\mathcal{H}_\textrm{eff}$ has discrete translational invariance and it can be written as:
\begin{align}
\mathcal{H}_{\rm eff} = \hbar \sum_{\kb} (J_\kb - \ii \Gamma_\kb/2 ) \, S^\dagger_\kb S_\kb.
\end{align}
Here $S^{(\dagger)}_\kb = N^{-1/2} \sum_j e^{-(+)\ii \kb \cdot \rb_j} \hat{\sigma}^j_{ge(eg)}$ represents the annihilation (creation) operator of one of the collective modes, which in this case  corresponds to an atomic spin wave with momentum $\kb$ (defined within the first Brillouin zone). The quantites $J_\kb$ and $\Gamma_\kb$ are real and correspond to the collective frequency shift and decay rate of the mode $\kb$, calculated by taking the discrete Fourier transform of Eq.(\ref{rates}), that is
\begin{subequations}
\begin{align}
J_\kb & =  \sum_{\rb_i - \rb_j} e^{\ii \kb \cdot (\rb_i-\rb_j)} J_{ji}\label{Jkdef},
\end{align}
\begin{align}
\Gamma_\kb &=  \sum_{\rb_i - \rb_j} e^{\ii \kb \cdot (\rb_i-\rb_j)} \Gamma_{ji}. \label{Gammakdef}
\end{align}
\end{subequations}

\subsection{Collective frequency shifts}
Here we evaluate the collective frequency shifts for an infinite 1D chain of atoms along the $\hat{z}$ direction. The collective frequency shifts when the atoms are polarized along or transversally to the chain (denoted by $J^{||}_{k_z}$ and $J^{\perp}_{k_z}$, respectively) are given by:
\begin{subequations}
\begin{align}
\frac{J_{k_z}^{\pl}}{\Gamma_0} &=\frac{3}{2 k_0^3 d^3}\textrm{Re} \sum_{\ell =\pm 1}^{\pm \infty} \frac{e^{\ii k_0 d |\ell|}}{|\ell|^3} e^{\ii k_z d \ell} \left(-1+\ii k_0 d |\ell|\right)  \\
\frac{J_{k_z}^{\perp}}{\Gamma_0} &= \frac{3}{4  k_0^3 d^3}\textrm{Re} \sum_{\ell =\pm 1}^{\pm \infty} \frac{e^{\ii k_0 d |\ell|}}{|\ell| ^3} e^{\ii k_z d \ell} \left(1 - \ii k_0 d |\ell| -  k_0^2 d^2 |\ell| ^2 \right).
\end{align}
\end{subequations}
Using the representation of the PolyLogarithmic function as an infinite sum: $\Li_s (z) = \sum_{\ell=1}^\infty z^\ell \, \ell^{-s}$ [for $z\in \mathbb{C}$, and $\Li_{s=1} (z)=-\ln(1-z)$], the previous expressions reduce to Eqs.(\ref{Eq:ShiftAnal1},\ref{Eq:ShiftAnal2}) in the main text. 

\subsection{Collective decay rates}
In order to compute the Fourier transform $\tilde{\GG}_0(\kb)$ it is useful to express the free space Green's tensor in terms of the spherical wave function:
\begin{equation}
\GG_0(\rb) = \left( k_0^2\mathbb{1} + \boldsymbol{\nabla} \otimes \boldsymbol{\nabla} \right) \, \frac{e^{\ii k_0 r}}{4\pi k_0^2 r}
\end{equation}
and make use of the spherical wave decomposition into plane waves: 
\begin{equation} 
\frac{e^{\ii k_0 r}}{r} = \frac{\ii}{2\pi} \int d^2\Qb_\pl \, \frac{1}{Q_x} \,e^{\ii \Qb_\pl \cdot \rb_\pl }e^{\ii Q_{x} |x|},
\end{equation}
with $Q_x = \sqrt{k_0^2-\Qb^2_\pl}$. Here, we have chosen the axis $\hat{x}$ to be perpendicular to the array, while the  components $\rb_{||}$ are parallel to a plane that contains the atomic chain (in 1D) or the atomic array (in 2D). Then, the free space Green's tensor can be expressed as
\begin{align}
\GG_0 (\rb)&= \frac{\ii}{8\pi^2 k_0^2}  \int d^2 \Qb_\pl \frac{(k_0^2\mathbb{1} - \bar{\Qb} \otimes \bar{\Qb})}{Q_x} e^{\ii \Qb_\pl \cdot \rb_\pl }e^{\ii Q_{x} |x|},\label{Tensor1}
\end{align}
where we have defined $\bar{\Qb}\equiv (Q_x \textrm{sign}(x),Q_y,Q_z)$. For 1D and 2D lattice geometries we can choose that the corresponding line or plane of atoms sits at $x=0$. Then, since we are only interested in evaluating $\GG_0$ at the atomic positions, we can set $x\rightarrow 0$ in the above expressions. Making use of the Dirac delta representation in $\textrm{D}$ dimensions,
 \begin{align}
\sum_{\rb_i \in \text{lattice}} e^{\ii \Qb \cdot \rb_i} &= \left(\frac{2\pi}{d}\right)^{\rm D} \sum_{\substack{\gb \in \text{reciprocal} \\ 
\text{lattice}}} \delta^{(\textrm{D})} (\Qb-\gb),\label{Eq:delta}
\end{align}
it is possible to express $\tilde{\GG}_0(\kb)$ as a sum over reciprocal lattice vectors $\gb$. For a 1D linear chain along $\hat{z}$, 
\begin{align}
\tilde{\GG}_0(\kb) = \frac{\ii}{8\pi^2}\frac{2\pi}{k_0^2 d}\sum_{\gb}  \int dQ_y  \frac{1}{q_x} \left[k_0^2\mathbb{1} -\qb \otimes  \qb \right],
\label{Gk_1D}
\end{align} 
where we have defined $\qb\equiv (q_x \textrm{sign}(x),Q_y,k_z+g_z)$, with $q_x= \sqrt{k_0^2-(k_z+g_z)^2-Q_y^2}$. In the case of a 2D square lattice in the $\hat{y}$-$\hat{z}$ plane this reads:
\begin{align}
\tilde{\GG}_0(\kb) = \frac{\ii}{8\pi^2}\left( \frac{2\pi}{k_0 d} \right)^2 \sum_{\gb}  \frac{1}{q_x} \left[k_0^2\mathbb{1}-\qb \otimes \qb \right],
\label{Gk_2D}
\end{align} 
with $\qb\equiv (q_x \textrm{sign}(x) , k_y+g_y, k_z+g_z)$ and $q_x = \sqrt{k_0^2-|\kb + \gb|^2}$.

Eqs.(\ref{Eq:Decay2D_parallel}) and (\ref{Eq:Decay2D_trans}) are ill-defined for the crossed components $xy$ and $xz$ of $\tilde{\GG}_0(\bf k)$. However, we note that $G_{0,x \alpha}(\rb_\perp, x \rightarrow 0) = 0$ ($\alpha=y,z$), since the electromagnetic field emitted by a dipole is always parallel to the dipole itself, at any point of its normal plane. Thus, also $\tilde{G}_{0,x \alpha}(\kb) = 0$ ($\alpha=y,z$). The fact that this crossed term vanishes is relevant, since it implies that the modes with transverse and in-plane polarization will not be mixed when dealing with multi-level atoms. This is true specifically for 1D and 2D lattices.\\

From Eq.(\ref{Gammakdef}) and Eqs.(\ref{Gk_1D},\ref{Gk_2D}) the collective decay rates can be evaluated. We first note that for a wave vector beyond the light line (\textit{i.e.}, $|\kb| > k_0$), $q_x$ is purely imaginary for any reciprocal lattice vector $\gb$. Thus, the imaginary part of all diagonal tensor components vanishes, and the decay rates are exactly zero. This mathematically demonstrates that any state beyond the light line is necessarily subradiant. In order to have states satisfying this condition, the maximum magnitude of the wave vectors defined in the first Brillouin zone must be larger than the one defining the light line, \textit{i.e.}, $k^\textrm{max} > k_0$. In a linear chain in 1D, one has $k^\textrm{max} = \pi / d$, and this sets the condition $d/\lambda_0 < 1/2$ for the existence of these states. In a 2D square lattice, for which the first Brillouin zone is a square, $k^\textrm{max} = \sqrt{2} \pi / d$, yielding instead the condition $d/\lambda_0 < 1/\sqrt{2}$.

For states with wave vector $|\kb| \leq k_0$ the only contribution in the decay rate is from reciprocal lattice vectors satisfying $|\kb + \gb| \leq  k_0$. In the 1D case we obtain, for parallel and transverse polarization, and after performing the integral in Eq.(\ref{Gk_1D}):
\begin{subequations}
\begin{align}
\frac{\Gamma_{k_z}^{||}}{\Gamma_0} &=\frac{3\pi}{2 k_0 d} \sum_{\substack{g_z \\ |k_z+g_z| \leq k_0}}\left(1-\frac{(k_z+g_z)^2}{k_0^2} \right),
\end{align}
\begin{align}
\frac{\Gamma_{k_z}^\perp}{\Gamma_0} &=\frac{3\pi}{4 k_0 d}  \sum_{\substack{g_z \\ |k_z+g_z| \leq k_0}} \left(1+\frac{(k_z+g_z)^2}{k_0^2} \right).
\end{align}
\end{subequations}
For the 2D square lattice one gets:
\begin{subequations}
\begin{align}
\frac{\Gamma_{\kb}^{||}}{\Gamma_0} &=\frac{3\pi}{ k_0^3 d^2}  \sum_{\substack{\gb \\ |\kb+\gb| \leq k_0}} \frac{k_0^2 - |(\kb+ \gb)\cdot \dbu|^2}{ \sqrt{k_0^2 -|\kb + \gb|^2}},
\end{align}
\begin{align}
\frac{\Gamma_{\kb}^\perp}{\Gamma_0} &=\frac{3\pi}{k_0^3 d^2}  \sum_{\substack{\gb \\ |\kb+\gb| \leq k_0}}  \frac{|\kb + \gb|^2}{ \sqrt{k_0^2 -|\kb + \gb|^2}}.
\end{align}
\end{subequations}

\section{Transfer Matrix Formalism}
\label{AppB}
In Sec.~\ref{SecIIIB}, we found that a linear chain of $N$ atoms has a set of subradiant single-excitation modes, whose decay rates scale like $\Gamma_{\xi} \sim \xi^2/N^3$. Here $\xi=1,2,3,...$ serves as an index for the subradiant modes. 

Here we present a simple one-dimensional model of light interacting with a periodic system of scatterers. It is important to note that one cannot establish a formal mapping from the original system to this one. Heuristically, however, one might hope that the simple model is sufficient to capture the salient features of a generic pseudo-1D system. In particular, we will find that the simple model also produces a set of resonances, whose decay rates scale like $\Gamma_{\xi} \sim \xi^2/N^3$.

\begin{figure}[t]
\centerline{\includegraphics[width=\linewidth]{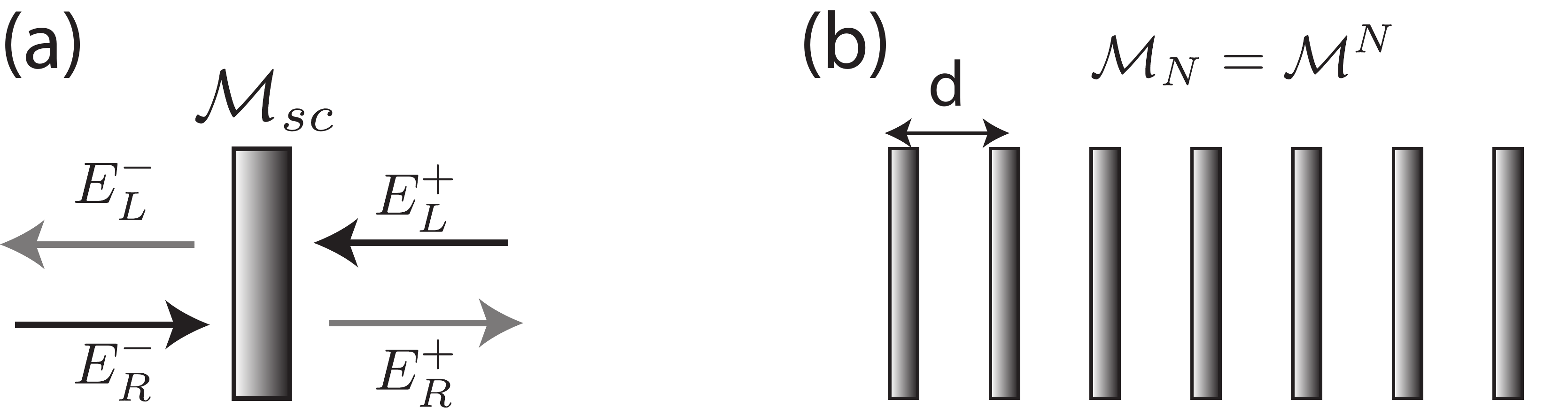}}
\caption{{\bf (a)} The transfer matrix $\mathcal{M}_{sc}$ relates the fields on one side and the other of the scatterer. The coefficients of $\mathcal{M}_{sc}$ are determined by imposing continuity on the fields. {\bf (b)} For a periodic array of scatterers the total transfer matrix is simply the product of matrices $\mathcal{M}_N =\mathcal{M}^N$.} \label{fsFigApp1}
\end{figure}

One-dimensional scattering through several optical elements (such as an array of scatterers) can be efficiently described using the Transfer Matrix formalism \cite{BDS96,JMW95}. This method takes advantage of the fact that in a one-dimensional scattering model there are only two propagation directions (left and right). The transfer matrix $\mathcal{M}_\textrm{sc}$ (see Fig.~\ref{fsFigApp1}) relates the fields on one side ($E_R^-$, $E_L^-$) and on the other side ($E_R^+$, $E_L^+$) of a point scatterer:
\begin{align}
\left( \begin{array}{c}
E_R^+ \\
E_L^+
\end{array} \right) = 
\mathcal{M}_\textrm{sc} \left( \begin{array}{c}
E_R^- \\
E_L^-
\end{array} \right).
\end{align}
Propagation through a unit cell can also be described via a transfer matrix, which itself is a product of transfer matrices describing interaction with the point scatterer (described by reflection and transmission coefficients $r$ and $t$), and the one-dimensional free-space propagation at frequency $\omega=c k$ over a distance $d$. That is, $\mathcal M = \mathcal{M}_\text{sc} \cdot \mathcal{M}_\text{free}$, with
\begin{subequations}
\begin{align}
\mathcal{M}_\textrm{sc}&=
\frac{1}{t} \left( \begin{array}{cc}
t^2-r^2 & r \\
-r & 1\\
\end{array} \right) = \left( \begin{array}{cc}
1+\ii\zeta & \ii\zeta \\
-\ii\zeta & 1-\ii\zeta\\
\end{array} \right), \label{Eq:M_sc} 
\end{align}
\begin{align}
\mathcal{M}_\textrm{free}&=
\left( \begin{array}{cc}
e^{\ii \omega d /c} & 0\\
0 & e^{-\ii \omega d /c}
\end{array} \right),
\end{align}
\end{subequations}
and $\zeta = -\ii r/t$. The relation of $\mathcal{M}_\textrm{sc}$ to $\zeta$ is determined by the additional constraint that $1+r=t$, which states that the field should be continuous on each side of the point scatterer.

It is useful to decompose the matrix $\mathcal M$ as:
\begin{equation}
\mathcal M = e^{\ii q d \mathcal{A} } = \cos(qd) \mathbb{1} + \ii\sin(qd) \mathcal{A} \label{Eq:M_dec}
\end{equation}
with 
\begin{align}
\mathcal{A}&=
 \left( \begin{array}{cc}
c / v_g  & \frac{\zeta}{\sin (qd)}  e^{-\ii \omega d/c} \\
-\frac{\zeta}{\sin (qd)}  e^{\ii \omega d/c} & -c / v_g\\
\end{array} \right),
\end{align}
since $\textrm{Tr} \mathcal A = 0$ and $\mathcal{A}^2=\mathbb{1}$. Here $v_g$ is the group velocity for an infinite lattice, and it will be given by Eq.(\ref{Eq:GroupVel}).\\

{\bf \textit{Dispersion relation.--}} The two eigenvalues that diagonalize the transfer matrix $\mathcal{M}$ are necessarily of the form $\lambda_{\pm}=e^{\pm \ii q d}$, since $\det \mathcal{M} = 1$. As we will see, $\pm q$ corresponds to the Bloch index or quasi-momentum. Moreover, since the trace of $\mathcal M$ is independent of the basis, one can obtain the dispersion relation (the relationship between $\omega$ and $q$):
\begin{align}
\frac{1}{2} \textrm{Tr} \mathcal{M} = \cos (qd) = \cos(\omega d / c) -\zeta \sin(\omega d / c). \label{Eq:Dispersion}
\end{align}

Given the quasi-momentum value $|q|$, there are two possible solutions (or branches) that fulfill this relation. In particular,
\begin{align}
\frac{\omega d}{c} &=  \notag \\
\cos^{-1} &\left[\frac{\cos(qd)}{1+\zeta^2} \left[1\pm \sqrt{1-(1+\zeta^2)(1-\zeta^2 / \cos^2(qd)) } \right]\right].
\end{align}
Moreover, from Eq.(\ref{Eq:Dispersion}) the group velocity for an infinite system can be derived:
\begin{align}
v_g \equiv \frac{d\omega}{d q} =  \frac{c \sin(qd)}{\sin(\omega d/c) +\zeta \cos(\omega d /c) }. \label{Eq:GroupVel} 
\end{align}\\

{\bf \textit{Group velocity close to the band edge.--}} 
At the band edge of the Brillouin zone ($q = \pi / d$) the dispersion relation exhibits a band gap, with the lower ($\omega_{0-}$) and upper ($\omega_{0+}$) frequencies of the gap given by:
\begin{subequations}
\begin{align}
\omega_{0+} d / c &=\pi, 
\end{align}
\begin{align}
\omega_{0-} d/c &= \cos^{-1} \left(\frac{\zeta^2-1}{\zeta^2+1}\right).
\end{align}
\end{subequations}
Near the band edge the dispersion relation can be approximated by 
\begin{equation}
\omega \sim \omega_{0 \pm} \mp \frac{c}{2\zeta d} (\pi-qd)^2, \label{Eq:Dispersion_Quad}
\end{equation}
and the group velocity can be identified as $v_g = \mp c (\pi-qd) /\zeta$.\\

{\bf \textit{Finite array. Transmission coefficient and resonances.--}} 
In an ordered array of $N$ point scatterers (separated by the distance $d$) the total transfer matrix is simply the product of matrices $\mathcal{M}_N = \mathcal{M}^N$. Thus, the eigenvectors of $\mathcal M$ are also eigenvectors of $\mathcal{M}_N$. Eq.(\ref{Eq:M_dec}) is very useful to compute $\mathcal{M}_N = e^{\ii N qd \mathcal{A}} = \cos(Nqd) \mathbb{1} + \ii \sin{(Nqd)} \mathcal{A}$ \cite{DSR95}.

As any transfer matrix, $\mathcal{M}_N$ can be written as:
\begin{align}
\mathcal{M}_N&=
\frac{1}{t_N} \left( \begin{array}{cc}
t_N^2-r_N^2 & r_N \\
-r_N & 1\\
\end{array} \right),
\end{align}
where now $t_N$ and $r_N$ represent the reflection and transmission coefficients throughout the whole array. Thus, one can obtain the transmission and reflection coefficients from the elements $\mathcal{M}_N^{22}$ and $\mathcal{M}_N^{12}$:
\begin{subequations}
\begin{align}
t_N^{-1}&= \cos(Nqd) + \ii \sin(Nqd) (c /v_g),
\end{align}
\begin{align}
r_N &= \frac{\ii \zeta \sin(Nqd) e^{-\ii \omega d /c}}{\sin(qd)} t_N.
\end{align}
\end{subequations}
The previous expressions allow us to identify resonances of the finite array. Indeed, for particular values of the Bloch index, namely $q_\xi d = \pi (N-\xi) /N$ (where $\xi$ is an integer number), the transmission coefficient $t_N(q_\xi) \rightarrow (-1)^{N-\xi}$. That is, the transmission probability through the array is maximum and equal to one.

It is also interesting to analyze the physics close to those resonances. As discussed, the transmission spectra exhibits peaks at each value $q_\xi$. Around each peak $\xi$, it can readily be shown that the transmission spectrum behaves approximately as a Lorentzian, $t_N \approx (-1)^{(N-\xi)} (\ii\Gamma_\xi/2)/(\ii \Gamma_\xi/2 - \delta\omega_\xi)$, where $\delta\omega_\xi$ denotes the detuning from the resonance frequency of mode $\xi$ and $\Gamma_\xi$ its linewidth. For small values of $\xi$, one finds approximately that 
\begin{align}
\Gamma_{\xi} \sim \frac{2 \xi^2 \pi^2 c}{\zeta^2 N^3 d}.
\label{Eq:Gamma_Scaling}
\end{align}

\section{Green's function of a nanofiber}\label{AppC}
In this section we provide the expressions for the radial components of the Green's function of an infinite nanofiber of radius $r$ directed along $\hat{z}$, following Klimov and Ducloy~\cite{KD04}. The total field produced by a dipole near a fiber can be expressed in terms of a free field (solution in vacuum) and a field re-scattered by the fiber. To exploit separation of variables, the free field can be written in cylindrical coordinates as an expansion in longitudinal wave vector $k_{\parallel}$ and angular momentum $e^{im\phi}$. The full Green's function $\textbf{G}(\rb,\rb',\omega_0)$ is a rather complicated expression. Here, we are interested in the case where all atoms sit at identical distances from the fiber and a fixed azimuthal angle around the cylinder. Since we only need the Green's function at the atomic positions themselves, we can construct a simplified version of the scattering Green's function, only evaluated at the atomic positions $z_j$:
\begin{align}\label{Grr}
G_{\rho\rho}(z_j,z_k)=\frac{1}{4\pi k_0^2}\sum_{m=-\infty}^{\infty}\int_{-\infty}^{\infty} dk_{\parallel} \,\tilde{G}_m(k_{\parallel}) \,e^{\ii k_{\parallel} (z_j-z_k)},
\end{align}
where 
\begin{equation}
    \begin{split}
    \tilde{G}_m(k_{\parallel})=\frac{1}{k_\perp^2}\left[\ii k_{\parallel}k_\perp\partial_x H^{(1)}_m(x) \,a_m(k_{\parallel})\right. \\
    \left.-\frac{k_0 m}{\rho_a}H^{(1)}_m(x) \,b_m(k_{\parallel})\right]_{x=k_\perp\rho_a}.
    \end{split}
\end{equation}
In the above expression, $\rho_a>r$ is the radial position of the atoms (assumed to be identical), $k_\perp=\sqrt{k_0^2-k_{\parallel}^2}$ is the perpendicular component of the wave vector outside the fiber, and $H^{(1)}_m$ is the Hankel function of the first kind. The coefficients $\,a_m(k_{\parallel})$ and $\,b_m(k_{\parallel})$ are found by matching boundary conditions for the electromagnetic field at the surface of the fiber, and can be taken from Eqs.~(47)-(50) in Ref.~\cite{KD04} (by choosing the value of the dipole moment equal to unity). 

\begin{figure}
\centerline{\includegraphics[width=\linewidth]{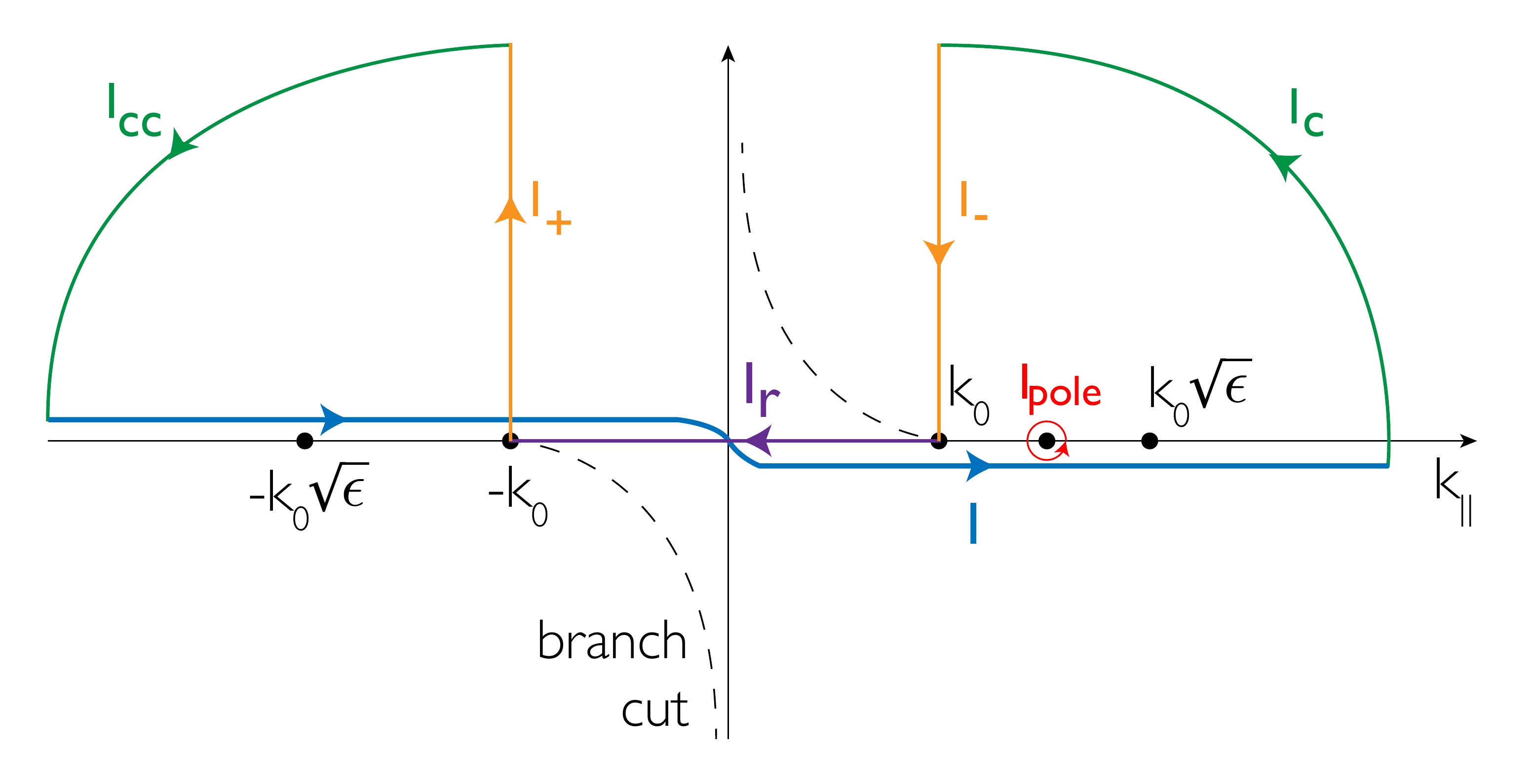}}
\caption{Integration contour for Eq.~(\ref{Grr}) depicting the pole and branch cuts.} \label{contour}
\end{figure}

In the following, we detail how to perform the integral in Eq.~(\ref{Grr}). We only need to do it for the case where $z_j>z_k$, as $G_{\rho\rho}(z_j,z_k)=G_{\rho\rho}(z_k,z_j)$, due to Lorentz reciprocity. In the complex plane, the integrand $\tilde{G}_m(k_{\parallel})$ has two branch cuts (due to the square root form of $k_\perp$) and two poles, that correspond to the guided mode of the fiber (for a small enough nanofiber, all the guided modes with $|m|\neq1$ are cut-off). This guided mode, the so-called HE$_{11}$\cite{YS08_2}, does not have a cut-off frequency and corresponds to the $m=\pm 1$ pole. In particular, the variation in the position of the pole with $\omega_0$ gives rise to the dispersion relation. In order to avoid the poles when integrating along the real line, we perform a contour integration and employ Cauchy's theorem. The contour that we choose is shown in Fig.~\ref{contour}. Based on this image, we have $I=I_{\rm pole}-I_r-I_+-I_--I_{c}-I_{cc}$, where $I$ is given by Eq.~(\ref{Grr}), and the other integrals are performed along the contours of Fig.~\ref{contour}. After performing the integrals, we find that the total Green's function can be separated into guided, $G^{\rm 1D}$, and non-guided, $G'$, contributions as
\begin{widetext}
\begin{subequations}\label{Gguided}
\begin{align}
&G^{\rm 1D}_{\rho\rho}(z_j,z_k)=\frac{e^{\ii k_{\rm 1D} (z_j-z_k)}}{2\pi k_0^2}\oint_{C_{pole}} dk_{\parallel} \,\tilde{G}_1(k_{\parallel}),\\\
&G'_{\rho\rho}(z_j,z_k)=G_{0,\rho\rho}(z_j,z_k)+\frac{1}{4\pi k_0^2}\sum_{m=-\infty}^{\infty}\left[\int_{-k_0}^{k_0} dk_{\parallel} \,\tilde{G}_m(k_{\parallel}) \,e^{\ii k_{\parallel} (z_j-z_k)}+2\int^{\infty}_0  d\gamma \,\text{Im}\left\{\tilde{G}_m(-k_0+\ii\gamma) \,e^{\ii (-k_0+\ii\gamma) (z_j-z_k)}\right\}\right.\\\nonumber
&\left.-2\text{Re}\left\{\int_{C_{cc}}  dk_{\parallel} \,\tilde{G}_m(k_{\parallel}) \,e^{\ii k_{\parallel} (z_j-z_k)}\right\}\right].
\end{align}
\end{subequations}
\end{widetext}
In the above expressions, $k_{\rm 1D}$ is the wave vector of the guided mode, $G_{0,\rho\rho}(z_j,z_k)$ is the vacuum's Green's function, and $C_{pole}$ and $C_{cc}$ are the red and left-most green contours in Fig.~\ref{contour}, respectively. For the non-guided Green's function, the integral $I_r$ produces both frequency shifts and decay, whereas the integrals $I_{\pm}, I_\text{c},I_\text{cc}$ only contribute to the frequency shifts. The integrals $I_\text{c}+I_\text{cc}$ do not contribute for $z_j\neq z_k$. For the local Green's function (\textit{i.e.}, $z_j= z_k$), both $I_++I_-$ and $I_\text{c}+I_\text{cc}$ are divergent individually, but the infinity is cancelled when they are added. 

\section{Linear optics for two-level atoms in the mirror configuration}\label{AppD}
\begin{figure*}
\centerline{\includegraphics[width=\linewidth]{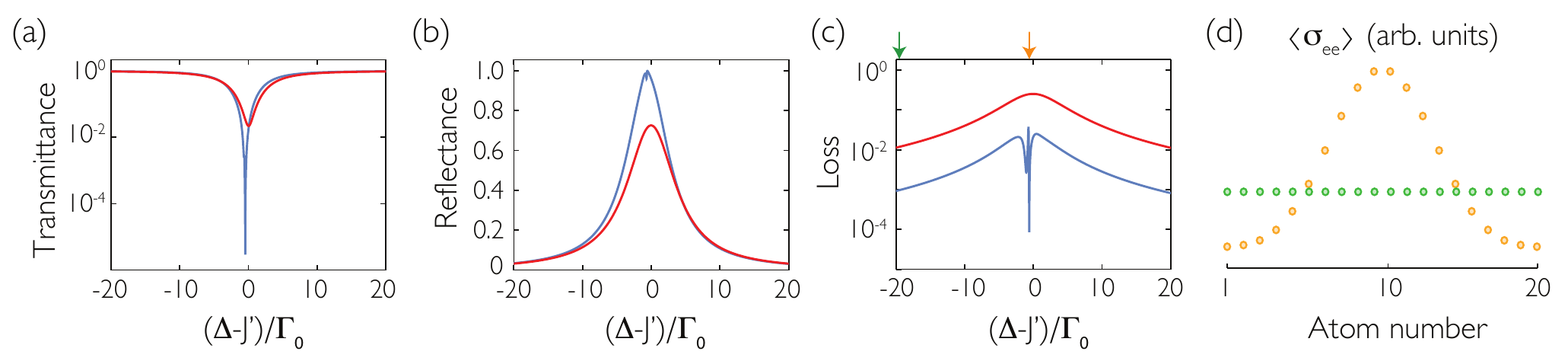}}
\caption{Linear optics for a chain of $N=20$ atoms coupled to a nanofiber, in the mirror configuration ($k_\text{1D}d=\pi$). \textbf{(a)} Transmittance, \textbf{(b)} reflectance, and \textbf{(c)} loss probability as a function of the atom-probe detuning, within the collective (blue curves) and independent emission (red curves) models. \textbf{(d)} Spatial profile of the excited state population (in arbitrary units) at two different detunings, as indicated by the arrows in (c). The green circles have been re-scaled for the sake of clarity. The parameters characterizing the nanofiber are given in Fig.~\ref{nFig1}.} \label{mirror}
\end{figure*}

Within the independent emission model, and when the atoms are spaced at distances such that $\kg d=n\pi$, with $n$ being an integer, only a single collective atomic mode couples to the fiber. This case constitutes the so-called ``mirror configuration", as it has been shown that the ensemble behaves as a nearly perfect mirror with increasing atom number $N$. As we will show in this section, the resultant physics when suppression of emission into non-guided modes is accounted for becomes significantly more complicated. In particular, while one can dramatically enhance the reflectance of the atomic chain, we find that the atoms no longer respond as a single mode, and that the impedance mismatch between the atomic chain and the photonic guided mode of the fiber is a major issue. 

For atoms spaced by a distance $\kg d=2\pi$, the guided Hamiltonian of Eq.~\eqref{hguided} simply reads
\begin{align}
\mathcal{H}_{\rm 1D}=-\ii\frac{\hbar\ga}{2} \sum_{i,j=1}^N \heg^{i}\hge^j.
\end{align} 
In the single excitation manifold there is only one superradiant mode, with decay rate $N\ga$, while all the others are completely dark (the same physics would be observed for any other separations of the form $\kg d=n\pi$). In this configuration, the transmittance and reflectance coefficients can be found analytically, and read \cite{CJG12}
\begin{subequations}
\begin{align}\label{trans}
&T_\text{indep}=\frac{\Gamma'^2+4(\Delta-J')^2}{(N\ga+\Gamma')^2+4(\Delta-J')^2},
\end{align}
\begin{align}\label{ref}
&R_\text{indep}=\frac{(N\ga)^2}{(N\ga+\Gamma')^2+4(\Delta-J')^2},
\end{align}
\end{subequations}
and the photon-loss probability is $\kappa_\text{indep}=1-T_\text{indep}-R_\text{indep}=2R_\text{indep}\Gamma'/N\ga$. The transmittance spectrum is a Lorentzian whose linewidth $N\ga+\Gamma'$ grows linearly with number of atoms, for sufficiently large $N$. On resonance (when $\Delta-J'=0$), the atomic chain becomes a very good mirror, and the only relevant quantity that determines how much light is reflected is the ratio $D=2N\ga/\Gamma'$, which is in fact the optical depth of the system. In particular, on resonance and in the limit of large optical depth, the transmittance, reflectance, and loss become $T_\text{indep}=4/D^2$, $R_\text{indep}=1-4/D$, and $\kappa_\text{indep}=4/D$, respectively.

For atoms at close distances from each other, the independent emission model is not valid, and one cannot find analytical expressions for the transmission and reflection coefficients. Figures~\ref{mirror}(a-c) show the transmission, reflection, and loss probability spectra of a chain of $N=20$ atoms coupled to the nanofiber, for both the independent emission model (red curves) and the collective emission calculation (blue curves). The differences are striking. For instance, close to resonance, the transmission and the loss decrease about four orders of magnitude when collective suppression into free-space is taken into consideration. 

Repeating these calculations for chains with different number of atoms, we extract the scalings of the minimum transmittance and maximum reflectance within the collective emission model. In particular we find $T\sim 1/N^8$ and $1-R\sim 1/N^6$. Moreover, at the detuning that minimizes emission into free space, the loss scales as $\kappa\sim 1/N^6$. It thus seems apparent that collective emission cannot be captured by some trivial modification of the independent emission model [\textit{e.g.}, one cannot simply replace $\Gamma'$ by some $\Gamma'_\text{eff}(N)$ in Eqs.~(\ref{trans},\ref{ref})]. In other words, the atom-light coupling can no longer be described by a single collective atomic eigenstate, but has become instead a multimode problem. This idea is further confirmed by a careful analysis of the lineshapes of Figs.~\ref{mirror}(a-c). Although hard to appreciate in the figures, the transmittance and reflectance spectra are not smooth Lorentzians due to interference between different atomic eigenstates, all of which contribute to the optical response of the chain.

Beyond the added complexity, there is another important issue to notice: Fig.~\ref{mirror}(c) shows that off-resonant losses are still quite large. We find that far away from resonance, the loss is independent of the atom number. Figure~\ref{mirror}(d) sheds light on the reasons for such behavior, as it compares the excited state profile at the detuning where the loss is minimal [orange arrow and circles in (c) and (d), respectively] and far off resonance, at $\Delta-J'=-20\Gamma_0$ [green arrow and circles in (c) and (d), respectively]. At the detuning where the loss is minimal, the atoms at the end of the chain are negligibly excited, and the atomic population smoothly increases toward the middle of the chain. In other words, the response of the atoms to the incoming field appears ``impedance matched," in that the smooth excitation profile reduces the amount of spin-wave components that sit within the light line and couple to free-space radiation. In contrast, far off resonance, the prepared state mostly builds upon a single eigenstate with a very uniform spatial profile. This further supports the argument in the main text, that the response of light at the interface between the atoms and the bare fiber plays an important role in the observed scattering losses.

\section{Polariton model}\label{AppE}
In this appendix, we develop a model for the dynamics of the dark- and bright-state polaritons where losses into free space are completely neglected. We extend previous theory for EIT in continuous atomic media~\cite{FL02}, in order to study the effect of spatially-dependent control fields on the dynamics of the bright polariton. Instead of focusing on the spin-model equations only, and reconstructing fields using an input-output equation, we return to explicitly keeping track of the wave equation of the electric field as it propagates through the fiber.  Employing continuous atomic operators (denoted by tildes), the equations of motion are:
\begin{align}
&\partial_t\tilde{\sigma}_{ge}=-\frac{\ga}{2}\tilde{\sigma}_{ge}+\ii\Omega_c\tilde{\sigma}_{gs}+\ii\sqrt{\frac{c\ga}{2}}E,\\
&\partial_t\tilde{\sigma}_{gs}=\ii\Omega_c\tilde{\sigma}_{ge},\\
&\left(\partial_t+c\partial_z\right)E=\ii n\sqrt{\frac{c\ga}{2}}\tilde{\sigma}_{ge}.
\end{align}
In the above equations, $n=1/d$ is the smoothed-out linear density associated with atoms spaced at distance $d$, and a phase $e^{\ii \kg z}$ has been incorporated into the field and atomic coherence operators to make them slowly varying in space. All operators depend on $z$ and $t$, and $\Omega_c=\Omega_c(z)$ is taken to be real. We now respectively define the dark and bright-state polaritons as 
\begin{align}
&\Psi=\cos\theta E-\sqrt{n}\sin\theta\tilde{\sigma}_{gs},\\
&\Phi=\sin \theta E+\sqrt{n}\cos\theta \tilde{\sigma}_{gs},
\end{align}
where the mixing angle is given by $\tan\theta=\sqrt{c n\ga/2\Omega_c^2}$. In the adiabatic and slow-light ($v_g\ll c$) limits, the equations of motion for these polaritons are
\begin{align}
&\left[\partial_t + v_g(z)\partial_z\right]\Psi(t,z)=\\\nonumber
&-\frac{1}{2}\left[\Psi(t,z)-\Phi(t,z)\right]\partial_z v_g(z)-\sqrt{c v_g(z)}\,\partial_z \Phi (t,z),\\
& \Phi (t,z)=\frac{1}{n\sqrt{c v_g(z)}}\partial_t\Psi(t,z)- \frac{1}{nc}\partial_t\Phi (t,z).\label{darkpol}
\end{align}
We consider that the bright-state polariton only perturbatively affects the dynamics of the dark-state polariton \cite{DCC16}. Therefore, as a first approximation, we can set $\Phi (t,z)=0$ and solve for $\Psi (t,z)$. Then, the equation of motion of the dark-state polariton reads
\begin{align}
\left[\partial_t + v_g(z)\partial_z\right]\Psi(t,z)=-\frac{1}{2}[\partial_z v_g(z)]\Psi(t,z).
\end{align}
Plugging the above expression into Eq.~\eqref{darkpol}, we readily find that the bright-state polariton follows
\begin{align}
\Phi (t,z)=-d\sqrt{\frac{v_g(z)}{c}}\partial_z\Psi(t,z)- \frac{d}{2\sqrt{c v_g(z)}}\Psi (z,t) \partial_z v_g(z).
\end{align}
The formal solution of the equation for the dark polariton is $\Psi(t,z)=\sqrt{c/v_g(z)}\,\tilde{f}\left(t-\int_{0}^z v_g^{-1}(z') dz'\right)$, where $\tilde{f}$ is a function that fulfils the condition $\tilde{f}\left(-\int_{0}^z v_g^{-1}(z') dz'\right)=\sqrt{v_g(z)/c}\,\Psi (0,z)$. In the slow-light limit, the dark-state polariton is nearly a pure spin-wave excitation, and thus corresponds to the spatial profile of the s-state spin wave at t=0. In order to obtain $\tilde{f}$ we need to perform an inversion. How hard this function inversion is depends on the profile of $v_g(z)$ and on the initial dark-state polariton shape. Introducing this expression into the equation for the bright-state polariton, we find
\begin{align}
\Phi (t,z)=-d\partial_z\tilde{f}\left(t-\int_0^z\frac{1}{v_g(z')}dz'\right).
\end{align}

\bibliography{refs_all}

\end{document}